\documentclass[10pt]{article}
\usepackage{fullpage}
\usepackage{setspace}
\usepackage{parskip}
\usepackage{titlesec}
\usepackage[section]{placeins}
\usepackage{cite}
\usepackage{graphicx}
\usepackage{xcolor}
\usepackage{mathtools}
\usepackage{fancyhdr}
\usepackage{subcaption}
\usepackage{amsfonts}
\usepackage{amssymb}
\usepackage{amsmath}
\usepackage[T1]{fontenc}
\usepackage{color}	
\usepackage{siunitx}
\usepackage{multirow}
\usepackage{multicol}
\usepackage{placeins}
\usepackage{booktabs}
\usepackage{rotating}
\usepackage{array}
\usepackage{dsfont}
\usepackage{bbm}
\usepackage{bbold}
\usepackage[hyphens]{url}
\usepackage{tikz}
\usepackage{enumitem}
\usetikzlibrary{backgrounds}
\usetikzlibrary{arrows.meta, positioning, shapes.geometric, calc, decorations.pathmorphing, decorations.markings, patterns}
\newlength{\rw}     
\newlength{\threerw} 
\usepackage{epsfig}
\usepackage[toc,page]{appendix}
\usepackage{pgfplots}
\usepackage{pgfplotstable}
\pgfplotsset{compat=newest}
\usepgfplotslibrary{units} 
\usepgfplotslibrary{external} 
\tikzset{external/mode=graphics if exists}
\usepackage{epsfig}
\usepackage{epstopdf} 
\usepackage{algorithm}
\usepackage{algorithmic}

\newcommand{\mc}[1]{\mathcal{#1}}
\newcommand{\R}{\mathbb{R}}
\newcommand{\N}{\mathbb{N}}
\newcommand{\bs}[1]{\boldsymbol{#1}}
\newcommand{\mf}[1]{\mathbf{#1}}

\newcommand{\tup}[1]{\textup{#1}}

\usepackage[acronym]{glossaries}
\newacronym{DR}{DR}{Demand Response}
\newacronym{DSO}{DSO}{Distribution System Operator}
\newacronym{TSO}{TSO}{Transmission System Operator}
\newacronym{BLO}{BLO}{BiLevel Optimization Problem}
\newacronym{GNE}{GNE}{Generalized Nash Equilibrium}
\newacronym{GNEP}{GNEP}{Generalized Nash Equilibrium Problem}
\newacronym{vGNE}{\textit{v}-GNE}{\textit{variational} Generalized Nash Equilibrium}
\newacronym{lSE}{$\ell$-SE}{local Stackelberg equilibrium}
\newacronym{MPEC}{MPEC}{Mathematical Program with Equilibrium Constraints}
\newacronym{MIP}{MIP}{Mixed-Integer Program}
\newacronym{KKT}{KKT}{Karush–Kuhn–Tucker}
\newacronym{IoT}{IoT}{Internet of Things}
\newacronym{DeePC}{DeePC}{Data-enabled Predictive Control}
\newacronym{MPC}{MPC}{Model Predictive Control}
\newacronym{MFD}{MFD}{Macroscopic Fundamental Diagram }
\newacronym{SUMO}{SUMO}{Simulation of Urban MObility}
\newacronym{CAV}{CAV}{Connected Autonomous Vehicle}
\newacronym{HBEFA}{HBEFA}{Handbook of Emissions Factors for Road Transport }
\newacronym{NC}{No Control}{No Control baseline }
\newacronym{ITSCP}{ITSCP}{Intersection Traffic Signal Control Problem}
\newacronym{SIMOPT}{SimOpt}{Simulation-based Optimization}
\newacronym{TRANSYT}{TRANSYT}{TRAffic Network StudY Tool}
\newacronym{SCOOT}{SCOOT}{Split-cycle and Offset Optimization Technique}
\newacronym{RL}{RL}{Reinforcement Learning}
\newacronym{MAS}{MAS}{Multi Agent System}
\newacronym{row}{r.o.w.}{right of way}
\newacronym{PCA}{PCA}{Principal Component Analysis}
\newacronym{LTI}{LTI}{Linear Time-Invariant}
\newacronym{EVR}{EVR}{Explained Variance Ratios }

\newtheorem{lemma}{Lemma}

\newcommand*{\rom}[1]{\expandafter\@slowromancap\romannumeral #1@}

\newcommand{\Image}{\operatorname{im}} 
\newcommand{\transpose}{\mathsf{T}} 

\newcommand{\col}{\operatorname{col}}
\newcommand{\rank}{\operatorname{rank}} 
\newcommand {\T}{\mathbb{T}} 	
\newcommand {\W}{\mathbb{W}} 	
\newcommand {\B}{\mathcal{B}} 	
\newcommand{\Li}{\mathcal{L}} 	
\newcommand {\bma}{\left[}
\newcommand {\ema}{\right]}

\newcommand{\Tini}{{T_{\textup{ini}}}}
\newcommand{\Tf}{{T_{\textup{f}}}}
\newcommand{\uini}{{u_{\textup{ini}}}}
\newcommand{\yini}{{y_{\textup{ini}}}}
\newcommand{\wini}{{w_{\textup{ini}}}}

\newcommand{\Up}{{U_{\mathrm{p}}}}
\newcommand{\Uf}{{U_{\mathrm{f}}}}
\newcommand{\Yp}{{Y_{\mathrm{p}}}}
\newcommand{\Yf}{{Y_{\mathrm{f}}}}

\newcommand {\nn}{\nonumber}
\newcommand{\beq}{\begin{equation}}
\newcommand{\eeq}{\end{equation}}
\newcommand {\bseq}{\begin{subequations}}
\newcommand {\eseq}{\end{subequations}}

\newlength\myindent
\newcommand\bindent{
  \begingroup
  \setlength{\itemindent}{\myindent}
  \addtolength{\algorithmicindent}{\myindent}
}
\newcommand\eindent{\endgroup}

\PassOptionsToPackage{hyphens}{url}
\usepackage[colorlinks = true,
            linkcolor = blue,
            urlcolor  = blue,
            citecolor = blue,
            anchorcolor = blue]{hyperref}
\usepackage{etoolbox}


\renewenvironment{abstract}
  {{\bfseries\noindent{\abstractname}\par\nobreak}\footnotesize}
  {\bigskip}

\titlespacing{\section}{0pt}{*3}{*1}
\titlespacing{\subsection}{0pt}{*2}{*0.5}
\titlespacing{\subsubsection}{0pt}{*1.5}{0pt}

\usepackage{authblk}

\usepackage{mwe}
\usepackage[space]{grffile}
\usepackage{latexsym}
\usepackage{textcomp}
\usepackage{longtable}
\usepackage{tabulary}
\providecommand\citet{\cite}
\providecommand\citep{\cite}

\newif\iflatexml\latexmlfalse

\AtBeginDocument{\DeclareGraphicsExtensions{.pdf,.PDF,.eps,.EPS,.png,.PNG,.tif,.TIF,.jpg,.JPG,.jpeg,.JPEG}}

\usepackage[utf8]{inputenc}
\usepackage[english]{babel}
\usepackage{float}
\usepackage{authblk}

\pgfplotsset{compat = 1.10}

\begin{document}

\title{Data-driven generalized perimeter control: Z\"urich case study}

\author[1]{Alessio Rimoldi
    \thanks{
    This work was supported as a part of NCCR Automation, a National Center of Competence in Research, funded by the Swiss National Science Foundation (grant number 51NF40\_225155)\\
    A. Padoan acknowledges the support of the Natural Sciences and Engineering Research Council of Canada (NSERC). Grant numbers: RGPIN-2025-06895 and DGECR-2025-00382
    }
}
\author[1,2]{Carlo Cenedese}
\author[3]{Alberto Padoan}
\author[1]{Florian D\"orfler}
\author[1]{John Lygeros}

\affil[1]{Automatic Control Laboratory, ETH Zürich ({\tt\small arimoldi,jlygeros,fdorfler@ethz.ch})}
\affil[2]{Delft Center for Systems and Control, TU Delft ({\tt\small ccenedese@tudelft.nl})}
\affil[3]{Dep. of Electrical and Computer Engineering, UBC ({\tt\small alberto.padoan@ubc.ca})}

\vspace{-1em}
  
\date{February 28, 2026}

\begingroup
\let\center\flushleft
\let\endcenter\endflushleft
\maketitle
\endgroup

\selectlanguage{english}
\begin{abstract}
Urban traffic congestion is a key challenge for the development of modern cities, requiring advanced control
techniques to optimize existing infrastructures usage. 
Despite the extensive availability of data, modeling such complex systems remains an expensive and time 
consuming step when designing model-based control approaches. 
On the other hand, machine learning approaches require simulations to bootstrap models, 
or are unable to deal with the sparse nature of traffic data and enforce hard constraints. 
We propose a novel formulation of traffic dynamics based on behavioral systems theory and apply data-enabled 
predictive control to steer traffic dynamics via dynamic traffic light control. 
A high-fidelity simulation of the city of Zürich,
the largest closed-loop microscopic simulation of urban traffic in the literature to the best of our knowledge, 
is used to validate the performance of the proposed method in terms of total travel time and CO$_2$ emissions. 
\end{abstract}

\sloppy

\section{Introduction}\label{sec:Intro}
As cities worldwide grow at an ever-increasing pace, infrastructures struggle under the stress of increasing demand. 
The United Nations forecasts that by 2050, 68\% of the total world population will live in urban centers~\cite{un2018}. 
Roads remain by far the preferred mode of transport for people and goods both in the United States and the European Union \cite{eutransportfigures},
accounting for 74\% of land freight transport in the European Union during 2021 \cite{transport-in-eu}. 
The combination of these factors will make the problem of reducing urban traffic congestion even more challenging.
Estimates show that road transport generates 12\% of human CO$_2$ emissions and urban air pollution 
has been linked to a host of negative health effects~\cite{aripollution-health}. 
Traffic congestion also leads to economic loss, due to unrealized productivity, valued between 1-2\% of 
the European Union total gross domestic product (80 to 160 billion euros per year)~\cite{eufututure-of-transport}. 
Infrastructural changes aimed at solving this problem are limited in scale by 
prohibitive cost and their effects are difficult to predict due to counterintuitive phenomena such as 
the Braess paradox~\cite{braess-paradox}. For this reason optimization of existing infrastructure through 
traffic management systems is often a preferable solution.

Adaptive traffic signal control is the most widespread form of urban traffic 
management~\cite{traffic-control-review-papageorgiu}, due to its effectiveness and relative
low cost. 
The most common approach is to derive a model of the city's traffic dynamics and then apply model-based control 
algorithms such as \gls{MPC} to compute optimal inputs for the traffic lights \cite{gerolimins:2013:optimal_periter_two_regions}. 
This methodology was proven to be effective \cite{kouvelas-mpc,gerolimins:2013:optimal_periter_two_regions} and easily integrates 
prior knowledge about the system into the model. 
However, modeling such large systems is expensive and time consuming due to the dynamic large-scale nature of a city, 
where the road network is often subject to changes (e.g., road closures) and demand changes over time. 
The availability of data is an opportunity to approach the problem differently. Thanks to data-informed 
decision-making algorithms, it is now possible to identify and exploit complex patterns to efficiently control traffic. 
Data-driven methods are some of the most promising approaches when dealing with large-scale systems as they have been shown to produce robust control 
policies without requiring explicit modeling~\cite{rimoldi2023urban, arel-rl,balaji-rl}.
Current research in this area focuses on machine learning techniques, such as deep reinforcement learning, which have been proven to be 
effective when controlling traffic systems \cite{arel-rl,mas-bazzan}. 
Training models directly on traffic data can nonetheless be challenging due to its sparse nature and the effects of an input being delayed in time.
Deep reinforcement learning is known to require large datasets to produce satisfactory policies. 
Synthetic data produced by simulations can be used, but this requires a model to simulate in the first place. 

Recently, \gls{DeePC} has been applied to the \gls{ITSCP} showing promising results through \gls{SIMOPT} on a lattice network~\cite{rimoldi2023urban}. 
This method has the advantage of not needing an explicit model, as it bases its representation 
of the traffic dynamics directly
on data collected from the system, requiring less parameters and retaining higher interpretability  
compared to machine learning procedures. 

Even though the \gls{ITSCP} has been a subject of research for over 70 years, 
there still exists a gap between the 
literature and practical implementations~\cite{traffic-control-review-papageorgiu}.
One of the main reasons behind this gap is the lack of a common benchmark in the field. 
Most of the traffic research so far has focused on several different small networks due to computational cost (see Section~\ref{subsub:simopt}). 
These networks are often too simple to allow conclusions to be drawn about real urban traffic networks 
and their differences make comparing different approaches difficult.
With this in mind, this work builds on previous results~\cite{rimoldi2023urban}, showing that solving the \gls{ITSCP} via \gls{DeePC} 
is a general approach able to handle large-scale networks and generic intersections. 
We corroborate our findings through closed-loop microscopic simulations based on the \gls{SUMO} package, 
using a real-scale high-fidelity simulation of the city of Z\"urich.

\par\null\selectlanguage{english}

\par\null

\subsection{Literature Review}\label{subsec:literature}
The \gls{ITSCP} is a complex problem involving many factors, such as human behavior, vehicle interactions 
within the network, stochastic traffic demand, and unpredictable events (e.g. traffic accidents). 
For these reasons, it has been the subject of extensive research since the seminal work in~\cite{webster} and the development of 
\gls{TRANSYT}~\cite{transyt}. 
In this section, we give a brief overview of the relevant literature on traffic signal control, for a more in-depth overview we refer the 
interested reader to~\cite{traffic-control-review-papageorgiu, traffic-light-review-eom, SimOpt-review}.

\subsubsection{Rule-Based}
Many early approaches employed rule-based methods to solve the \gls{ITSCP}. 
Rule-based methods work by defining states and a set of appropriate actions associated with each state to optimize a performance criterion. 
An example of these early methods can be found in~\cite{potts-algorithm, Panos-oversaturatedI,Panos-oversaturatedII}, where the authors developed 
the saturation flow algorithm.
In \cite{fuzzy-sets} a rule-based algorithm is combined with fuzzy logic. 
Fuzzy rule-based systems derive actions from given inputs by defining if-then rules which represent the relationship between the 
variables~\cite{fuzzy-application, fuzzy-adaptive,fuzzy-controller,fuzzy-multi-phased}. 
In~\cite{dynamic-transit} the authors devise a dynamic rule-based system that changes 
the rules depending on traffic conditions to handle problems in real time. 
However, as noted by~\cite{fuzzy-niitty}, it can be difficult to generate effective rules, 
especially for general intersections with a large number of possible phases.

\subsubsection{Dynamic Programming}
Dynamic programming has been applied to the \gls{ITSCP} to develop flexible control algorithms applicable 
to a variety of traffic conditions. 
The authors of \cite{dp-sen-head} first applied dynamic programming using phases and stages as control variables. 
In \cite{dp-mirchandani} an algorithm based on dynamic programming capable of controlling an arterial network was developed. 
In \cite{dp-cai} was shown that an adaptive traffic signal controller using approximate dynamic programming and reinforcement learning 
is capable of reducing vehicle delays. In~\cite{dp-zheng}, the authors developed a recursive data pipeline comprising data processing, 
flow prediction, parameter optimization, and signal control. 
The main drawback of these methods is their computational complexity, which makes scaling to large scale urban traffic 
networks difficult. Complexity can be mitigated by using techniques such as approximate dynamic programming at the expense of performance. 

\subsubsection{Distributed and Multi-Agent Methods}
Recently, there has been multiple attempts to control traffic networks in a distributed 
way~\cite{white-distributed-control,mas-bazzan,villabos-game,arel-rl,choi-mas,balaji-rl,varaiya_max_pressure,wongpiromsarn_max_pressure}.
A popular distributed algorithm in traffic light control is Max-pressure, derived from
the backpressure scheduling framework \cite{original_max_pressure}. The algorithm maximizes network
throughput by selecting, at each intersection, the phase with the largest upstream-downstream queue differential.
The authors of \cite{wongpiromsarn_max_pressure} proved that a fully distributed implementation of Max-pressure
stabilizes any feasible demand distribution without requiring forecasts. In \cite{varaiya_max_pressure} 
the store-and-forward model is derived and the authors show that Max-pressure control is able to maximize the 
stability region using only local measurements and turn ratios. A practical drawback of classical
Max-pressure theory is that it may induce highly variable cycle lengths and rapid phase switching, moreover,
it does not incentivize coordination across the network.
Multi-agent learning and game-theoretic approaches are also used in this context.
The authors of~\cite{white-distributed-control} were able to deal with computational challenges
by devising a distributed control approach and tested it on a simulated $9$ by $7$ grid network. 
The authors of \cite{villabos-game} model an intersection as a non-cooperative game in which each player 
(traffic signal) aims at minimizing its queue and 
the solution of the \gls{ITSCP} is the Nash equilibrium of the game. 
In~\cite{arel-rl,choi-mas,mas-bazzan,balaji-rl}, the authors used reinforcement learning 
to implement cooperative hierarchical multi-agent systems for real-time signal timing control of 
complex traffic networks.
A drawback of these methodologies is the inherent difficulty to provide theoretical guarantees 
such as stability or convergence. This is, however, a very active area of research.  

\subsubsection{Model Predictive Control}
\gls{MPC} is a powerful control methodology providing strong theoretical guarantees on the stability, feasibility, 
and optimality of the closed-loop system, under the assumption of having a model that well describes the
real system~\cite{mpc-guarantees}. 
An example of its application to the \gls{ITSCP} can be found in~\cite{kouvelas-mpc}, where a linear-parameter 
varying formulation of \gls{MPC} for perimeter flow control is presented. 
This line of work is based on the partitioning of the network into regions described by the 
\gls{MFD}~\cite{mfd,existencemfd,geroliminis:2011:properties_MFD}. 
In~\cite{geroliminig:2018:MPC_perimeter_control}, a similar approach is presented using economic \gls{MPC} and 
regional route-guidance actuation-based control.
The authors in~\cite{lin2012efficient} propose two different macroscopic models of urban traffic to design 
an \gls{MPC} algorithm able to 
compute a structured network-wide traffic controller.
These approaches are computationally efficient and allow for centralized control of large networks as shown 
in~\cite{kouvelas-mpc}. 
However, model estimation procedures for such large systems can be computationally demanding.

\subsubsection{Data-enabled Predictive Control}
Recently, there has been a surge of interest for data-driven control approaches~\cite{markovsky2021behavioral}.  
In this regard, behavioral system theory~\cite{willems1986timeI} offers a powerful tool to obtain systems descriptions directly from data. 
The main idea is that the set of all possible trajectories of a dynamical system, called the \textit{behavior}, 
contains all the information necessary to describe the system. Behaviors of linear systems can be 
represented by non-parametrized models. For linear systems, such representations are images of matrices 
containing data sampled from the system; these can then be leveraged by control algorithms.
A particularly successful direct data-driven control algorithm obtained in this way is \gls{DeePC}~\cite{coulson2019data}, 
which has been applied to numerous practical studies~\cite{coulson2019data,carlet2020data,huang2019data}. 
\gls{DeePC} has been also applied to urban traffic control, in the contexts of connected and autonomous vehicles coordination~\cite{wang2023deep} and vehicle 
rebalancing in mobility-on-demand systems~\cite{zhu2023data}. 
We have previously shown \gls{DeePC} to be effective in tackling the \gls{ITSCP} on a symmetric lattice network comprising 64 intersections~\cite{rimoldi2023urban}. 
In this work we show that \gls{DeePC} is suitable for controlling much larger networks with general intersection layouts.

\subsubsection{Simulation-based Optimization}\label{subsub:simopt}
\gls{SIMOPT} is a field in which optimization techniques are integrated with simulation analysis~\cite{deng-phd}. 
Simulation tools are frequently used in \gls{ITSCP} research, approximately 77\% of the papers in the field use a simulation framework \cite{SimOpt-review}. 
\textit{AIMSUN}~\cite{AimsunManual}, \textit{CORSIM}~\cite{corsim}, \textit{MATSim}~\cite{matsim}, \textit{Paramics}~\cite{paramics}, \gls{SUMO}~\cite{sumo}, 
\textit{VISSIM}~\cite{vissim} are some of the most widely used simulation packages in research and industry. The majority of papers analyzed in 
\cite{SimOpt-review} uses \textit{VISSIM} as a simulation platform, closely followed by \gls{SUMO}. Most works focus on isolated single intersections 
\cite{simopt-singleinter-1, simopt-singleinter-2, simopt-singleinter-3,simopt-singleinter-4, simopt-singleinter-5} and regular lattice networks 
\cite{simopt-lattice-1,simopt-lattice-2, simopt-lattice-3, simopt-lattice-4, simopt-lattice-5}. Some papers consider controlled roundabouts 
\cite{simopt-roundabout-1, simopt-roundabout-2} and arterial networks \cite{simopt-arterial-1, simopt-arterial-2}.
In recent years, more general networks have been the focus of attention; for example, the authors of \cite{simopt-changsha-china} considered a part of the city 
of Changsha, China as a simulation setting. In \cite{simopt-skopje,simopt-bologna, simopt-lausanne}, the traffic networks of Skopje, Northern Macedonia; Bologna, Italy; 
and Lausanne, Switzerland were simulated. The largest online closed-loop simulation, to the best of the authors knowledge is described in 
\cite{linearMPC,kouvelas-mpc} where a part of Barcelona, Spain is used as a network, serving 79,143 vehicles throughout the simulation. 
It is important to note that all these simulations make the unrealistic assumption that all intersections are four-legged intersections or 
do not regulate traffic flow at intersections; for example, \cite{linearMPC} directly controls the flows between regions. 

\subsection{Contributions}\label{subsec:clations montributions}

The main contributions of this paper are three: 
\begin{enumerate}[label=(\roman*)]
    \item We introduce a flexible formulation of the \gls{ITSCP} based on behavioral theory that can seamlessly incorporate different kinds 
          of data and generalize classical perimeter control \cite{gerolimins:2013:optimal_periter_two_regions}, the 
          formulation accomodates other forms of actuation, such as dynamic speed limits,
    \item we conjecture a linear relationship between the traffic density and dynamic traffic light signals. We validate this conjecture using extensive high-fidelity simulations,
    \item we perform closed loop control on what is, to the best of our knowledge, the largest microsimulation in the literature equipped with demand estimated from real data. We propose the simulation as a standard 
          benchmark in traffic control research and use it to validate the performance of the proposed methodology in terms of travel time and emissions metrics.
\end{enumerate}

The paper is organized as follows. Section \ref{sec:methodology} describes the methodological framework, casting the \gls{ITSCP} problem 
in the framework of behavioral system theory~\cite{willemsBehavioural}. Section \ref{sec:DeePC} presents the \gls{DeePC} algorithm.
Section \ref{sec:simulations} presents the simulation framework, the traffic networks used to test our 
method and numerical results showcasing the effectiveness of the closed-loop control. Section \ref{sec:discussion} summarizes our findings 
and future research directions.

\paragraph{Notation}  
$\N$ denotes the set of positive integer numbers, $\R$, $\R^n$ and $\R^{p \times m}$  denote the set of real numbers, 
the set  of $n$-dimensional vectors with real entries, and the set of $p \times m$ matrices with real entries, respectively. 
For every ${p\in\N}$, the set of positive integers $\{1, 2, . . . , p\}$ is denoted by $\mf p$. 
$I$ denotes the identity matrix.
A map $f$ from $X$ to $Y$ is denoted by $f:X \to Y$; $(Y)^{X}$ denotes the collection of all such maps.  
The restriction of $f:X \to Y$ to a set $X^{\prime}$ such that $X' \cap X \neq \emptyset$ is denoted by $f|_{X^{\prime}}$ and defined as $f|_{X^{\prime}}(x)= f(x)$ 
for  ${x \in X \cap X^{\prime}}$. 
For $\mathcal{F} \subset (Y)^{X}$, then  $\mathcal{F}|_{X^{\prime}}$ denotes ${\{ f|_{X^{\prime}} \, : \, f \in \mathcal{F}\}}$.

\section{Behavioral formulation of urban traffic dynamics}\label{sec:methodology}

\subsection{Traffic network, sensing and demand}\label{ssec:urban_td}

We consider the traffic network of a city composed of intersections and roads, each divided into one or more lanes.   
Following~\cite{geroliminig:2018:MPC_perimeter_control}, we partition a city in $p \in \N$ regions of 
homogeneous average traffic density to leverage the concept of \gls{MFD}, see Figure \ref{fig:MFD}. We use the \gls{MFD} to obtain 
a reference trajectory for \gls{DeePC} to track, as will be discussed in Section~\ref{subsec:MFD}. 
Several methods can be used to generate such a partitioning~\cite{review-clustering}, here we use a parallelized 
version of the \textit{snake clustering} algorithm~\cite{gerolinimis2016CsnakeClustering}, detailed in Appendix \ref{app:partitioning}. 
The algorithm finds connected clusters of roads with homogeneous average traffic density, yielding a low-scatter \gls{MFD}.  
The set of resulting regions is defined as $\mf{p} \coloneqq \{1,\cdots,p\}$.

\begin{figure}[h!]
\centering
\includegraphics[width=\textwidth]{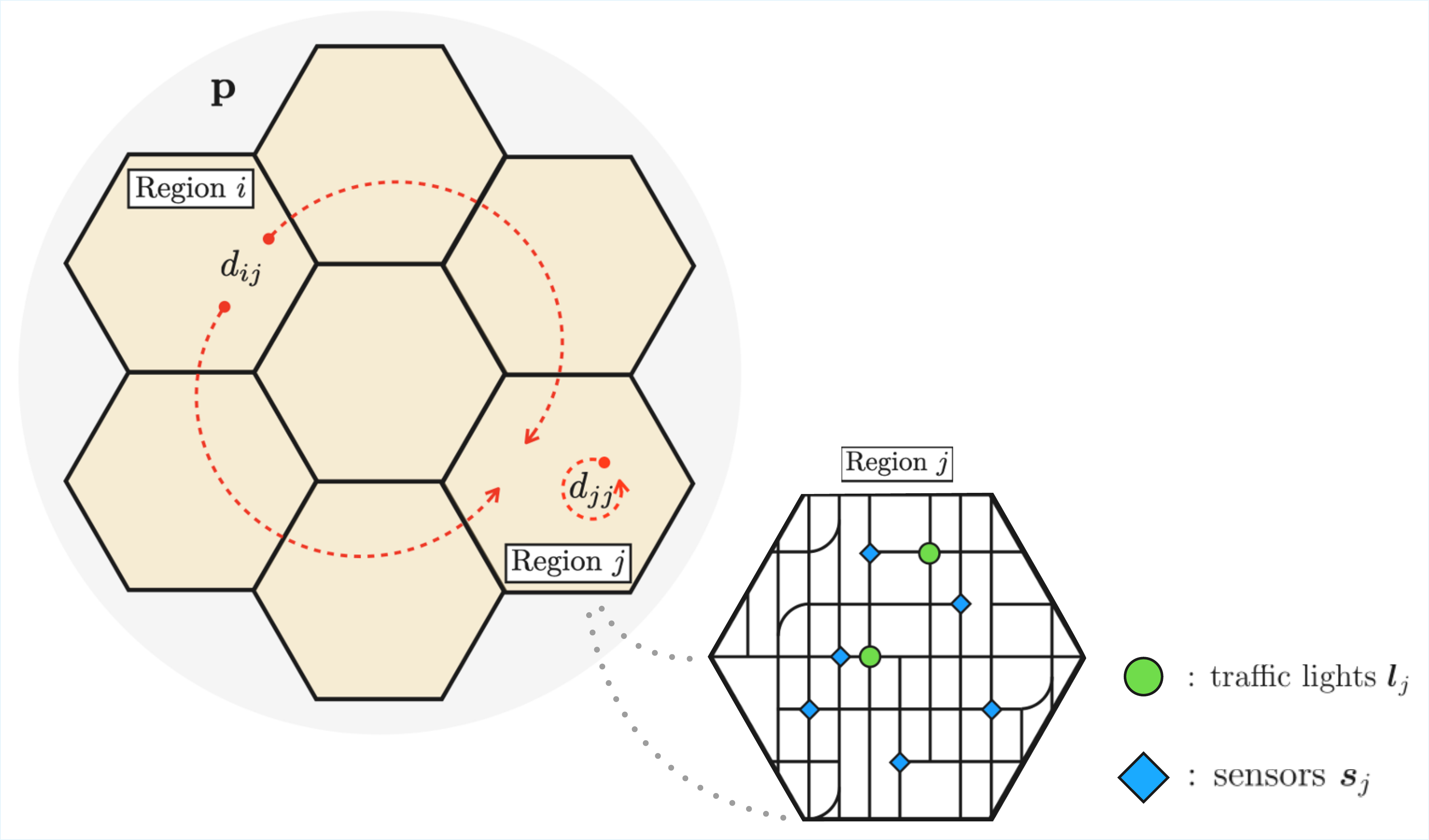}
\centering
\caption{The city is divided into $p$ regions homogeneous in average traffic density (the hexagons) and 
the demand $\bs d$ among them. A detail of sensors $\mf s_j$ (blue diamonds) and traffic lights 
$\mf l_j$ (green circles) locations within region $j\in\mf p$ is shown.}
\label{fig:info_traffic}
\end{figure}

The most common sensors used by municipalities to measure traffic conditions are \textit{Eulerian} 
sensors~\cite{zhanfeng:2001:single_loop_detector}, which are stationary devices, 
as opposed to \textit{Lagrangian}~\cite{lagrangian-estimation} sensors which can be mobile. 
Throughout this paper, we assume the use of \textit{Eulerian} sensors, capable of measuring traffic density
and flow. However, proposed procedure is agnostic to the type of sensors used and 
Lagrangian sensors could be integrated~\cite{claudel:2008:lagrangian-eulerian_sensing}.  
We denote the set of sensors in region ${i\in \mf p}$ as $\mf s_i \coloneqq \{1,\cdots,s_i\}$, where $s_i\in \N$
is the number of sensors in the region. 
From each sensor ${j\in\mathbf{s}_i}$, we obtain the traffic density $\rho^{j}(t)$\,[veh/km] 
and flow $\phi^j(t)$\,[veh/h], at time ${t\in \T \subseteq \N}$, on the 
road where the sensor is located, normalized by the number of lanes. 
We then average the measurements over each region $i \in \mf p$, to reduce the variability and capture macroscopic 
variations~\cite{geroliminis:2011:properties_MFD}.
We denote the average density and flow in region $i\in\mf p$ at time $t \in \T$ as   
\begin{subequations}
\begin{align}
 \rho_i(t) \coloneqq \frac{1}{s_i}\sum_{j=1}^{s_i}\rho^j(t) \\
 \phi_i(t) \coloneqq \frac{1}{s_i}\sum_{j=1}^{s_i}\phi^j(t) 
\end{align}
\end{subequations}

The vectors of average densities and flows in the city at time $t$ are then,
\begin{subequations}
\begin{align}
\bs \rho(t) &\coloneqq [\rho_i(t)]_{i=1}^p \in \R^p\\
\bs \phi(t) &\coloneqq [\phi_i(t)]_{i=1}^p \in \R^p
\end{align}
\end{subequations}

The evolution of $\bs \rho$ and $\bs \phi$ depends on the flow of vehicles entering the traffic 
network during every time interval, known as the \textit{demand}. The flow of vehicles starting their trip in region ${i \in\mf p}$ at time $t$ 
with final destination in region ${j \in\mf p}$, is denoted by $d_{ij}(t)$\ [veh/h]; $d_{ii}(t)$ is the internal demand 
of region $i$, as seen in Figure \ref{fig:info_traffic}. The vector of demands starting in region $i$ is denoted by 
$\bs d_i(t) \coloneqq [d_{ij}(t)]_{j=1}^p\in\R^{p}$, while the vector of all the demands among all regions is 
$\bs d(t) \coloneqq [\bs d_{i}(t)]_{i=1}^p\in\R^{p^2}$. 
Despite being influenced by many factors such as commuters choices and weather conditions, $\bs d(t)$ 
presents seasonal patterns at different time scales, weekly, monthly, and yearly, that make it possible to reliably estimate it~\cite{kumar_short-term_2015}. 
Demand estimation, in the form of time dependent origin-destination pairs, is a well-known problem 
and has been extensively studied~\cite{foulds:2013:origin_destination_matrix,olmos:2018:urban_traffic}.
In this work, we do not consider the problem of predicting the demand and instead we consider it as 
a given exogenous input to the system.

\subsection{Intersection Control}\label{subsec:IntersectionControl}
\begin{figure}[h!]
    \centering
        \begin{subfigure}[h]{0.45\textwidth}
            \begin{tikzpicture}[
            scale = 0.5,  
            center/.style     = {fill=black!20, draw=none},
            approach/.style   = {fill=black!60, draw=none},
            median/.style     = {white, line width=1pt},
            lanesep/.style    = {white, dash pattern=on 8pt off 4pt, line width=1pt},
            signal/.style     = {line width=6pt, rounded corners=1pt}
            ]

            \fill[center] (-\rw,-\rw) rectangle ++(2*\rw,2*\rw);

            \foreach \ang in {0,90,180,270}{
                \begin{scope}[rotate=\ang]
                \fill[approach] (-\threerw,-\rw) rectangle ++(\rw,2*\rw);
                \fill[approach] ( \rw,-\rw) rectangle ++(\rw,2*\rw);

                \draw[median] (-\threerw,0) -- (-\rw,0);
                \draw[median] ( \rw,0)   -- ( \threerw,0);

                \draw[lanesep] ( \rw,1cm) -- ( \threerw,1cm);
                \draw[lanesep] (-\threerw,-1cm) -- (-\rw,-1cm);
                \end{scope}
            }

            \fill[fill=cyan!60, opacity=0.6] ( \rw,\rw) rectangle ++(2*\rw,-1cm);
            \fill[fill = red!80, opacity=0.6] (-\rw,-\rw) rectangle ++(2*\rw,2*\rw);
            \fill[fill = yellow!80, opacity=0.8] (-\rw,-\rw) rectangle ++(-2*\rw,\rw);

            \draw[signal,green!80!black]  (-\rw,\rw)   -- (-1.05cm,\rw);
            \draw[signal,green!80!black]  (-0.95,\rw)   -- (0,\rw);
            \draw[signal,green!80!black]  (0,-\rw)  -- (0.95cm,-\rw);
            \draw[signal,green!80!black]  (1.05cm,-\rw)  -- (\rw,-\rw);
            \draw[signal,red!80!black]    (\rw,0cm)   -- (\rw,0.95);
            \draw[signal,red!80!black]    (\rw,1.05cm)   -- (\rw,\rw);
            \draw[signal,red!80!black]    (-\rw,-0.95cm)  -- (-\rw,0);
            \draw[signal,red!80!black]    (-\rw,-\rw)  -- (-\rw,-1.05cm);

            \draw (0,0) -- (-3cm, \threerw - 2cm);
            \draw (-3cm, \threerw - 2cm) -- (-6cm, \threerw -2cm ) node[above] {Crossing area};
            \draw (2*\rw  ,1.5cm) -- (2.5*\rw, 4cm);
            \draw (2.5*\rw  ,4cm) -- (3*\rw, 4cm) node[above] {Lane};
            \draw (-2*\rw  ,-1.5cm) -- (-2.5*\rw, -4cm);
            \draw (-2.5*\rw  ,-4cm) -- (-3*\rw, -4cm) node[below] {Approach};
            \draw (\rw-0.5cm  ,-2cm) -- (1.5*\rw, -4cm);
            \draw (1.5*\rw  ,-4cm) -- (2.5*\rw, -4cm) node[below] {Signal head};

            \end{tikzpicture}
        \end{subfigure}
        \hfill
        \begin{subfigure}[h]{0.45\textwidth}
            \begin{tikzpicture}
            \node[draw] (fig-1) {
                    \begin{tikzpicture}[
            scale = 0.25,  
            center/.style     = {fill=black!60, draw=none},
            approach/.style   = {fill=black!60, draw=none},
            median/.style     = {white, line width=0.5pt},
            lanesep/.style    = {white, dash pattern=on 2pt off 1pt, line width=0.5pt},
            signal/.style     = {line width=1.5pt, rounded corners=1pt}
            ]

            \fill[center] (-\rw,-\rw) rectangle ++(2*\rw,2*\rw);

            \foreach \ang in {0,90,180,270}{
                \begin{scope}[rotate=\ang]
                \fill[approach] (-\threerw,-\rw) rectangle ++(\rw,2*\rw);
                \fill[approach] ( \rw,-\rw) rectangle ++(\rw,2*\rw);

                \draw[median] (-\threerw,0) -- (-\rw,0);
                \draw[median] ( \rw,0)   -- ( \threerw,0);

                \draw[lanesep] ( \rw,1cm) -- ( \threerw,1cm);
                \draw[lanesep] (-\threerw,-1cm) -- (-\rw,-1cm);
                \end{scope}
            }

            \draw[signal,green!80!black]  (-\rw,\rw)   -- (-1.05cm,\rw);
            \draw[signal,green!80!black]  (-0.95,\rw)   -- (0,\rw);
            \draw[signal,green!80!black]  (0,-\rw)  -- (0.95cm,-\rw);
            \draw[signal,green!80!black]  (1.05cm,-\rw)  -- (\rw,-\rw);
            \draw[signal,red!80!black]    (\rw,0cm)   -- (\rw,0.95);
            \draw[signal,red!80!black]    (\rw,1.05cm)   -- (\rw,\rw);
            \draw[signal,red!80!black]    (-\rw,-0.95cm)  -- (-\rw,0);
            \draw[signal,red!80!black]    (-\rw,-\rw)  -- (-\rw,-1.05cm);

            \draw [thick,arrows = {-Stealth[reversed, reversed]}, color=green] (-1cm,\rw-0.5cm) -- (-1cm,-\rw);

            \draw [thick,arrows = {-Stealth[reversed, reversed]}, color=green] (1cm,-\rw+0.5cm) -- (1cm,\rw);
            \draw [thick,arrows = {-Stealth[reversed, reversed]}, color=green] (0.5cm,-\rw+0.5cm) .. controls (0.25,0.25\rw)  .. (-2cm,0.5*\rw);

            \end{tikzpicture}
            } node[above = 0.2 cm of fig-1] {Compatible flows};
            \node[draw, right= of fig-1] (fig-2) {
                    \begin{tikzpicture}[
            scale = 0.25,  
            center/.style     = {fill=black!60, draw=none},
            approach/.style   = {fill=black!60, draw=none},
            median/.style     = {white, line width=0.5pt},
            lanesep/.style    = {white, dash pattern=on 2pt off 1pt, line width=0.5pt},
            signal/.style     = {line width=1.5pt, rounded corners=1pt}
            ]

            \fill[center] (-\rw,-\rw) rectangle ++(2*\rw,2*\rw);

            \foreach \ang in {0,90,180,270}{
                \begin{scope}[rotate=\ang]
                \fill[approach] (-\threerw,-\rw) rectangle ++(\rw,2*\rw);
                \fill[approach] ( \rw,-\rw) rectangle ++(\rw,2*\rw);

                \draw[median] (-\threerw,0) -- (-\rw,0);
                \draw[median] ( \rw,0)   -- ( \threerw,0);

                \draw[lanesep] ( \rw,1cm) -- ( \threerw,1cm);
                \draw[lanesep] (-\threerw,-1cm) -- (-\rw,-1cm);
                \end{scope}
            }

            \draw[signal,green!80!black]  (-\rw,\rw)   -- (-1.05cm,\rw);
            \draw[signal,green!80!black]  (-0.95,\rw)   -- (0,\rw);
            \draw[signal,red!80!black]  (0,-\rw)  -- (0.95cm,-\rw);
            \draw[signal,red!80!black]  (1.05cm,-\rw)  -- (\rw,-\rw);
            \draw[signal,green!80!black]    (\rw,0cm)   -- (\rw,0.95);
            \draw[signal,green!80!black]    (\rw,1.05cm)   -- (\rw,\rw);
            \draw[signal,red!80!black]    (-\rw,-0.95cm)  -- (-\rw,0);
            \draw[signal,red!80!black]    (-\rw,-\rw)  -- (-\rw,-1.05cm);

            \draw [thick,arrows = {-Stealth[reversed, reversed]}, color=red] (-1cm,\rw-0.5cm) -- (-1cm,-\rw);

            \draw [thick,arrows = {-Stealth[reversed, reversed]}, color=red] (\rw-0.5cm,1cm) -- (-\rw,1cm);

            \end{tikzpicture}
            } node[above = 0.2 cm of fig-2] {Antagonistic flows};
            \end{tikzpicture}
        \end{subfigure}
    \caption{Example of a four-legged intersection and compatibility of traffic flows. On the left the structural components of the intersection are highlighted 
for clarity. On the right the difference between compatible and antagonistic flows can be observed. }
\label{fig:intersection}
\end{figure}
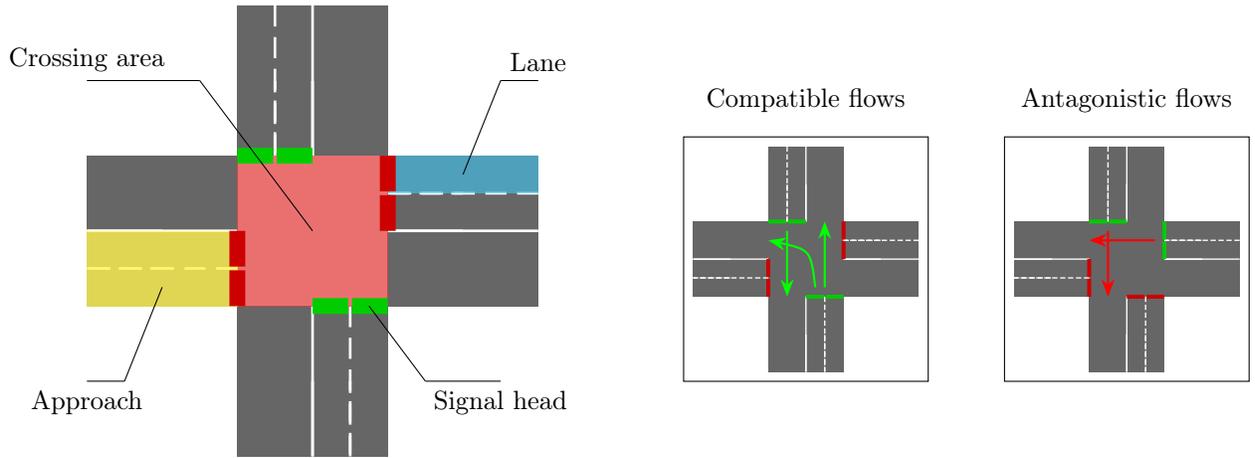

Following \cite{traffic-control-review-papageorgiu}, we define an \textit{intersection} as a set of \textit{approaches} 
and a \textit{crossing area}. The \textit{crossing area} is the portion of space delimited by the stop lines of the approaches, where 
vehicles and pedestrians are allowed to cross intermittently under right of way. An \textit{approach} is a road leading to the crossing area, 
comprising one or more \textit{lanes}, each equipped with a \textit{signal head}. Signal heads can be in 
one of three states: \textit{green} (g), \textit{yellow} (y), or \textit{red} (r); the set of all the signal heads in the intersection makes
up the \textit{traffic signal} of the intersection. 
Approaches are used by \textit{traffic flows} to reach the crossing area. Two flows are said to be \textit{compatible} 
when the vehicles composing them can pass the crossing area simultaneously under right of way; 
otherwise they are called \textit{antagonistic}. Figure \ref{fig:intersection} shows a simplified drawing of
an intersection. On the left the crossing area, approach, lane and signal heads are highlighted for clarity. 
On the right the difference between compatible and antagonistic flows can be seen.

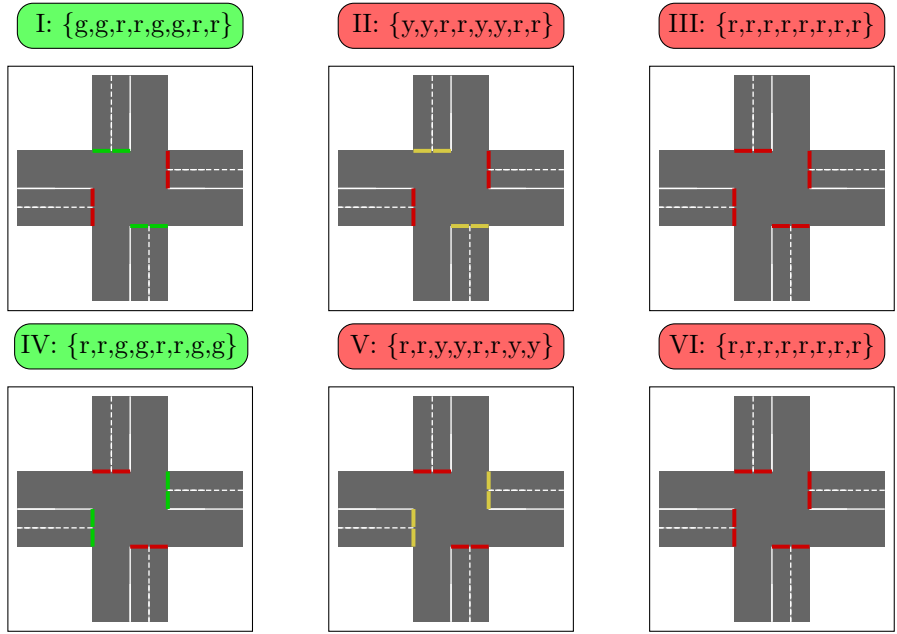
\begin{figure}[h!]
    \centering
            \begin{tikzpicture}
            \node[draw] (fig-1) {
                \begin{tikzpicture}[
            scale = 0.25,  
            center/.style     = {fill=black!60, draw=none},
            approach/.style   = {fill=black!60, draw=none},
            median/.style     = {white, line width=0.5pt},
            lanesep/.style    = {white, dash pattern=on 2pt off 1pt, line width=0.5pt},
            signal/.style     = {line width=1.5pt, rounded corners=1pt}
            ]

            \fill[center] (-\rw,-\rw) rectangle ++(2*\rw,2*\rw);

            \foreach \ang in {0,90,180,270}{
                \begin{scope}[rotate=\ang]
                \fill[approach] (-\threerw,-\rw) rectangle ++(\rw,2*\rw);
                \fill[approach] ( \rw,-\rw) rectangle ++(\rw,2*\rw);

                \draw[median] (-\threerw,0) -- (-\rw,0);
                \draw[median] ( \rw,0)   -- ( \threerw,0);

                \draw[lanesep] ( \rw,1cm) -- ( \threerw,1cm);
                \draw[lanesep] (-\threerw,-1cm) -- (-\rw,-1cm);
                \end{scope}
            }

            \draw[signal,green!80!black]  (-\rw,\rw)   -- (-1.05cm,\rw);
            \draw[signal,green!80!black]  (-0.95,\rw)   -- (0,\rw);
            \draw[signal,green!80!black]  (0,-\rw)  -- (0.95cm,-\rw);
            \draw[signal,green!80!black]  (1.05cm,-\rw)  -- (\rw,-\rw);
            \draw[signal,red!80!black]    (\rw,0cm)   -- (\rw,0.95);
            \draw[signal,red!80!black]    (\rw,1.05cm)   -- (\rw,\rw);
            \draw[signal,red!80!black]    (-\rw,-0.95cm)  -- (-\rw,0);
            \draw[signal,red!80!black]    (-\rw,-\rw)  -- (-\rw,-1.05cm);

            \end{tikzpicture}
            } node[above = 0.1 cm of fig-1] {
                \begin{tikzpicture}
                     \draw[rectangle, fill = green!60!white, rounded corners = 6pt] (-1.5cm, 0.4cm) rectangle (1.5cm,1cm) node[below left] {\rom{1}: \{g,g,r,r,g,g,r,r\}};
                \end{tikzpicture}
            }; 
            \node[draw, right= of fig-1] (fig-2) {
                    \begin{tikzpicture}[
            scale = 0.25,  
            center/.style     = {fill=black!60, draw=none},
            approach/.style   = {fill=black!60, draw=none},
            median/.style     = {white, line width=0.5pt},
            lanesep/.style    = {white, dash pattern=on 2pt off 1pt, line width=0.5pt},
            signal/.style     = {line width=1.5pt, rounded corners=1pt}
            ]

            \fill[center] (-\rw,-\rw) rectangle ++(2*\rw,2*\rw);

            \foreach \ang in {0,90,180,270}{
                \begin{scope}[rotate=\ang]
                \fill[approach] (-\threerw,-\rw) rectangle ++(\rw,2*\rw);
                \fill[approach] ( \rw,-\rw) rectangle ++(\rw,2*\rw);

                \draw[median] (-\threerw,0) -- (-\rw,0);
                \draw[median] ( \rw,0)   -- ( \threerw,0);

                \draw[lanesep] ( \rw,1cm) -- ( \threerw,1cm);
                \draw[lanesep] (-\threerw,-1cm) -- (-\rw,-1cm);
                \end{scope}
            }

            \draw[signal,yellow!80!black]  (-\rw,\rw)   -- (-1.05cm,\rw);
            \draw[signal,yellow!80!black]  (-0.95,\rw)   -- (0,\rw);
            \draw[signal,yellow!80!black]  (0,-\rw)  -- (0.95cm,-\rw);
            \draw[signal,yellow!80!black]  (1.05cm,-\rw)  -- (\rw,-\rw);
            \draw[signal,red!80!black]    (\rw,0cm)   -- (\rw,0.95);
            \draw[signal,red!80!black]    (\rw,1.05cm)   -- (\rw,\rw);
            \draw[signal,red!80!black]    (-\rw,-0.95cm)  -- (-\rw,0);
            \draw[signal,red!80!black]    (-\rw,-\rw)  -- (-\rw,-1.05cm);

            \end{tikzpicture}
            } node[above = 0.1 cm of fig-2] {
                \begin{tikzpicture}
                     \draw[rectangle, fill = red!60!white, rounded corners = 6pt] (-1.5cm, 0.4cm) rectangle (1.5cm,1cm) node[below left] {\rom{2}: \{y,y,r,r,y,y,r,r\}};
                \end{tikzpicture}
                };
            \node[draw, right= of fig-2] (fig-3) {
                    \begin{tikzpicture}[
            scale = 0.25,  
            center/.style     = {fill=black!60, draw=none},
            approach/.style   = {fill=black!60, draw=none},
            median/.style     = {white, line width=0.5pt},
            lanesep/.style    = {white, dash pattern=on 2pt off 1pt, line width=0.5pt},
            signal/.style     = {line width=1.5pt, rounded corners=1pt}
            ]

            \fill[center] (-\rw,-\rw) rectangle ++(2*\rw,2*\rw);

            \foreach \ang in {0,90,180,270}{
                \begin{scope}[rotate=\ang]
                \fill[approach] (-\threerw,-\rw) rectangle ++(\rw,2*\rw);
                \fill[approach] ( \rw,-\rw) rectangle ++(\rw,2*\rw);

                \draw[median] (-\threerw,0) -- (-\rw,0);
                \draw[median] ( \rw,0)   -- ( \threerw,0);

                \draw[lanesep] ( \rw,1cm) -- ( \threerw,1cm);
                \draw[lanesep] (-\threerw,-1cm) -- (-\rw,-1cm);
                \end{scope}
            }

            \draw[signal,red!80!black]  (-\rw,\rw)   -- (-1.05cm,\rw);
            \draw[signal,red!80!black]  (-0.95,\rw)   -- (0,\rw);
            \draw[signal,red!80!black]  (0,-\rw)  -- (0.95cm,-\rw);
            \draw[signal,red!80!black]  (1.05cm,-\rw)  -- (\rw,-\rw);
            \draw[signal,red!80!black]    (\rw,0cm)   -- (\rw,0.95);
            \draw[signal,red!80!black]    (\rw,1.05cm)   -- (\rw,\rw);
            \draw[signal,red!80!black]    (-\rw,-0.95cm)  -- (-\rw,0);
            \draw[signal,red!80!black]    (-\rw,-\rw)  -- (-\rw,-1.05cm);

            \end{tikzpicture}
            } node[above = 0.1 cm of fig-3] {
                \begin{tikzpicture}
                     \draw[rectangle, fill = red!60!white, rounded corners = 6pt] (-1.5cm, 0.4cm) rectangle (1.5cm,1cm) node[below left] {\rom{3}: \{r,r,r,r,r,r,r,r\}};
                \end{tikzpicture}
            };
            \node[draw, below= of fig-1] (fig-4) {
                    \begin{tikzpicture}[
            scale = 0.25,  
            center/.style     = {fill=black!60, draw=none},
            approach/.style   = {fill=black!60, draw=none},
            median/.style     = {white, line width=0.5pt},
            lanesep/.style    = {white, dash pattern=on 2pt off 1pt, line width=0.5pt},
            signal/.style     = {line width=1.5pt, rounded corners=1pt}
            ]

            \fill[center] (-\rw,-\rw) rectangle ++(2*\rw,2*\rw);

            \foreach \ang in {0,90,180,270}{
                \begin{scope}[rotate=\ang]
                \fill[approach] (-\threerw,-\rw) rectangle ++(\rw,2*\rw);
                \fill[approach] ( \rw,-\rw) rectangle ++(\rw,2*\rw);

                \draw[median] (-\threerw,0) -- (-\rw,0);
                \draw[median] ( \rw,0)   -- ( \threerw,0);

                \draw[lanesep] ( \rw,1cm) -- ( \threerw,1cm);
                \draw[lanesep] (-\threerw,-1cm) -- (-\rw,-1cm);
                \end{scope}
            }

            \draw[signal,red!80!black]  (-\rw,\rw)   -- (-1.05cm,\rw);
            \draw[signal,red!80!black]  (-0.95,\rw)   -- (0,\rw);
            \draw[signal,red!80!black]  (0,-\rw)  -- (0.95cm,-\rw);
            \draw[signal,red!80!black]  (1.05cm,-\rw)  -- (\rw,-\rw);
            \draw[signal,green!80!black]    (\rw,0cm)   -- (\rw,0.95);
            \draw[signal,green!80!black]    (\rw,1.05cm)   -- (\rw,\rw);
            \draw[signal,green!80!black]    (-\rw,-0.95cm)  -- (-\rw,0);
            \draw[signal,green!80!black]    (-\rw,-\rw)  -- (-\rw,-1.05cm);

            \end{tikzpicture}
            } node[above = 0.1 cm of fig-4] {
                \begin{tikzpicture}
                     \draw[rectangle, fill = green!60!white, rounded corners = 6pt] (-1.55cm, 0.4cm) rectangle (1.55cm,1cm) node[below left] {\rom{4}: \{r,r,g,g,r,r,g,g\}};
                \end{tikzpicture}
            };
            \node[draw, below= of fig-2] (fig-5) {
                    \begin{tikzpicture}[
            scale = 0.25,  
            center/.style     = {fill=black!60, draw=none},
            approach/.style   = {fill=black!60, draw=none},
            median/.style     = {white, line width=0.5pt},
            lanesep/.style    = {white, dash pattern=on 2pt off 1pt, line width=0.5pt},
            signal/.style     = {line width=1.5pt, rounded corners=1pt}
            ]

            \fill[center] (-\rw,-\rw) rectangle ++(2*\rw,2*\rw);

            \foreach \ang in {0,90,180,270}{
                \begin{scope}[rotate=\ang]
                \fill[approach] (-\threerw,-\rw) rectangle ++(\rw,2*\rw);
                \fill[approach] ( \rw,-\rw) rectangle ++(\rw,2*\rw);

                \draw[median] (-\threerw,0) -- (-\rw,0);
                \draw[median] ( \rw,0)   -- ( \threerw,0);


                \draw[lanesep] ( \rw,1cm) -- ( \threerw,1cm);
                \draw[lanesep] (-\threerw,-1cm) -- (-\rw,-1cm);
                \end{scope}
            }

            \draw[signal,red!80!black]  (-\rw,\rw)   -- (-1.05cm,\rw);
            \draw[signal,red!80!black]  (-0.95,\rw)   -- (0,\rw);
            \draw[signal,red!80!black]  (0,-\rw)  -- (0.95cm,-\rw);
            \draw[signal,red!80!black]  (1.05cm,-\rw)  -- (\rw,-\rw);
            \draw[signal,yellow!80!black]    (\rw,0cm)   -- (\rw,0.95);
            \draw[signal,yellow!80!black]    (\rw,1.05cm)   -- (\rw,\rw);
            \draw[signal,yellow!80!black]    (-\rw,-0.95cm)  -- (-\rw,0);
            \draw[signal,yellow!80!black]    (-\rw,-\rw)  -- (-\rw,-1.05cm);

            \end{tikzpicture}
            } node[above = 0.1 cm of fig-5] {
                \begin{tikzpicture}
                     \draw[rectangle, fill = red!60!white, rounded corners = 6pt] (-1.5cm, 0.4cm) rectangle (1.5cm,1cm) node[below left] {\rom{5}: \{r,r,y,y,r,r,y,y\}};
                \end{tikzpicture}
            };
            \node[draw, below= of fig-3] (fig-6) {
                    \begin{tikzpicture}[
            scale = 0.25,  
            center/.style     = {fill=black!60, draw=none},
            approach/.style   = {fill=black!60, draw=none},
            median/.style     = {white, line width=0.5pt},
            lanesep/.style    = {white, dash pattern=on 2pt off 1pt, line width=0.5pt},
            signal/.style     = {line width=1.5pt, rounded corners=1pt}
            ]

            \fill[center] (-\rw,-\rw) rectangle ++(2*\rw,2*\rw);

            \foreach \ang in {0,90,180,270}{
                \begin{scope}[rotate=\ang]
                \fill[approach] (-\threerw,-\rw) rectangle ++(\rw,2*\rw);
                \fill[approach] ( \rw,-\rw) rectangle ++(\rw,2*\rw);

                \draw[median] (-\threerw,0) -- (-\rw,0);
                \draw[median] ( \rw,0)   -- ( \threerw,0);

                \draw[lanesep] ( \rw,1cm) -- ( \threerw,1cm);
                \draw[lanesep] (-\threerw,-1cm) -- (-\rw,-1cm);
                \end{scope}
            }

            \draw[signal,red!80!black]  (-\rw,\rw)   -- (-1.05cm,\rw);
            \draw[signal,red!80!black]  (-0.95,\rw)   -- (0,\rw);
            \draw[signal,red!80!black]  (0,-\rw)  -- (0.95cm,-\rw);
            \draw[signal,red!80!black]  (1.05cm,-\rw)  -- (\rw,-\rw);
            \draw[signal,red!80!black]    (\rw,0cm)   -- (\rw,0.95);
            \draw[signal,red!80!black]    (\rw,1.05cm)   -- (\rw,\rw);
            \draw[signal,red!80!black]    (-\rw,-0.95cm)  -- (-\rw,0);
            \draw[signal,red!80!black]    (-\rw,-\rw)  -- (-\rw,-1.05cm);

            \end{tikzpicture}
            } node[above = 0.1 cm of fig-6] {
                \begin{tikzpicture}
                     \draw[rectangle, fill = red!60!white, rounded corners = 6pt] (-1.5cm, 0.4cm) rectangle (1.5cm,1cm) node[below left] {\rom{6}: \{r,r,r,r,r,r,r,r\}};
                \end{tikzpicture}
            };

            \end{tikzpicture}
    \caption{Example of a duty cycle comprising six phases divided in active and passive. Active phases (\rom{1},\rom{4}) are grouped in the leftmost column and highlighted in green, while passive phases 
    (\rom{2},\rom{3},\rom{5},\rom{6}) can be found in the two rightmost columns highlighted in red. Active phases have at least one 'g' state in the tuple. }
\label{fig:phases}
\end{figure}

A traffic signal \textit{duty cycle}, is an ordered set $\Pi$ of \textit{phases}, its duration is called \textit{cycle time} $\Delta_{\textrm{dc}}$. 
A \textit{phase} is a tuple containing the state value for each signal head, and is designed to allow compatible traffic flows to cross the intersection. 
Figure~\ref{fig:phases} illustrates a duty cycle for a four legged intersection, where a phase of only red states is used to clear the intersection.

By influencing the flow of vehicles traveling through an actuated intersection, it is possible to indirectly influence 
$\phi_i(t)$ and $\rho_i(t)$. Given two regions ${i,j\in\mf p}$, if one intersection sitting at the boundery between 
$i$ and $j$ can be actuated, one can regulate the flow of vehicles moving between the two regions at that point.
If one can control all the access points from region $i$ to $j$, then the flow of vehicles moving between the two regions can be 
directly controlled, that is, classical perimeter control \cite{geroliminig:2018:MPC_perimeter_control}.
Remarkably, our approach does not assume the actuators to be placed at the boundaries of the regions, as this is rarely the case in practice. 
Instead, we allow the actuators to be placed anywhere within the network, see Figure \ref{fig:info_traffic}. 
According to \cite{traffic-control-review-papageorgiu}, there are four possible approaches when controlling traffic light operations:

\newpage

\begin{itemize}
    \item \textit{Phase specification}: The control input is the state composition of the phases. 
    When dealing with complex intersections involving a multiple approaches, the specification of the optimal 
    composition of the phases is non-trivial and has a major impact on the intersection's efficiency, often giving rise to mixed-integer problems~\cite{mixed-integer-phase-specs}.
    \item \textit{Split control}: The control input is the duration of each phase as a portion of the duty cycle time $\Delta_{\textrm{dc}}$. 
    This duration can be optimized based on current network conditions. 
    \item \textit{Cycle time}: The control input is the cycle time $\Delta_{\textrm{dc}}$. 
    Longer cycle times lead to an increase in efficiency due to the smaller proportion that the constant lost times occupy, 
    however, this may increase vehicle delays in underutilized intersections due to the longer red phase.
    \item \textit{Offset}: The control input is the phase difference between the duty cycles of multiple successive intersections. 
    In contrast to previous schemes, this approach assumes multiple intersections to be actuated. 
    This can be used, for example, to generate a \textit{green wave} along an arterial.
\end{itemize}

In this work, we focus on \textit{split control} with fixed duty cycle time $\Delta_{\textrm{dc}}$, making our method a fixed-time adaptive 
strategy in the classification of \cite{traffic-control-review-papageorgiu}. Under this scheme, the optimization problem can be readily 
formulated as a quadratic program with continuous bounded inputs. This problem can be solved efficiently, enabling practical 
implementations where the response time of the system is constrained. 

\begin{figure}[h!]
\centering
\begin{tikzpicture}[scale=2]
    \draw[black, thick, fill=black!10!white] (0,0) rectangle (4,2.1);
    
    \begin{scope}[shift={(0,1.1)}]
        \draw[draw=none,fill=white] (0.75,0.8) rectangle (3.75,0.2);
        \draw [arrows = {-Stealth[inset=0pt, angle=30:3pt]}, color=black] (0.75,0.2) -- (3.85,0.2);
        \draw [arrows = {-Stealth[inset=0pt, angle=30:3pt]}, color=black] (0.75,0.2) -- (0.75,0.9);
        \node at (0.65,0.2) {$\delta_p$};
        \node at (0.65,0.8) {$\delta_a$};
        \draw[dotted,dash pattern=on 1pt off 1pt] (1.75,0.2) -- (1.75,0.75);
        \node at (1.8,0.08) {$\Delta_{\textrm{dc}}$};
        \draw[dotted,dash pattern=on 1pt off 1pt] (2.75,0.2) -- (2.75,0.75);
        \node at (2.8,0.08) {$2\Delta_{\textrm{dc}}$};
        \node at (3.75,0.08) {$3\Delta_{\textrm{dc}}$};

        \draw[green!80!white] (1.25,0.75) -- (1.75,0.75);
        \draw[red!80!white] (0.75,0.25) -- (1.25,0.25);

        \draw[green!80!white] (2.5,0.75) -- (2.75,0.75);
        \draw[red!80!white] (1.75,0.25) -- (2.5,0.25);

        \draw[green!80!white] (3.05,0.75) -- (3.75,0.75);
        \draw[red!80!white] (2.75,0.25) -- (3.05,0.25);
    \end{scope}
    \begin{scope}[shift={(0,0.1)}]
        \draw[draw=none,fill=white] (0.75,0.2) rectangle (3.75,0.8);
        \draw [arrows = {-Stealth[inset=0pt, angle=30:3pt]}, color=black] (0.75,0.2) -- (3.85,0.2);
        \draw [arrows = {-Stealth[inset=0pt, angle=30:3pt]}, color=black] (0.75,0.2) -- (0.75,0.9);
        \node at (0.375,0.5) {$\lambda_\ell(k)$};
        \node at (0.65,0.2) {0};
        \node at (0.65,0.8) {1};
        \draw[thick,cyan] (0.75,0.525) -- (1.75,0.525);
        \draw[dotted,dash pattern=on 1pt off 1pt] (1.75,0.2) -- (1.75,0.525);
        \node at (1.8,0.08) {$\Delta_{\textrm{dc}}$};
        \draw[thick,cyan] (1.75,0.375) -- (2.75,0.375);
        \draw[dotted,dash pattern=on 1pt off 1pt] (2.75,0.2) -- (2.75,0.65);
        \node at (2.8,0.08) {$2\Delta_{\textrm{dc}}$};
        \draw[thick,cyan] (2.75,0.65) -- (3.75,0.65);
        \node at (3.75,0.08) {$3\Delta_{\textrm{dc}}$};
    \end{scope}
\end{tikzpicture}
\caption{Different control input values $\lambda_\ell$ result in different active and passive phase durations, that is $\delta_a$ and $\delta_p$, 
for a traffic light $\ell\in\mf l$. The bottom graph shows a possible input trajectory for the $\ell$-th traffic light spanning three duty cycles, the top graph shows proportion of duty cycle time
assigned to the active and passive phases as a result of the input.}
\label{fig:control_input}
\end{figure}
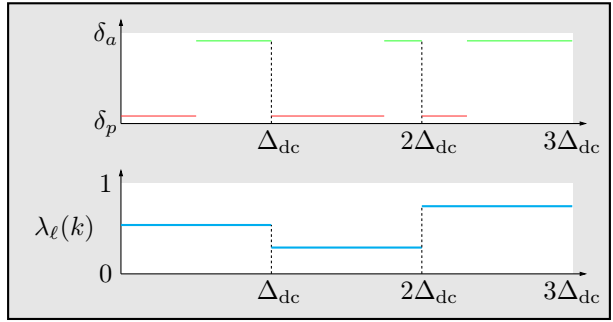

We consider the \textit{Split control} scheme with fixed duty cycle time $\Delta_{\textrm{dc}}$ for each controlled intersection. 
The actuated intersections are acting under the coordination of a centralized controller. 
The duty cycle of $\ell$-th actuated intersection, is a finite ordered set $\Pi_\ell$ whose elements are referred to as \textit{phases}. 
Given an intersection with $n$ approaches each equipped with a signal head that can be red (\textrm{r}), yellow (\textrm{y}) or green (\textrm{g}) for any time interval, 
we define a \textit{phase} as the $n$-tuple $y\in\{\textrm{r},\textrm{y},\textrm{g}\}^n$, where the $i$-th element represents the state of the $i$-th signal head.
We associate to each phase $i \in \Pi_\ell$ a duration $\delta_i \in \N$ in seconds and we define the duty cycle time $\Delta^\ell_{\textrm{dc{}}} \in \N$ as
\begin{equation*}
    \Delta^\ell_{\textrm{dc}}=\sum_{i\in\Pi_\ell} \delta_i.
\end{equation*} 
When a phase contains at least one '\textrm{g}' value we call it an \textit{active phase}; otherwise, we call it a \textit{passive phase}. 
Note that we include the duration of yellow phases in the duration of passive phases, see Figure~\ref{fig:phases}. 
The set of active phases for the $\ell$-th traffic light is denoted as $\mathcal{A_\ell}\subset \Pi_\ell$, 
the set of passive phases corresponds to $\mathcal{A}^\textrm{c}_\ell$. 
Intuitively, the main difference between an active phase and a passive one is that during an active phase, some
vehicles are allowed to cross the intersection. The duration of the \textit{active} and \textit{passive} phases are
\begin{equation*}
\delta_\textrm{a} = \sum_{i \in \mathcal{A_\ell}}\delta_i \ , \hspace{0.3cm }\text{ and } \hspace{0.3cm} \delta_\textrm{p} = \sum_{i \in \mathcal{A}^\textrm{c}_\ell}\delta_i = \Delta^\ell_\textrm{dc} - \delta_\textrm{a}\ , 
\end{equation*}
respectively.\\
We define the percentage of time, with respect to the duty cycle time, for which an intersection is active as 
$\lambda_\ell=\delta_\textrm{a}/\Delta^\ell_{\textrm{dc}} \in [0,1)$, $0\leq \delta_{\textrm{a}}\leq\Delta^\ell_{\textrm{dc}}$. 
By controlling the value $\lambda_\ell$ one controls the fraction of $\Delta_{\textrm{dc}}$ for which the intersection is active, 
where a value $\lambda_\ell(t) = 0$ corresponds to an intersection with all the signal heads red for the whole duty cycle.
Note that while the value $\lambda_\ell(t)=0$ is achievable, the value $\lambda_\ell(t) = 1$ is not due to the constant duration 
of the yellow phases embedded in the passive phase duration.
For the sake of simplicity, we consider the values $\lambda_\ell(t)$ to be real numbers, and we recover the corresponding $\delta_a$ 
by means of a floor operation. Figure \ref{fig:control_input} shows an example of input trajectory and for the $\ell$-th traffic light
and the resulting values of $\delta_a$ and $\delta_p$.

We define the control input given to our system at time $t$ as the vector $\bs \lambda(t) \coloneqq [\lambda_i (t)]_{i =1}^l \in \R^l$, with $\mf l \coloneqq  \{1,\cdots,l\}$ 
denoting all the actuated traffic lights. This formulation allows one to control a general intersection with a single scalar bounded input 
$\lambda_\ell (t)$. However, it comes with the disadvantage of relinquishing control over the durations of the individual phases, 
which determine traffic flows with different directionalities. To mitigate this shortcoming, one could introduce additional inputs to distinguish between 
different phases, at the cost of a more computationally expensive optimization problem. 
We assume that the traffic demand on one side of the intersection has greater magnitude than the opposite direction, 
which is often the case during peak hours. Under this assumption, the control input $\lambda_\ell(t)$ is sufficient to decongest the intersection.

\subsection{Macroscopic fundamental diagram}\label{subsec:MFD}
\begin{figure}[h!]
    \centering
    \begin{tikzpicture}
        \begin{axis}[black,
        width = \textwidth,
        height= 8 cm,
        grid = both,
        mark color= gray!50,
        legend style={cells={align=right}},
        legend columns = 1,
        legend entries={Data, MFD}, 
        ylabel={$\phi_i$},
        xlabel={$\rho_i$},
        xmin= 0,
        ymin=0,
        xmax=105,
        ymax=260
        ]
        \coordinate (a) at (16.24,225);
        \coordinate (b) at (16.24,15);
        \draw[dashed] (a) -- (b);
        \addplot [only marks,gray!50] table[x=density,y=flow,col sep = comma]{./csv/ZRH_NEW/mfd_reference_reduced_20.csv};    
        \addplot [smooth,green, ultra thick] table[x=reference x,y=reference y,col sep = comma]{./csv/ZRH_NEW/approximated_mfds.csv};
        \addplot [ only marks, 
                        mark= star, 
                        mark size=5pt,
                        point meta= explicit symbolic,
                        nodes near coords
                        ] coordinates {(16.24,0) [$\rho_{\tup{cr},i}$] };
        \addplot [ only marks, 
                        mark= star, 
                        mark size=5pt,
                        point meta= explicit symbolic,
                        nodes near coords
                        ] coordinates {(101.8,0) [$\rho_{\tup{max},i}$] };
        \end{axis}
    \end{tikzpicture}
    \caption{The \gls{MFD} for the region of Wiedikon, Zürich (solid green line) simulated below. 
    The critical density $\rho_{\tup{cr},i}$ is the density where the flow is maximum, and the maximal density of region $\rho_{\tup{max},i}$
    is the density where the flow becomes zero and region $i$ gridlocks.} 
    \label{fig:MFD}  
\end{figure}
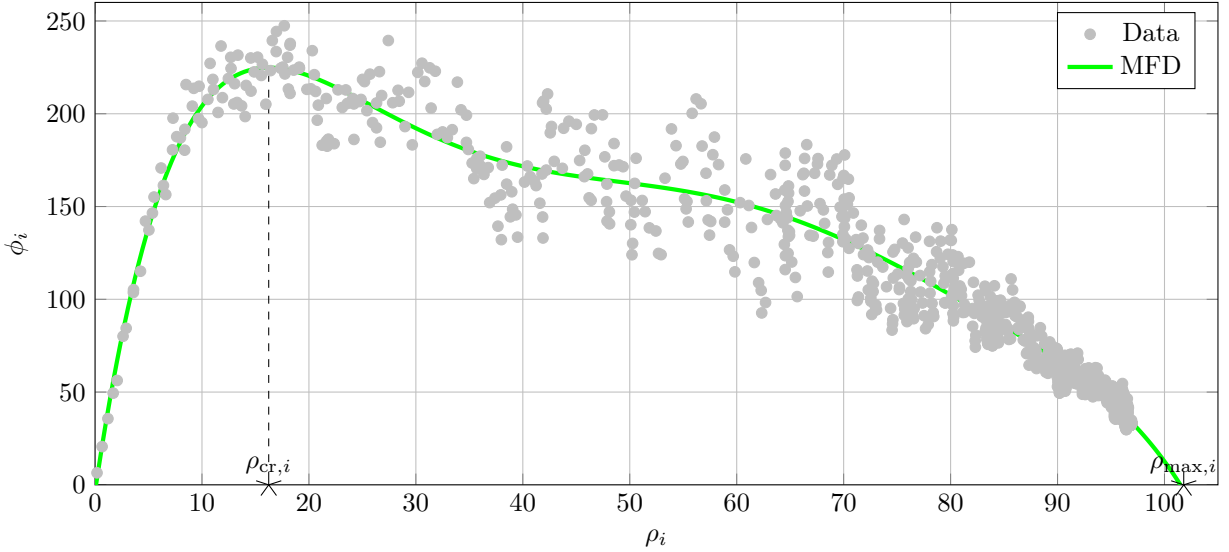

The concept of \gls{MFD} captures the relation between the flow $\phi_i$ and density 
$\rho_i$ within a region $i\in\bf p$. For each region $i \in \bs p$ we plot the measurements of $(\rho_i,\phi_i)$ against each other and 
estimate the \gls{MFD} using a 4-th degree polynomial, see Figure~\ref{fig:MFD}. We then use the estimated \gls{MFD} to identify 
each region's critical density and maximal density, denoted as $\rho_{\tup{cr},i}$, and $\rho_{\tup{max},i}$, respectively.
The critical density $\rho_{\tup{cr},i}$ corresponds to the abscissa of the maximum point for the \gls{MFD} 
of region $i$, where the region reaches its maximum possible flow and throughput. 
This observation will be used later in Section \ref{sec:DeePC} to define the reference trajectory to be tracked.
We define the vector of critical densities as 
$\bs \rho_{\tup{cr}} \coloneqq [\rho_{\tup{cr},i}]_{i=1}^p\in \R^p$.
The maximal density $\rho_{\max,i}$ is the smallest positive root 
of the polynomial used to estimate the \gls{MFD}, where the network reaches its maximum capacity and the 
flow equals zero. This state is often referred to as grid-lock, and we define the vector of the maximal 
densities as $\bs \rho_{\max}\coloneqq [\rho_{\max,i}]_{i=1}^p\in \R^p$.

The \gls{MFD} captures the non-linear and non-injective relation
that exists between the density $\rho_i$ and flow $\phi_i$ in each region $i \in \bs p$.
As we will discuss in Section \ref{sec:DeePC}, \gls{DeePC} works best under an approximately linear relationship 
between the input and output measurements of the system. 
Despite the non-linear dependencies encoding the system dynamics through the \gls{MFD}, as shown in Section \ref{sec:simulations} and supported by \cite{rimoldi2023urban}, the \gls{DeePC} algorithm is able to find effective control
policies using $\bs \lambda$ as the control input and $\bs \rho$ as the output measurement. This observation leads us to conjecture that the relationship between the input
$\bs \lambda$ and the density $\bs \rho$ must be approximately linear.
Moreover, empirical trials conducted by the authors using only the flow $\bs \phi$ or a combination of flow and density 
$\bs y (t) := \text{col}(\bs \phi (t), \bs \rho (t))$ as output of the system resulted in degraded performance 
of the algorithm (data not shown). This makes us conjecture that the relationship between $\bs \lambda$ and $\bs \phi$ must be non-linear, a fact that is not
surprising given the non-linear relationship between density and flow. 
The two conjectures are summarised in the commutative diagram in Figure \ref{fig:commutative-diagram}.

\begin{figure}[ht!]      
\centering
\begin{tikzpicture}[->,>=stealth,
                    auto,
                    node distance=4cm,
                    thick,
                    main node/.style={draw=none,font=\sffamily\Large\bfseries},
                    decoration = {snake, 
                    pre length=3pt,post length=7pt,
                    }]
    \node[main node] (1) {$\lambda$};
    \node[main node] (2) [right of=1] {$\rho$};
    \node[main node] (3) [below of=2] {$\phi$};

    \draw[-latex,line width=1pt, decorate] (1) -- (2) node[midway, above] {approx. linear};
    \draw[-latex,line width=1pt, dashed] (1) -- (3) node[midway, below,sloped] {non-linear};
    \draw[-latex,line width=1pt, dashed] (2) -- (3) node[midway, above,sloped] {non-linear} node[midway, below,sloped] {non-injective};
    
\end{tikzpicture}
\caption{Commutative diagram of the relationships between $\lambda$, $\rho$ and $\phi$. The relationship between $\rho$ and $\phi$
is assumed to be a \gls{MFD} similar to Figure \ref{fig:MFD}. }
\label{fig:commutative-diagram}
\end{figure}
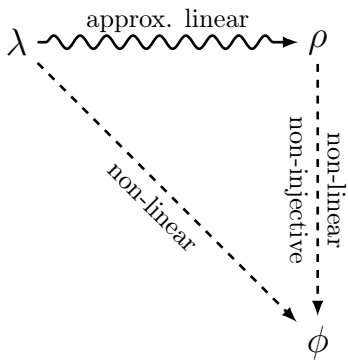

\subsection{Behavioral traffic dynamics}
We model traffic as a dynamical system using behavioral system theory \cite{willems1986timeI,willems1986timeII,willems1986timeIII}. 
The resulting formulation naturally generalizes perimeter traffic control by allowing the actuators to be 
placed anywhere in the network. A behavioral formulation is flexible, as it can in principle integrate any kind of measurement 
data. For example, in a traffic system, one could consider CO$_2$ emissions or noise as alternative outputs of the system. 
Below we give a brief introduction to behavioral systems theory, the interested reader is referred to \mbox{Appendix \ref{app:preliminaries}}
for more details. 

In behavioral systems theory a \textit{dynamical system} (or, briefly, \textit{system}) is a triple 
$\Sigma=(\T,\W,\B)$ \cite{willems1986timeI}, where $\T$ is the \textit{time set}, $\W$ is the \textit{signal set}, and 
$\B \subseteq (\W)^{\T}$ is the \textit{behavior} of the system.
The behavior characterises the set of all signal trajectories over the time set $\T$ that are compatible with the system.
More specifically, if the signal set is $\W=\R^q$ the set of finite trajectories $w=(w(1),\ldots,w(T))$ of length ${T\in\N}$, with ${w(t)\in \W}$ for ${t\in\T \coloneqq [1,\dots,T]\subset \N}$, 
is $\W^\T= {(\R^q)}^\T$ and the behavior is a subset $\mathcal{B} \subseteq (\R^q)^\T$. 
We identify every finite trajectory ${w\in(\R^q)}^\T$ with the corresponding vector ${\col(w(1),\ldots,w(T))\in\R^{qT}}$. 
In this case, a system $\Sigma$ is called \textit{linear} if the corresponding behavior $\B$ is a linear subspace, and \textit{time-invariant} if $\B$ is 
shift-invariant, \textit{i.e.},  if ${w\in \B}$ implies ${\sigma w\in \B}$, where $\sigma$ is the \textit{shift operator} 
defined as $(\sigma w)(t)\coloneqq w(t+1)$. The class of all complete \gls{LTI} systems is denoted by $\Li^q$, and for simplicity 
we write ${\B \in \Li^{q}}$ to denote a linear behavior.

To model the traffic dynamics outlined above in this framework, we consider behaviors $\B \in \Li^{m+p}$ over a signal $ w = \col(u, y) \in \R^q$ divided into an
input $u$ of dimension $m = l + p^2$ and an output $y$ of dimension $p$. 
As output we consider the traffic density $\bs \rho \in \R^p$ in the $p$ regions of the traffic network. The inputs $u$ are further subdivided into controllable
inputs $\bs \lambda \in \R^l$ (the fractions of active phase of the controlled intersections) and uncontrollable inputs $\bs d \in \R^{p^2}$ (the demands of all possible
origin-destination region pairs). This leads to 
\begin{flalign*}
    & u(t) \coloneqq  \col(\bs \lambda(t),\, \bs d(t)) \in \R^m,\, \\
    & y(t) \coloneqq  \bs \rho(t) \in \R^p,
\end{flalign*}

and a signal $w = \col(u,y)\in \R^q$ of dimension $q = l + p + p^2$. The dynamics of the signal are described by a non-linear function $h$ relating, for example, the density or flow in one
time instant with those in the next time instant. Common choices for $h$ include partial differential equations, discretised by first or
second order methods~\cite{first-order-traffic-model,second-order-traffic-model}. 
In this work, we do not concern ourselves with the choice of a specific model for $h$, but consider trajectories of the corresponding quantities and construct
a high-dimensional linear approximation $f$ of $h$ based solely on data.
Let $T,\Tini \in \N$ with $\Tini < T$ and a fixed $t\in \T$, consider a finite output trajectory
$y|_{[t-\Tini,t]}=(y(t-\Tini),\dots,y(t))$, and input trajectory
$u|_{[t-\Tini,t]}=(u(t-\Tini),\dots,u(t))$. We then consider a linear function $f: (\R^p)^\Tini\times (\R^m)^\Tini \rightarrow \R^p$ of the following form
\begin{equation}
    \label{eq:rho_dyn}
    y(t+1) = f(y|_{[t-\Tini,t]},u|_{[t-\Tini,t]}),
\end{equation} 

and introduce the behavior associated with the traffic density as
\begin{equation}\label{eq:behavior}
    \B_{\bs \rho} = \left\{ (y,u) \in  \mathbb{R}^{(l+p+p^2)\Tini}| \text{(\ref{eq:rho_dyn}) holds} \right\}.
\end{equation}

Note that in what follows we never explicitly construct the behavior in \ref{eq:behavior}, but encode the constraint in \ref{eq:rho_dyn}
implicitly using data collected from the system.

To ensure that the trajectories included in the behavior are realistic, additional constraints need to be imposed. 
Let $\Tf \in \N$ be a multiple of the duty cycle time $\Delta_{\tup{dc}}$ and such that $\Tini +\Tf \leq T$.
We assume that once the input $\bs \lambda$ has been chosen, it is kept constant during the duty cycle, that is for a 
duration $\Delta_{\tup{dc}}$. 
This translates into the linear constraint ${M\bs u =\bs 0}$, where 
$M\in\R^{m\Tf\times m\Tf}$ is used to impose for every $\ell \in \bf l$ and 
$k \leq \Tf$ the constraint $\lambda_{\ell}(k+1)=\lambda_{\ell}(k+2)=
\cdots=\lambda_{\ell}(k+\Delta_{\tup{dc}})\,$. 
We further assume that at every time instant the controllable input has to satisfy the box constraints $\bs \lambda(k)\in[\underline{\bs \lambda},
\overline{\bs \lambda}]\subset [0,1)^{l}$, where $\underline{\bs \lambda}=\col(\underline{\lambda}_1,
\cdots , \underline{\lambda}_{l})$ and $\overline{\bs \lambda}\col(\overline{\lambda}_1,
\cdots , \overline{\lambda}_{l})$  
As mentioned before, the values are not allowed to reach the boundary value 1 due to the yellow phase constant duration.

The uncontrollable input $\bs d$ is assumed to be known and equal to an estimate of the
demand \mbox{$\overline{\bs d}=\col((\bs d(t),\cdots,\bs d(t+\Tf)))\in\R^{p^2\Tf}$ }
over the prediction horizon, i.e., $Du=\overline {\bs d}$ where the matrix \mbox{$D\in\R^{m\Tf\times p^2\Tf}$} 
projects to the entries of the input corresponding to the demand. Therefore, the set of constraints for the inputs over the whole prediction horizon reads as
\begin{equation}\label{eq:cal_U}
    \mc U=\left\{u\in \big([\underline{\bs \lambda},\overline{\bs \lambda}]\times\R_+^{p^2}\big)^{T_{\tup f}} \,|\, Mu=0,\, Du=\overline {\bs d}\right\}.
\end{equation}

Similarly, box constraints on the output are selected to ensure that the density remains non negative and below 
grid-lock conditions $\bs \rho_{\max}\in\R_+^{p}$, leading to $y(t)\in[0,\bs \rho_{\max}]$ and a constraint set
\mbox{$\mc Y\coloneqq[0,\,\bs \rho_{\max}]^{T_{\tup f}} $}.

\section{Data-enabled predictive control}\label{sec:DeePC}

Consider a LTI system ${\B\in \Li ^{m+p}}$, with $m$ inputs and $p$ outputs, assume that data 
recorded offline from system $\B$ is available. 
Specifically, assume that an input sequence \mbox{$u_{\tup{d}} =\col(u_{\tup{d}}(1),\cdots,u_{\tup{d}}(T)) \in \R^{mT}$} 
of given length $T\in \N$ is applied to the system $\B$ and that the corresponding output sequence 
\mbox{$y_{\tup{d}}=\col(y_{\tup{d}}(1),\cdots,y_{\tup{d}}(T))\in \R^{pT}$} is recorded.  
The subscript ``$\textup{d}$'' is used to indicate that these are trajectories of data samples collected offline.  
Finally, let $\Tini\in \N$ and $\Tf\in \N$, with $\Tini + \Tf \le T,$ and  
assume that the sequence $w_{\tup{d}}= \col(u_{\tup{d}}(t),y_{\tup{d}}(t))\in \R^{(m+p)T}$, for ${t\in \T \coloneqq [1,\dots,T]\subset \N}$, 
satisfies the generalized persistency of excitation condition (see Appendix \ref{app:preliminaries}).
Intuitively, the persistency of excitation condition requires the sequence $w_{\tup{d}}$ to contain enough information 
to allow for the systems' behavior to be reconstructed.
We define the \textit{Hankel matrix} of depth ${L\in\T}$ associated with the finite sequence ${w\in {\R^{qT}}}$ as 
\begin{equation} \label{eq:Hankel} 
\! H_{L}(w) \! = \!
\scalebox{0.85}{$
\bma  \nn
\begin{array}{ccccc}
w(1) & w(2)  & \cdots &  w(T-L+1)   \\
w(2) & w(3)  & \cdots &   w(T-L+2)   \\
\vdots  & \vdots  & \ddots & \vdots  \\
w(L) & w(L+1)  & \cdots  & w(T)
\end{array}
\ema
$} . \! \!
\end{equation}

Next, we partition the input-output data into two parts, \emph{past data} $(\Up,\Yp)$ and \emph{future data} $(\Uf,\Yf)$. 
Formally, given the time horizons $\Tini \in \N$ and $\Tf \in \N$, we define 
\begin{equation}\label{eq:UpUfYpYf}
\begin{pmatrix}
\Up \\ \Uf 
\end{pmatrix}= H_{\Tini+\Tf}(u_{\textup{d}}), \quad
\begin{pmatrix}
\Yp \\ \Yf 
\end{pmatrix}= H_{\Tini+\Tf}(y_{\textup{d}}),
\end{equation}
where ${\Up\in\R^{(m\Tini)\times (T-\Tini+1)}}$ is composed of the first $\Tini$ block-rows of the matrix 
$H_{\Tini+\Tf}(u_{\textup{d}})$ and ${\Uf \in\R^{(m\Tf)\times (T-\Tf+1)}}$ of the last $\Tf$ block-rows, $\Yp$ and $\Yf$ are attained similarly from $H_{\Tini+\Tf}(y_{\textup{d}})$. 
In the remainder, past data, denoted by the subscript ``$\mathrm{p}$'',  is used to implicitly estimate the initial condition of the underlying state, 
whereas the future data, denoted by the subscript ``$\mathrm{f}$'',  is used to predict future trajectories. 

Given ${\Tf \in \N}$, a reference trajectory for the output $\hat y$ and the input $\hat u$, most recent past input/output data $
\wini = \col(\uini,\yini)$, an input constraint set $\mathcal{U}\subseteq \R^{mT_{\tup f}}$, an output constraint set 
$\mathcal{Y}\subseteq \R^{pT_{\tup f}}$, an output cost matrix $Q \geq 0 \in \R^{p\times p}$, a control cost matrix 
$R \geq 0 \in \R^{m\times m}$, a regularization function $\psi:\R^{T-\Tini-\Tf+1}\to \R$, and parameters $\lambda_y,\lambda_1,\lambda_2\in\R$, 
the \gls{DeePC} algorithm solves the optimisation problem:

\begin{subequations}\label{eq:DeePC}

\begin{align}
\underset{u,y,g, \sigma_y}{\text{min}}\,
&  
\displaystyle
\sum_{k=1}^{\Tf}\left\|y(k)-\hat{y}(t+k)\right\|_Q^2 
+\left\|u(k)-\hat{u}(t+k)\right\|_R^2 +\psi(g) + \lambda_y\left\|\sigma_y\right\|_1 \label{eq:DeePC-cost} \\  
\text{s.t.\,}
& \begin{pmatrix}
\Up \\ \Yp \\ \Uf \\ \Yf
\end{pmatrix}g
=\begin{pmatrix}
\uini \\ \yini \\ u \\ y
\end{pmatrix} +
\begin{pmatrix}
    0 \\ \sigma_y \\ 0 \\ 0
\end{pmatrix}, \label{eq:DeePC-hankel-constraints}\\
& u\in \mathcal{U}, \: y\in\mathcal{Y}. \label{eq:DeePC-set-constraints}
\end{align}
\end{subequations}

Solving \eqref{eq:DeePC} generates a trajectory of future inputs , $u$, and outputs, $y$, for the next $\Tf$ time steps.
The equality constraint ensures that the trajectories are compatible with the system dynamics (encoded in the data matrix) 
and the most recent data collected from the system and stored in $\uini$ and $\yini$.  
The regularization term $\psi(g)$ is defined as 
\begin{equation*}
    \psi(g): = \lambda_1||(I-\Pi)g||_2^2 + \lambda_2||g||_1 \text{ where } \Pi = \begin{bmatrix}\Up\\ \Yp \\ \Uf \end{bmatrix}^\dagger \begin{bmatrix}\Up\\ \Yp \\ \Uf \end{bmatrix}.
\end{equation*}
The first addend $||(I-\Pi)g||_2^2$ computes the orthogonal projection on the kernel of the first three block equations of \eqref{eq:DeePC-hankel-constraints}, while the
second added $||g||_1$ computes a low-rank approximation on the column space of $H$.
Together the two terms greatly improve the performance of \gls{DeePC} in non-linear settings, for more details on the regularization action 
the interested reader is referred to \cite{DeePC-regularized}.
The slack variable $\sigma_y$ relaxes the constraint to ensure that it remains feasible; it is penalised by the regularization term $\lambda_y||\sigma_y||_1$ to 
keep the constraint violation small. The set membership constraints \eqref{eq:DeePC-set-constraints} ensure that the decision variables remain within the admissible
input domain $\mathcal{U}$ and the physically consistent output domain $\mathcal{Y}$.
The cost \eqref{eq:DeePC-cost} trades off tracking accuracy $||y(k) - \hat{y}(t+k)||^2_Q$ and control effort $||u(k) -\hat{u}(t+k)||^2_R$ with respect to
the reference values $\hat{y}$ and $\hat{u}$. Once the optimisation problem is solved, the first element of the trajectory $u$ is applied to the system, a new measurement 
is collected, the vectors $\uini$ and $\yini$ are updated with this new information and the process is repeated in a receding horizon fashion.
The process is summarised in Algorithm \ref{alg:robustdeepc}.
\setlength{\intextsep}{10pt}
\begin{algorithm}[h!]
	\caption{DeePC}
	\label{alg:robustdeepc}
	\textbf{Input:} 
    Future time horizon ${\Tf \in \N}$, a reference trajectory for the output $\hat y=(\hat y_{0},\hat y_{1},\cdots)\in (\R^{p})^{\N}$ 
    and the input $\hat u=(\hat u_{0},\hat u_{1},\cdots)\in (\R^{m})^{\N}$, past input/output data $\wini = \col(\uini,\yini)\in \B|_{ [1,\Tini] }$, 
    an input constraint set $\mathcal{U}\subseteq \R^{mT_{\tup f}}$, an output constraint set $\mathcal{Y}\subseteq \R^{pT_{\tup f}}$, 
    a output cost matrix $Q\in \R^{p\times p}$, a control cost matrix $R \in \R^{m\times m}$, a regularization function 
    $\psi:\R^{T-\Tini-\Tf+1}\to \R$, and parameter $\lambda_y\in\R$.
	\begin{algorithmic}[1]
		\STATE \label{step:deepcbegin} Compute the minimizer $g^*$ of ~\eqref{eq:DeePC}.
		\STATE Compute optimal input sequence $u^{\star} = \Uf {g}^{\star}$.
		\STATE Apply optimal input sequence $(u_t,\cdots,u_{t+j-1})=(u_1^{\star},\cdots,u_{j}^{\star})$ for some $j\leq \Tf$.
		\STATE Set $t$ to $t+j$ and update ${u}_{\textup{ini}}$ and ${y}_{\textup{ini}}$ to the $\Tini$ most recent input/output measurements.
		\STATE Return to~\ref{step:deepcbegin}.
	\end{algorithmic}
\end{algorithm}

For our traffic model we use the quantity $\bs \rho_{\tup{cr}}$ 
estimated from the regions \gls{MFD}s in Section~\ref{subsec:MFD} as constant output reference $\hat{y}$. 
In \eqref{eq:DeePC}, we minimize the distance between the regions' density $\rho_{i}$ and critical $\rho_{\tup{cr},i}$ at each time step. 
The \gls{MFD} (and hence the reference $\hat{y}$) can initially be estimated using data from the historical operation 
of the system. We note, however, that this data implicitly depends on the policy used to control the intersections. 
Since our goal is to change this policy, it is conceivable that when the system is operating under the DeePC 
algorithm, its MFD will change. This effect can be mitigated by re-estimating the \gls{MFD} online with data collected under the \gls{DeePC} policy,
 the reference $\hat{y}$ can be recomputed and the process repeated. One should be careful to ensure the stability of the algorithm as 
considering a non-continuous trajectory can be problematic. For the sake of a fair comparison among the different control policies explored in Section \ref{sec:simulations},
we decided to leave this online optimization of the control scheme for future research.

The reference value of the inputs $\hat u :=\col(\hat\lambda, \hat d)$ is composed by two parts, the first is 
the default green time to duty cycle time ratio $\hat \lambda$, which is the ratio that the traffic lights would 
exhibit given their phase specification if no dynamic control were to be applied. 
The latter is exogenous and set equal to $\overline {\bs d}$ according to \eqref{eq:cal_U}.
In the cost function in \eqref{eq:DeePC}, the matrices $R$ and $Q$ are chosen positive definite. 
Note that the components of $\lVert u - \hat u\rVert_R$ associated with $\overline {\bs d}$ are always equal to zero due to the chosen reference and the constraints on $u$. 
Finally, the regularization term $\psi$ is chosen as in \cite{coulson2019data}. 
Note that in general the optimization problem only requires convexity of the cost function.

\section{Simulation}\label{sec:simulations}  
We use \gls{SUMO} to simulate traffic dynamics at a microscopic level and generate realistic data. 
An implementation of the controller then interacts with the simulation in closed-loop through a custom interface built on TraCI.
The interface code as well as the simulations networks are openly available at \url{https://github.com/AlessioRimoldi/TrafficGym}.
The routing of each vehicle is recomputed periodically using the A* algorithm, accounting for current traffic conditions. 
Once a vehicle reaches its destination, it is removed from the simulation.
We consider two different traffic networks, a lattice network and the urban traffic network of Z\"urich.
We use a simulation where the traffic lights follow the standard static cycle specified by the 
\gls{SUMO} configuration as a baseline, in this case no input is given to the simulation.
We further compare \gls{DeePC} with the \gls{MPC} formulation detailed in \cite{linearMPC}, in this work the authors first 
devise a linear program formulation to solve the finite-time optimal perimeter flow control problem. 
The traffic dynamics are described first using a non-linear model derived from the \gls{MFD}, the linear formulation is then 
achieved as a linearization of this model, employing a piecewise approximation of the \gls{MFD}.
The linear program is then solved in a rolling horizon using a feedback \gls{MPC} framework, for completeness we
report the \gls{MPC} formulation in Appendix \ref{app:mpc}. 
The parameters used by the controllers can be found in \mbox{Appendix \ref{app:parameters}}.

\subsection{Lattice network simulation}\label{subsec:lattice network}
\begin{figure}[h!]
        \begin{subfigure}[t]{.45\textwidth}
        \centering
                \begin{tikzpicture}
                    \begin{axis}[black,
                    scale=0.75,
                    grid = both,
                    xtick distance= 600,
                    xticklabels = {0,0,\,,20,\,,40,\,,60,\,,80,\,,100,\,,120,\,,140},
                    xlabel = Time (min),
                    ylabel = Demand (veh/min),
                    xmin=0,
                    xmax=7200
                    ]
                        \addplot [blue,ultra thick] table[x=0,y= from 0 to 1,col sep = comma] {./csv/GRID/demand.csv};
                    \end{axis}
                \end{tikzpicture}
            \caption{}
            \label{fig:demandgrid}
        \end{subfigure}
        \hfill
        \begin{subfigure}[t]{.45\textwidth}
            \centering
            \begin{tikzpicture}[scale=1.25, every node/.style={scale=1}]
            
            \node at (-6.5,5) {};
            \node at (-1.5,5.5) {};
            \node at (-1.5,1) {};
            \node at (-6.5,1) {};
             
            \draw[thick, color = magenta!70!white]   (-5.5,5) rectangle (-2,1.5);
            \draw [thick, color = magenta!70!white] (-5,4.5) rectangle (-2.5,2);
            \draw [thick, color = magenta!70!white](-5,5) -- (-5,4.5);
            \draw [thick, color = magenta!70!white](-4.5,5) -- (-4.5,4.5);
            \draw [thick, color = magenta!70!white](-4,5) -- (-4,4.5);
            \draw [thick, color = magenta!70!white](-3.5,5) -- (-3.5,4.5);
            \draw [thick, color = magenta!70!white](-3,5) -- (-3,4.5);
            \draw [thick, color = magenta!70!white](-2.5,5) -- (-2.5,4.5);
            \draw [thick, color = magenta!70!white](-5,2) -- (-5,1.5);
            \draw [thick, color = magenta!70!white](-4.5,2) -- (-4.5,1.5);
            \draw [thick, color = magenta!70!white](-4,2) -- (-4,1.5);
            \draw [thick, color = magenta!70!white](-3.5,2) -- (-3.5,1.5);
            \draw [thick, color = magenta!70!white](-3,2) -- (-3,1.5);
            \draw [thick, color = magenta!70!white](-2.5,2) -- (-2.5,1.5);
            \draw [thick, color = magenta!70!white](-5.5,4.5) -- (-5,4.5);
            \draw [thick, color = magenta!70!white](-5.5,4) -- (-5,4);
            \draw [thick, color = magenta!70!white](-5.5,3.5) -- (-5,3.5);
            \draw [thick, color = magenta!70!white](-5.5,3) -- (-5,3);
            \draw [thick, color = magenta!70!white](-5.5,2.5) -- (-5,2.5);
            \draw [thick, color = magenta!70!white](-5,2) -- (-5.5,2);
            \draw [thick, color = magenta!70!white](-2.5,4.5) -- (-2,4.5);
            \draw [thick, color = magenta!70!white](-2.5,4) -- (-2,4);
            \draw [thick, color = magenta!70!white](-2.5,3.5) -- (-2,3.5);
            \draw [thick, color = magenta!70!white](-2.5,3) -- (-2,3);
            \draw [thick, color = magenta!70!white](-2.5,2.5) -- (-2,2.5);
            \draw [thick, color = magenta!70!white](-2,2) -- (-2.5,2);
            
            \draw [thick, color = magenta!70!white](-4,4) -- (-4,4.5);
            \draw [thick, color = magenta!70!white](-3.5,4) -- (-3.5,4.5);
            \draw [thick, color = magenta!70!white](-4,2) -- (-4,2.5);
            \draw [thick, color = magenta!70!white](-3.5,2) -- (-3.5,2.5);
            \draw [thick, color = magenta!70!white](-5,3.5) -- (-4.5,3.5);
            \draw [thick, color = magenta!70!white](-4.5,3) -- (-5,3);
            \draw [thick, color = magenta!70!white](-3,3.5) -- (-2.5,3.5);
            \draw [thick, color = magenta!70!white](-2.5,3) -- (-3,3);
            
            \draw [thick,dashed , color = green!80!black] (-4.5,4) rectangle (-3,2.5);
            \draw [thick, dashed, color = green!80!black](-4.5,3.5) -- (-3,3.5);
            \draw [thick, dashed, color = green!80!black](-4.5,3) -- (-3,3);
            \draw [thick, dashed, color = green!80!black](-4,4) -- (-4,2.5);
            \draw [thick, dashed, color = green!80!black](-3.5,4) -- (-3.5,2.5); 
            
            \node[draw, shape = circle, fill=black, minimum size = 0.1 cm, inner sep =0pt] at (-4,3.5) {};
            \node[draw, shape = circle, fill=black, minimum size = 0.1 cm, inner sep =0pt] at (-3.5,3.5) {};
            \node[draw, shape = circle, fill=black, minimum size = 0.1 cm, inner sep =0pt] at (-4.5,3.5) {};
            \node[draw, shape = circle, fill=black, minimum size = 0.1 cm, inner sep =0pt] at (-3,3.5) {};
            \node[draw, shape = circle, fill=black, minimum size = 0.1 cm, inner sep =0pt] at (-4,3) {};
            \node[draw, shape = circle, fill=black, minimum size = 0.1 cm, inner sep =0pt] at (-3.5,3) {};
            \node[draw, shape = circle, fill=black, minimum size = 0.1 cm, inner sep =0pt] at (-4.5,3) {};
            \node[draw, shape = circle, fill=black, minimum size = 0.1 cm, inner sep =0pt] at (-3,3) {};
            \node[draw, shape = circle, fill=black, minimum size = 0.1 cm, inner sep =0pt] at (-3.5,4) {};
            \node[draw, shape = circle, fill=black, minimum size = 0.1 cm, inner sep =0pt] at (-4,2.5) {};
            \node[draw, shape = circle, fill=black, minimum size = 0.1 cm, inner sep =0pt] at (-4,4) {};
            \node[draw, shape = circle, fill=black, minimum size = 0.1 cm, inner sep =0pt] at (-3.5,2.5) {};
            \draw [thick](-6.5,1.5998) -- (-6.5,1.5) -- (-6,1.5) -- (-6,1.5998);
            \node at (-6,1.400) {\scalebox{0.5}{$100$\,m}};
            \end{tikzpicture}
            \centering 
            \caption{}
            \label{fig:netgrid}
            \end{subfigure} 
            \caption{ Left panel shows the demand profile, the vehicles move from the outer to the inner region.
            The network is partitioned into the \textit{outer region} (solid magenta line) and in 
            \textit{inner region} (dashed green line). Black dots represent the controlled traffic lights.}
\end{figure}
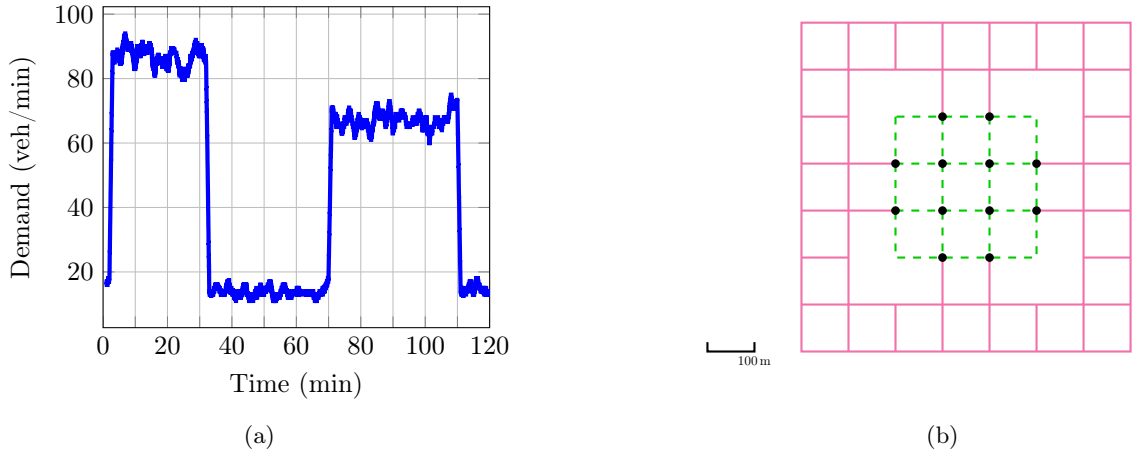

The lattice network is composed of 104 two-way roads and 64 intersections. 
Figure~\ref{fig:netgrid} shows the network partitioned into two regions emulating a city 
center and its suburban area, connected by 8 roads. We refer to these two regions as the inner and outer region respectively.

\paragraph{Actuators and demand profile}
We selected twelve traffic lights to be used as actuators, eight of which are at the boundary between the two 
regions and can be used to perform classical perimeter control, while the remaining are in the inner region to manage 
the internal dynamics, see Figure~\ref{fig:netgrid}. All controlled intersections have 
four approaches. We consider a standard duty cycle time of $\Delta_{\textrm{dc}} = 90\ s$, with two green phases 
lasting 42 seconds, plus two yellow phases lasting 3 seconds. 
The demand profile in Figure~\ref{fig:demandgrid} has two distinct peaks, spanning two hours with a total of 6149 vehicles.
Each vehicle is generated at the external boundary of the outer region, and assigned a destination edge inside the inner region.
This makes the demand unidirectional, from the outer to the inner region, no additional vehicles are generated.

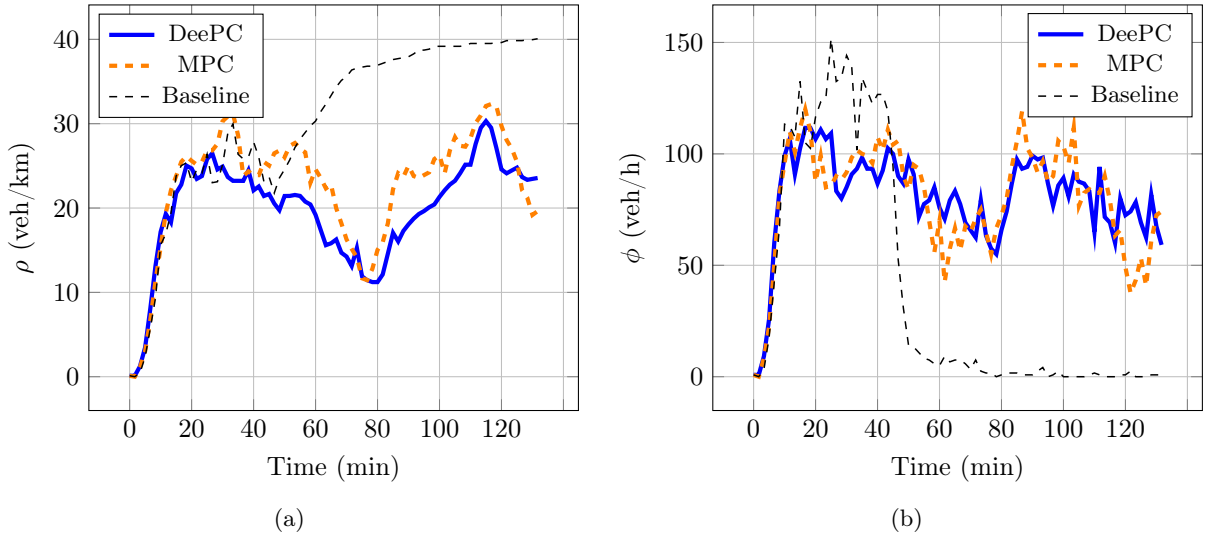
\begin{figure}[h!]
        \begin{subfigure}[t]{.49\textwidth}
        \centering
                \begin{tikzpicture}
                    \begin{axis}[black,
                    width = \textwidth,
                    grid = both,
                    x coord trafo/.code={
                        \pgfmathparse{\pgfmathresult/60}
                    },
                    x coord inv trafo/.code={
                        \pgfmathparse{\pgfmathresult*60}
                    },
                    xticklabels = {0,0,20,40,60,80,100,120},
                    xlabel = Time (min),
                    ylabel = $\rho$ (veh/km),
                    legend style={cells={align=right},at={(0.02,0.85)},anchor=west},
                    legend columns = 1,
                    legend entries={\small {\gls{DeePC}},\small \gls{MPC},\small {Baseline}},
                    ]  
                        \addplot [blue,ultra thick] table[x=0,y= Region 1,col sep = comma] {./csv/GRID/density/results/density.csv};
                        \addplot [orange,dashed,ultra thick] table[x=0,y= Region 1,col sep = comma] {./csv/GRID/mpc/results/density_results.csv}; 
                        \addplot [black,dashed, line width = 0.2 mm] table[x=0,y= Region 1,col sep = comma] {./csv/GRID/NoControl/results/density.csv};
                    \end{axis}
                \end{tikzpicture}
            \caption{}
            \label{fig:GRIDdensity}
        \end{subfigure}
        \begin{subfigure}[t]{.49\textwidth}
        \centering
                \begin{tikzpicture}
                    \begin{axis}[black,
                    width = \textwidth,
                    grid = both,
                    x coord trafo/.code={
                        \pgfmathparse{\pgfmathresult/60}
                    },
                    x coord inv trafo/.code={
                        \pgfmathparse{\pgfmathresult*60}
                    },
                    xticklabels = {0,0,20,40,60,80,100,120}, 
                    xlabel = Time (min),
                    ylabel = $\phi$ (veh/h),
                    legend style={cells={align=right}},
                    legend columns = 1,
                    legend entries={\small {\gls{DeePC}},\small \gls{MPC},\small {Baseline}}
                    ]
                        \addplot [blue,ultra thick] table[x=0,y= Region 1,col sep = comma] {./csv/GRID/density/results/flow.csv};
                        \addplot [orange,dashed,ultra thick] table[x=0,y= Region 1,col sep = comma] {./csv/GRID/mpc/results/flow_results.csv}; 
                        \addplot [black,dashed, line width = 0.2 mm] table[x=0,y= Region 1,col sep = comma] {./csv/GRID/NoControl/results/flow.csv};
                    \end{axis}
                \end{tikzpicture}
            \caption{}
            \label{fig:GRIDFlow}
        \end{subfigure} 
        \caption{Evolution of the density (left panel) and flow (right panel) in inner region of the lattice network.}
    \label{fig:GRID}
\end{figure}

\paragraph{Results}
The evolution of the density and flow in the inner region is shown in 
Figure~\ref{fig:GRID}. 
The baseline simulation reaches a state of grid-lock shortly before 
the end of the first peak of demand, leading to high travel times. 
Due to the congestion only 4636 vehicles complete their trip. 
On the other hand, control of the traffic lights via \gls{DeePC} and \gls{MPC}
prevents grid-lock, maintaining a stable flow throughout the length of the simulation. 
Table \ref{tab:avgmetricsGRID} shows the average travel time and emissions 
metrics over the number of vehicles that completed their trip, we observe
\gls{DeePC} performing better than \gls{MPC} on all metrics. We conjecture that this 
improvement in performance is due to the implicit model defined by the Hankel matrix included in 
\gls{DeePC} capturing the underlying traffic dynamics better than the linear formulation used by
\gls{MPC}, note that both models perform a linear approximation. 

\begin{table}[h!]
    \centering
        \begin{tabular}{@{}lllll@{}}
        \toprule
        {} & Baseline & MPC & \gls{DeePC} & \% variation\\
        \midrule
        Travel Time (min)    & 27.5 & 26.03 & $\boldsymbol{24.93}$ & - 9.34\%\\
        Waiting Time (min)   & 25.33 & 22.24 & $\boldsymbol{21.39}$ & - 15.55\%\\
        CO Emissions(g)     & 259.65 & 239.59 & $\boldsymbol{229.5}$ & - 10.92\%\\
        CO2 Emissions (g) & 4,351.16  & 4,084.12 & $\boldsymbol{3,917.79}$ & - 9.95\% \\
        HC Emissions (g) & 1.28 & 1.19 & $\boldsymbol{1.14}$ & - 10.37\%\\
        PMx Emissions (g) & 0.11 & $\boldsymbol{0.1}$ & $\boldsymbol{0.1}$ & - 8.95\%\\
        NOx Emissions (g) & 1.98 & 1.86 & $\boldsymbol{1.78}$ & - 10.10\%\\
        Fuel consumption (ml) & 1,387.91 & 1,302.73 & $\boldsymbol{1,249.67}$ & - 9.96\%\\
        Trips completed & 4,636 & 5,377 & $\boldsymbol{5,642}$ & + 21.69\%\\
        \bottomrule
    \end{tabular}
    \caption{Average metrics for the lattice network}
    \label{tab:avgmetricsGRID}
\end{table}

\subsection{Z\"urich simulation}\label{subsec:zurich}
\begin{figure}[h!]
\centering
     \begin{subfigure}[t]{.49\textwidth}
     \centering
        \begin{tikzpicture}

        \node[draw=none,fill=none] at (0,0) {\includegraphics[width=.95\linewidth,trim={300 250 400 450},clip]{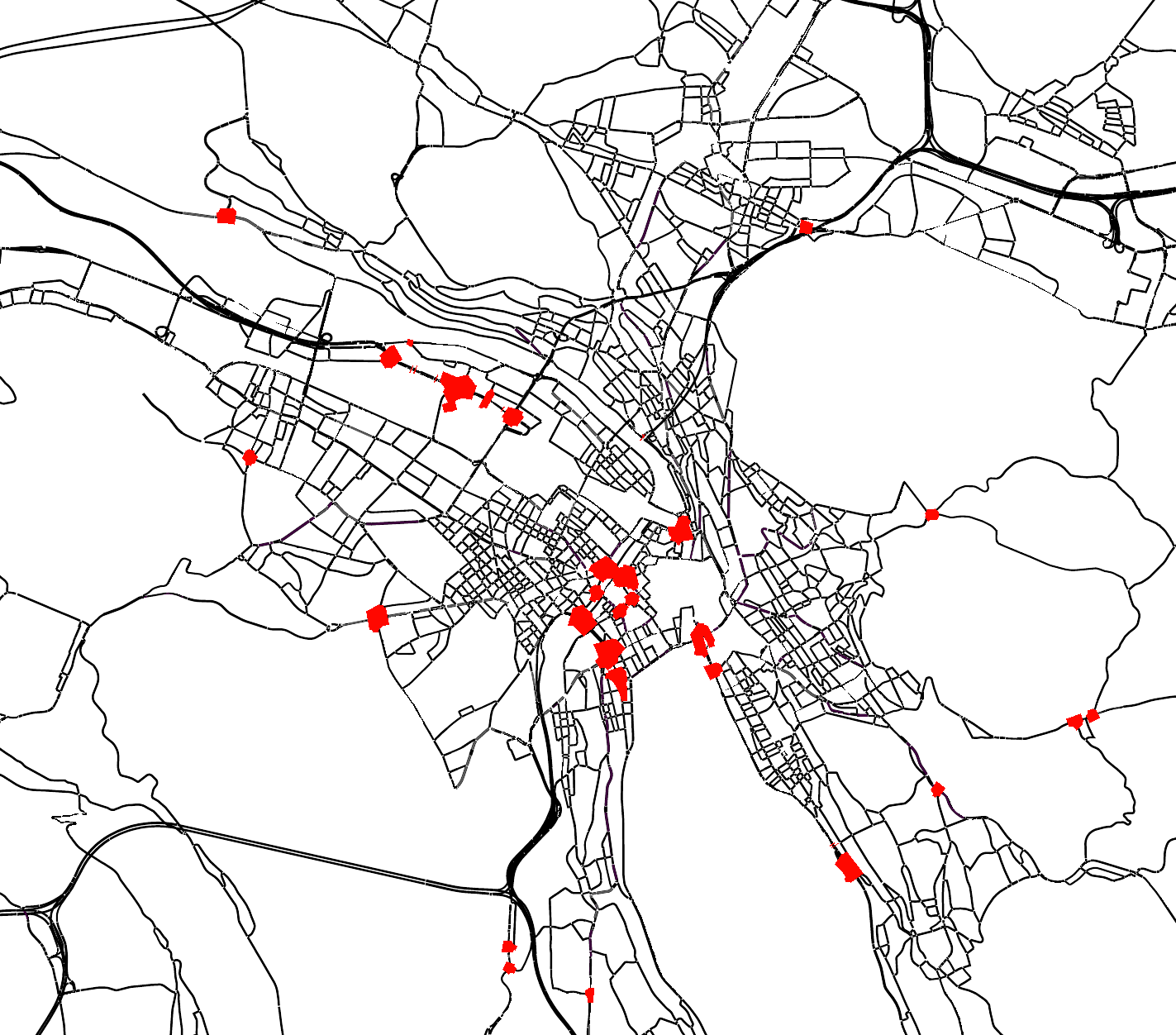}};
        
        \coordinate (overlay) at (-2.25,-2.25);
    \node[draw=none,fill=none] at (overlay) {\frame{\includegraphics[width=0.4\linewidth]{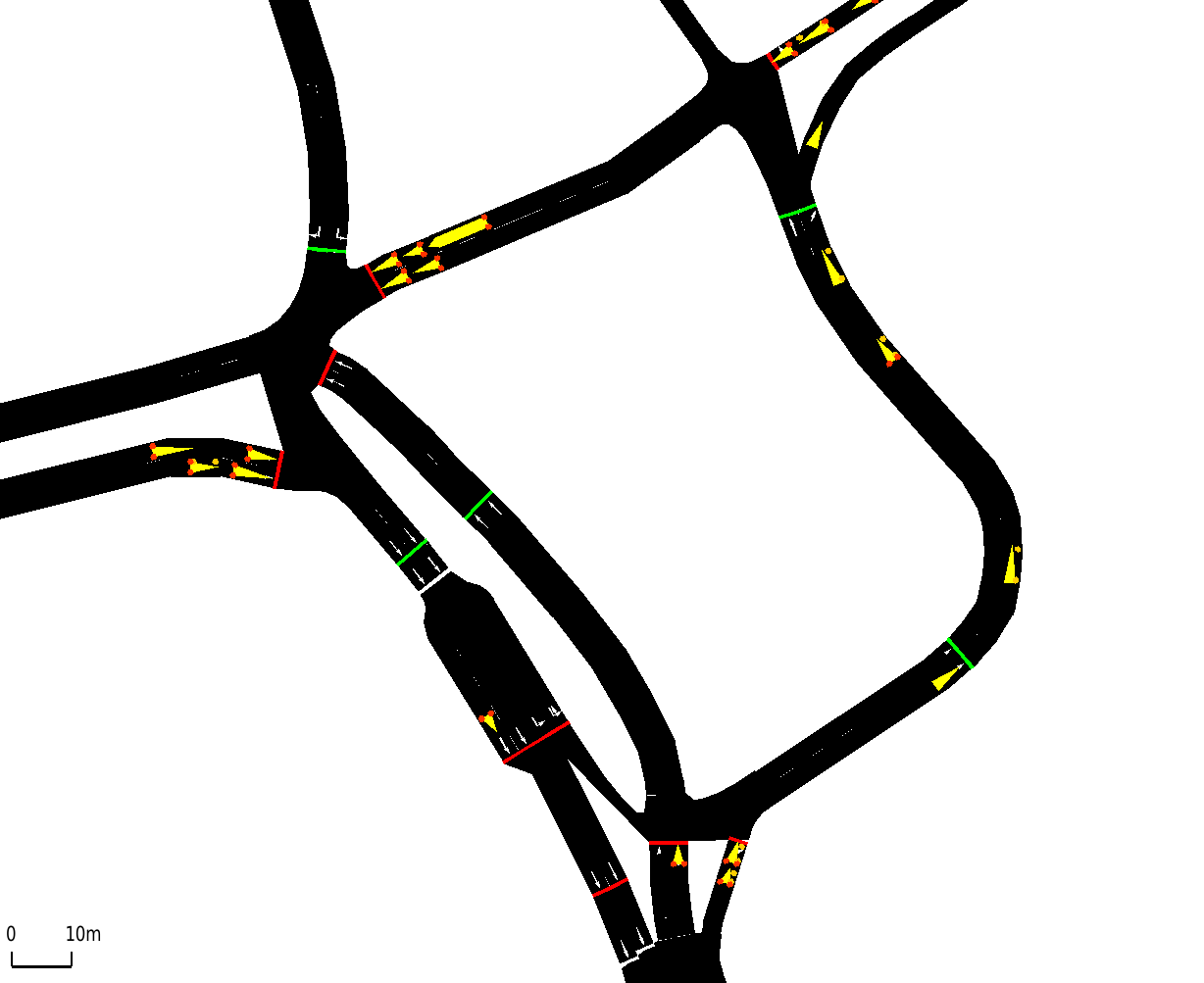}}};

    \end{tikzpicture}
    \caption{}
    \label{fig:net}
    \end{subfigure}%
    \hfill
     \begin{subfigure}[t]{.49\textwidth}
     \centering
    \includegraphics[width=.95\linewidth]{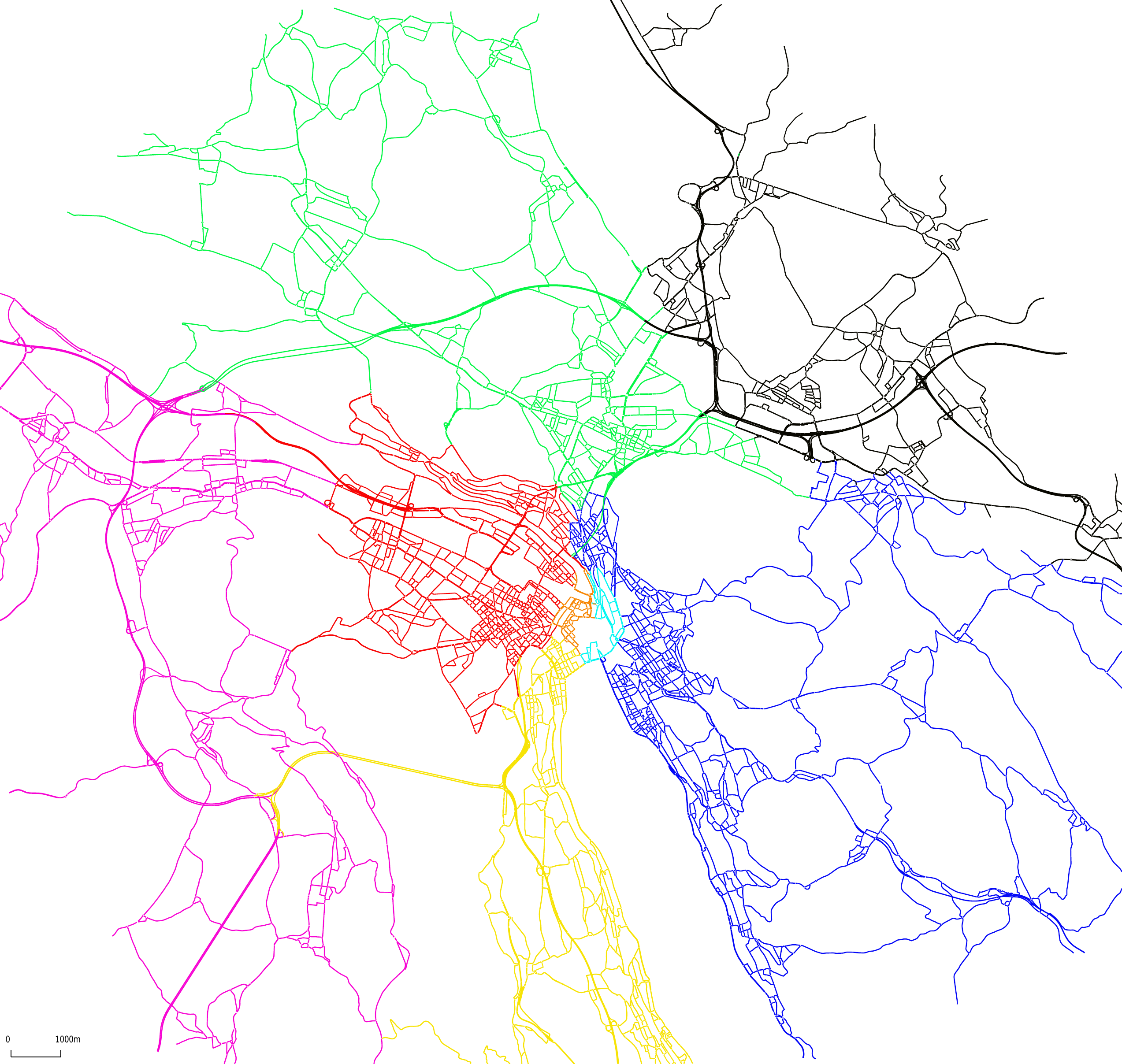}
    \caption{}
    \label{fig:partitioning}
    \end{subfigure}
  \caption{Left panel shows the Z\"urich urban traffic network embedded in \gls{SUMO}. 
  Actuated intersections are highlighted in red. In the corner, one of the controlled intersections, Bellevue.
  Right panel shows the region partitioning found using the snake clustering algorithm, the Wiedikon region is highlighted in red.} 
  \label{fig:NET}
\end{figure}

We now present the simulation of Z\"urich, Switzerland, built 
on a digital twin developed at Transcality~\cite{transcality}. 
The urban traffic network of the city comprises almost 15,000 roads 
with different number of lanes, connecting approximately 7,000 intersections, see Figure \ref{fig:NET}.
The demand profile is estimated using real data measured by the city loop detectors. 
In our experiments, we consider the period corresponding to the 
evening traffic peak, spanning the hours from 16:00 to 21:00. 
During this surge in demand, the network serves more than 170,000 vehicles. 
Figure~\ref{fig:partitioning} shows the traffic regions obtained 
by applying a parallelized version of the snake clustering algorithm  described in \cite{gerolinimis2016CsnakeClustering}, 
see Appendix \ref{app:partitioning}. 
Remarkably, the identified regions closely resemble the real districts of the city, 
for example, one can recognize the Wiedikon area in red. 
This similarity is due to the use of real data in demand and traffic network modeling. 

\begin{figure}[h!]
    \centering 
        \begin{tikzpicture}
            \begin{axis}[black,
            height=7cm,
            width=\textwidth,
            grid = both, 
            xlabel = Time (min),
            ylabel = $\overline{\bs d}$ (veh/min),
            xmin=7,
            xmax=310,
            ymin=0,
            legend style={cells={align=right},at={(0.635,0.9)},anchor=west},
                    legend columns = 3,
                    legend entries={\tiny {Oerlikon},\tiny {Wiedikon},\tiny {Züriberg},\tiny {Uetliberg},\tiny {Enge},\tiny {Opfikon},\tiny {Center 1}, \tiny {Center 2}},
            ]
                \addplot [green,smooth,ultra thick] table[x=0,y= from 0 to 1,col sep = comma] {./csv/ZRH_NEW/demand.csv};
                \addplot [red,smooth,ultra thick] table[x=0,y= from 1 to 1,col sep = comma] {./csv/ZRH_NEW/demand.csv};
                \addplot [blue,ultra thick] table[x=0,y= from 2 to 1,col sep = comma] {./csv/ZRH_NEW/demand.csv};
                \addplot [magenta,smooth,ultra thick] table[x=0,y= from 3 to 1,col sep = comma] {./csv/ZRH_NEW/demand.csv};
                \addplot [yellow,smooth,ultra thick] table[x=0,y= from 4 to 1,col sep = comma] {./csv/ZRH_NEW/demand.csv};
                \addplot [black,smooth,ultra thick] table[x=0,y= from 5 to 1,col sep = comma] {./csv/ZRH_NEW/demand.csv};
                \addplot [orange,smooth,ultra thick] table[x=0,y= from 6 to 1,col sep = comma] {./csv/ZRH_NEW/demand.csv};
                \addplot [cyan,smooth,ultra thick] table[x=0,y= from 7 to 1,col sep = comma] {./csv/ZRH_NEW/demand.csv};
            \end{axis}
        \end{tikzpicture} 
    \caption{The demand profile directed to the Wiedikon region broken down by region of origin.
    The color coding used matches the one in Figure \ref{fig:partitioning}.}
    \label{fig:demandZRH6}
\end{figure}
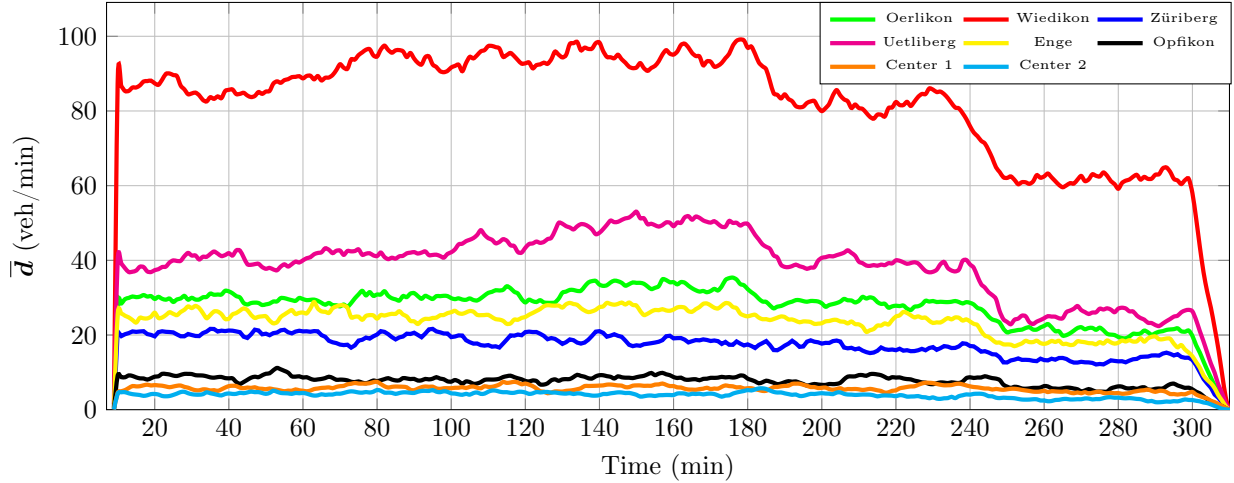

\paragraph{Actuators and demand profile}
The network comprises over 800 intersections with controllable traffic lights. 
To ensure the simulation is realistic, we took advantage of the expert knowledge of the traffic engineers 
of the city of Z\"urich and selected the 47 actuators that the municipality is currently using as 
actuators~\cite{lukas-two-layer}, the position of these actuators can be seen in Figure~\ref{fig:net}. 
Although the phase definition depends on the topology of the intersection, all actuated traffic lights have a duty cycle time of 
$\Delta_{\textrm{dc}}=72 \ s$.
One of the actuated intersections can be seen in Figure~\ref{fig:net}. 
The demand profile directed to the Wiedikon region 
(highlighted in red in Figure~\ref{fig:partitioning}) is shown in Figure~\ref{fig:demandZRH6}, where the color coding of the plot matches the one of Figure \ref{fig:partitioning}.
The data is collected during a day of average congestion in the city, therefore the baseline simulation does not 
reach a grid-lock. 

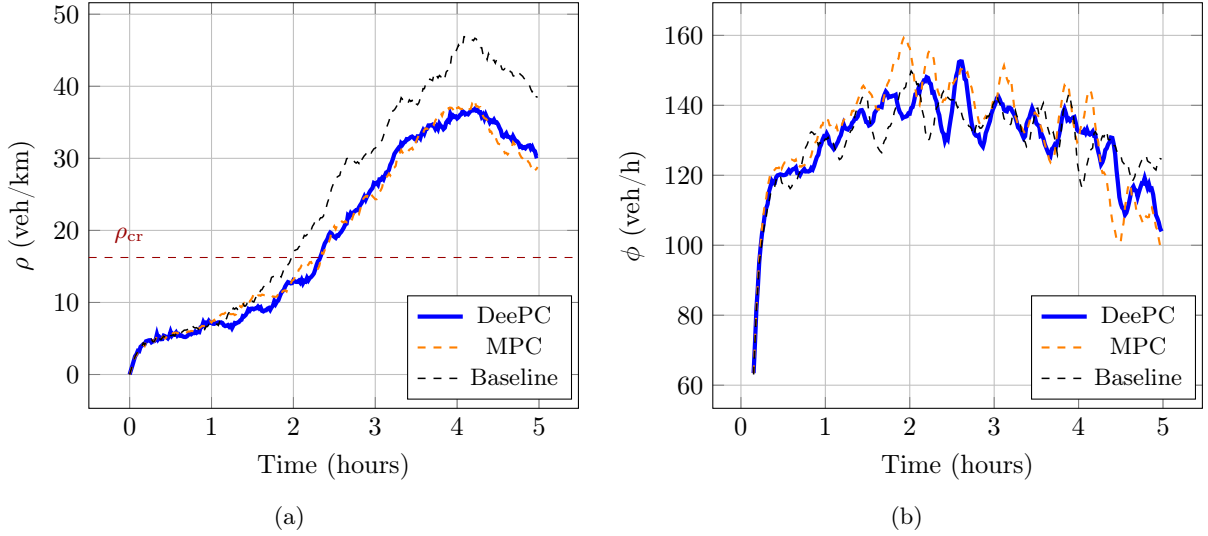
\begin{figure}[ht!]
        \begin{subfigure}[t]{.49\textwidth}
        \centering
                \begin{tikzpicture}
                    \begin{axis}[black,
                    width = \textwidth,
                    grid = both,
                    x coord trafo/.code={
                        \pgfmathparse{\pgfmathresult/60}
                    },
                    x coord inv trafo/.code={
                        \pgfmathparse{\pgfmathresult*60}
                    },
                    xticklabels = {0,0,1,2,3,4,5,6},
                    xlabel = Time (hours),
                    ylabel = $\rho$ (veh/km),
                    legend style={cells={align=right},at={(0.65,0.15)}, anchor=west},
                    legend columns = 1,
                    legend entries={\small {\gls{DeePC}},\small {\gls{MPC}}, \small {Baseline}},     
                    ]   

                        \addplot [blue,ultra thick] table[x=0,y= Region 1,col sep = comma] {./csv/ZRH_NEW/deepc-freq-1/results/density_results.csv};
                        \addplot [orange,dashed, thick] table[x=0,y= Region 1,col sep = comma] {./csv/ZRH_NEW/mpc/results/density_results.csv}; 
                        \addplot [black,dashed, line width = 0.2 mm] table[x=0,y= Region 1,col sep = comma] {./csv/ZRH_NEW/NoControl/results/density_results.csv};
                        \draw [red!60!black, dashed] (-50,162.39) -- (600,162.39);
                        \node[ above,color=red!60!black, font=\small] at (0, 170) {$\rho_{\tup{cr}}$};
                    \end{axis}
                \end{tikzpicture}
            \caption{}
            \label{fig:ZRHdensity}
        \end{subfigure}
        \begin{subfigure}[t]{.49\textwidth}
        \centering
                \begin{tikzpicture}
                    \begin{axis}[black,
                    width = \textwidth,
                    grid = both,
                    x coord trafo/.code={
                        \pgfmathparse{\pgfmathresult/60}
                    },
                    x coord inv trafo/.code={
                        \pgfmathparse{\pgfmathresult*60}
                    },
                    xticklabels = {0,0,1,2,3,4,5,6},
                    xlabel = Time (hours),
                    ylabel = $\phi$ (veh/h),
                    legend style={cells={align=right}, at={(0.65,0.15)}, anchor=west},
                    legend columns = 1,
                    legend entries={\small {\gls{DeePC}},\small {\gls{MPC}},\small {Baseline}},
                    ]
                        \addplot [blue,ultra thick] table[x=0,y= DeePC 1,col sep = comma] {./csv/ZRH_NEW/flow_mpc_deepc.csv};
                        \addplot [orange,dashed, thick] table[x=0,y= MPC,col sep = comma] {./csv/ZRH_NEW/flow_mpc_deepc.csv};
                        \addplot [black,dashed, line width = 0.2 mm] table[x=0,y= NoControl,col sep = comma] {./csv/ZRH_NEW/flow_mpc_deepc.csv};
                    \end{axis}
                \end{tikzpicture}
            \caption{}
            \label{fig:ZRHFlow}
        \end{subfigure}
        \caption{Evolution of density (left panel) and flow (right panel) under control
        of \gls{DeePC} and \gls{MPC} for the Wiedikon region. The critical density value $\rho_{\tup{cr}}$ (red dashed line) 
        is reported in the left panel, the line represents the threshold for congestion of the system.}
        \label{fig:ZRHcontrol}
\end{figure}

\paragraph{Results}
The comparison of density and flow under control of \gls{DeePC} and \gls{MPC} 
for the Wiedikon region, highlighted in red in Figure \ref{fig:partitioning},  
can be seen in Figure \ref{fig:ZRHcontrol}. 
Figure \ref{fig:ZRHdensity} additionally shows the value of the critical density $\rho_{\tup{cr}}$ of the region.
Due to the high variability of the 
flow, a moving average filter of window 10 has been applied to the flow time series to aid comprehension 
of plot \ref{fig:ZRHFlow}. 
The behavior of this region is representative as all the other regions in the network behave similarly.
Also in this network \gls{DeePC} is able to outperform \gls{MPC} and 
improve traffic conditions leading to a $18\%$ decrease 
in travel time, see Table \ref{tab:avgmetricsZRH}. 
This suggests that the data-generated approximation of the
relation between traffic light control and density succeeds in capturing traffic dynamics 
effectively. This supports the conjecture made in Figure \ref{fig:commutative-diagram} that the relation 
between the two quantities is linear.
The \gls{MPC} controller also achieves an improvement in travel time which, however, remains higher
than under the \gls{DeePC} policy. One reason for this is that the \gls{MPC} solves the perimeter control
problem by computing the optimal flows, referring back to Figure \ref{fig:commutative-diagram}, we know 
that the relation between the traffic light control and the flow to be non-linear, this
leads to an inaccurate linear approximation.
Moreover, both algorithms perform a linearization of the traffic dynamics providing a fair comparison, 
however, since the \gls{MPC} formulation computes first the optimal flows between regions, 
a mapping between these flows and the actuators is then needed to assign 
the inputs to the right actuators. If the actuators are placed on the borders between regions 
constructing this mapping is trivial. 
If, however, the actuators are allowed to be anywhere in the network, this makes constructing the mapping more difficult, 
as one actuator could be responsible for the modulation of multiple flows directed to different regions.
In practice, each actuator must receive just one input at a time, this requires it to be assigned to a single flow
which introduces an approximation error and degrade performance. 
\gls{DeePC} avoids this problem as the mapping is defined implicitly by the Hankel matrix 
leading to better performance.
Because updating the control at every duty cycle may be undesirable in practice 
due to the frequent changes in phase duration, we also evaluate \gls{DeePC} with longer control periods. 
Specifically, we simulate control with a period of 
1,3 and 6 $\Delta_{\textup{dc}}$, this means that we compute the optimal green 
times and keep them constant for 1,3 and 6 duty cycles before recomputing. 
The evolution of the density under this control can be seen below in 
Figure~\ref{fig:ZRHcontrol-period}.
Note that as we control less often, the performance of the algorithm degrades and 
approaches the baseline. Table \ref{tab:avgmetricsZRH} shows the effects on average travel time 
and emissions metrics of with the different control periods. 
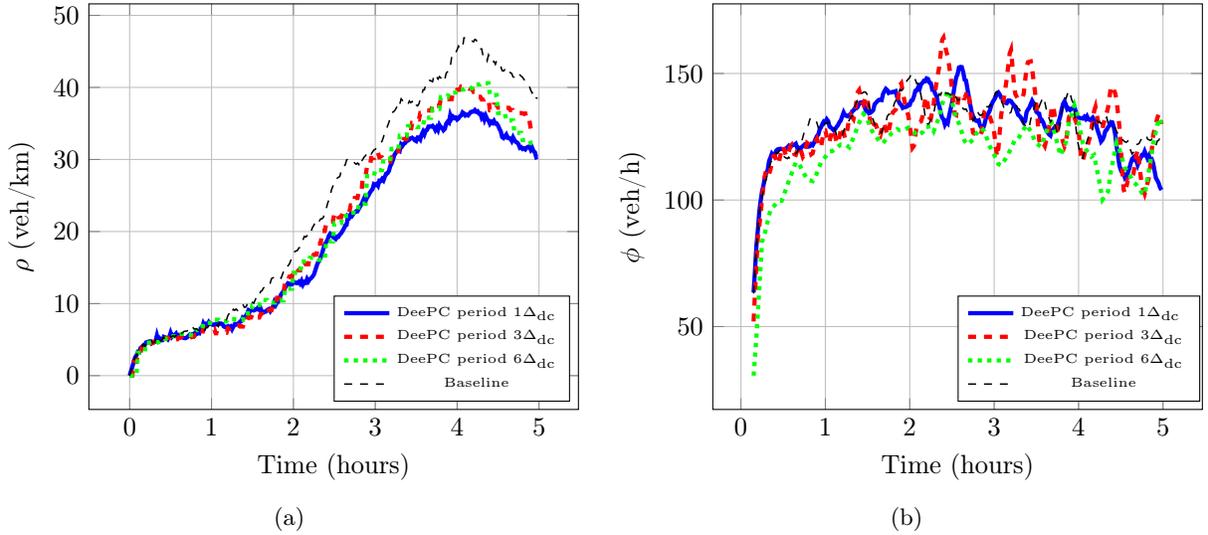
\begin{figure}[ht!]
        \begin{subfigure}[t]{.49\textwidth}
        \centering
                \begin{tikzpicture}
                    \begin{axis}[black,
                    width = \textwidth,
                    grid = both,
                    x coord trafo/.code={
                        \pgfmathparse{\pgfmathresult/60}
                    },
                    x coord inv trafo/.code={
                        \pgfmathparse{\pgfmathresult*60}
                    },
                    xticklabels = {0,0,1,2,3,4,5,6},
                    xlabel = Time (hours),
                    ylabel = $\rho$ (veh/km),
                    legend style={cells={align=right},at={(0.5,0.15)}, anchor=west},
                    legend columns = 1,
                    legend entries={\tiny {\gls{DeePC} period 1$\Delta_{\textrm{dc}}$},\tiny {\gls{DeePC} period 3$\Delta_{\textrm{dc}}$},\tiny {\gls{DeePC} period 6$\Delta_{\textrm{dc}}$}, \tiny {Baseline}}
                    ]   
                    \addplot [blue,ultra thick] table[x=0,y= Region 1,col sep = comma] {./csv/ZRH_NEW/deepc-freq-1/results/density_results.csv};
                    \addplot [red,dashed,ultra thick] table[x=0,y= Region 1,col sep = comma] {./csv/ZRH_NEW/deepc-freq-3/results/density_results.csv};
                    \addplot [green,dotted,ultra thick] table[x=0,y= Region 1,col sep = comma] {./csv/ZRH_NEW/deepc-freq-6/results/density_results.csv};
                    \addplot [black,dashed, line width = 0.2 mm] table[x=0,y= Region 1,col sep = comma] {./csv/ZRH_NEW/NoControl/results/density_results.csv};
                    \end{axis}
                \end{tikzpicture}
            \caption{}
            \label{fig:different_freq_density}
        \end{subfigure}
        \begin{subfigure}[t]{.49\textwidth}
        \centering
                \begin{tikzpicture}
                    \begin{axis}[black,
                    width = \textwidth,
                    grid = both,
                    x coord trafo/.code={
                        \pgfmathparse{\pgfmathresult/60}
                    },
                    x coord inv trafo/.code={
                        \pgfmathparse{\pgfmathresult*60}
                    },
                    xticklabels = {0,0,1,2,3,4,5,6},
                    xlabel = Time (hours),
                    ylabel = $\phi$ (veh/h),
                    legend style={cells={align=right}, at={(0.5,0.15)}, anchor=west},
                    legend columns = 1,
                    legend entries={\tiny {\gls{DeePC} period 1$\Delta_{\textrm{dc}}$ },\tiny {\gls{DeePC} period 3$\Delta_{\textrm{dc}}$},\tiny {\gls{DeePC} period 6$\Delta_{\textrm{dc}}$}, \tiny {Baseline}}
                    ]  
                    \addplot [blue,ultra thick] table[x=0,y= DeePC 1,col sep = comma] {./csv/ZRH_NEW/flow_deepc_frequencies.csv};
                    \addplot [red,dashed,ultra thick] table[x=0,y= DeePC 3,col sep = comma] {./csv/ZRH_NEW/flow_deepc_frequencies.csv};
                    \addplot [green,dotted,ultra thick] table[x=0,y= DeePC 6,col sep = comma] {./csv/ZRH_NEW/flow_deepc_frequencies.csv};
                    \addplot [black,dashed, line width = 0.2 mm,] table[x=0,y= NoControl,col sep = comma] {./csv/ZRH_NEW/flow_deepc_frequencies.csv};        
                    \end{axis}
                \end{tikzpicture}
            \caption{}
            \label{fig:different_freq_flow}
        \end{subfigure}
        \caption{Evolution of density (left panel) and flow (right panel) under control of \gls{DeePC} with different 
        periods for the Wiedikon region.}
        \label{fig:ZRHcontrol-period}
\end{figure}

\begin{table}[ht!]
\caption{Average Metrics Z\"urich}\label{tab:avgmetricsZRH}
    \centering
    \begin{tabular}{@{}lllllll@{}}
        \toprule
          {} & Baseline & \gls{DeePC} 6$\Delta_{\textrm{dc}}$& \gls{DeePC} 3$\Delta_{\textrm{dc}}$& MPC & \gls{DeePC} 1$\Delta_{\textrm{dc}}$ \\
        \midrule
        Travel Time (min)    & 45.23  & 39.07 & 37.94 & 38.51 &$\boldsymbol{37.05}$ & -18.08\% \\
        Waiting Time (min)   & 28.27  & 23.31 & 22.62  & 23.07 & $\boldsymbol{21.87}$ & -22.60\%\\
        CO Emissions(g)     & 16.91  & 15.84 & 15.41 & 16.14 & $\boldsymbol{15.67}$ & -7.33\%\\
        CO2 Emissions (g) & 4,992.15  & 4,379.17 & 4,253.91 & 4,338.89 & $\boldsymbol{4,184.46}$ & -16.17\%\\
        HC Emissions (g) & 0.11 & 0.11 & 0.10& 0.11 &0.11& -0.00\%\\
        PMx Emissions (g) & 0.22 & 0.21 & 0.21 & 0.21 & 0.21 & -4,54\%\\
        NOx Emissions (g) & 1.91 & 1.67 & 1.62 & 1.66 & $\boldsymbol{1.60}$ & - 16.23\%\\
        Fuel consumption (ml) & 1,618.41  & 1,419.69 & 1,379.07 &1,406.63& $\boldsymbol{1,356.56}$ & - 16.17\%\\
        Trips completed & 176,821 & 177,555 & $\boldsymbol{178,143}$ & 176,980 & 178,035 & +0.74\%  \\
        \bottomrule
    \end{tabular}
\end{table}

\begin{figure}[ht!]
    \begin{subfigure}[t]{.50\textwidth}
        \centering
            \begin{tikzpicture}
                \begin{axis}[
                black,
                grid= both,
                x coord trafo/.code={
                        \pgfmathparse{\pgfmathresult/60}
                    },
                    x coord inv trafo/.code={
                        \pgfmathparse{\pgfmathresult*60}
                    },
                xticklabels = {0,0,1,2,3,4,5,6},
                xlabel= Hour,
                ylabel= Principal component value,
                legend style={cells={align=right},at={(0.74,0.85)}, anchor=west},
                legend columns =1,
                legend entries={\small {PC1},\small {PC2},\small {PC3}}
                ]
                \addplot [cyan, smooth, ultra thick] table [x=0,y=PC1,col sep = comma] {./csv/ZRH_NEW/pca_components.csv};
                \addplot [green, dashed, ultra thick] table [x=0,y=PC2,col sep = comma] {./csv/ZRH_NEW/pca_components.csv};
                \addplot [orange, dotted, ultra thick] table [x=0,y=PC3,col sep = comma] {./csv/ZRH_NEW/pca_components.csv};
                \end{axis}
            \end{tikzpicture}
        \caption{}
        \label{fig:pca}
    \end{subfigure}
    \begin{subfigure}[t]{.50\textwidth}
        \centering
        \begin{tikzpicture}
             \begin{axis}[
                ybar,
                ymajorgrids,
                x= 0.14cm,
                bar width=0.05cm,
                xlabel={Traffic Light (sorted)},
                ylabel={Loading},
                xtick = data,
                xticklabels = {34,39,35,15,29,21,17,32,25,11,19,7,27,16,31,33,12,13,22,38,28,43,37,1,26,30,40,23,5,18,3,24,10,9,42,46,45,14,2,0,20,41,4,44,8,36,6}, 
                xticklabel style={rotate = 90, anchor = east},
                ticklabel style = {font=\tiny},
                enlarge x limits=0.1
            ]
            \addplot table[ y=0, col sep = comma] {./csv/ZRH_NEW/sorted_loadings.csv};
            \end{axis}
        \end{tikzpicture}
        \caption{}
        \label{fig:loadings}
    \end{subfigure}
    \caption{Left panel shows the evolution of the first three PCA components. Right panel shows the loadings value associated 
    to PC1.}
\end{figure}
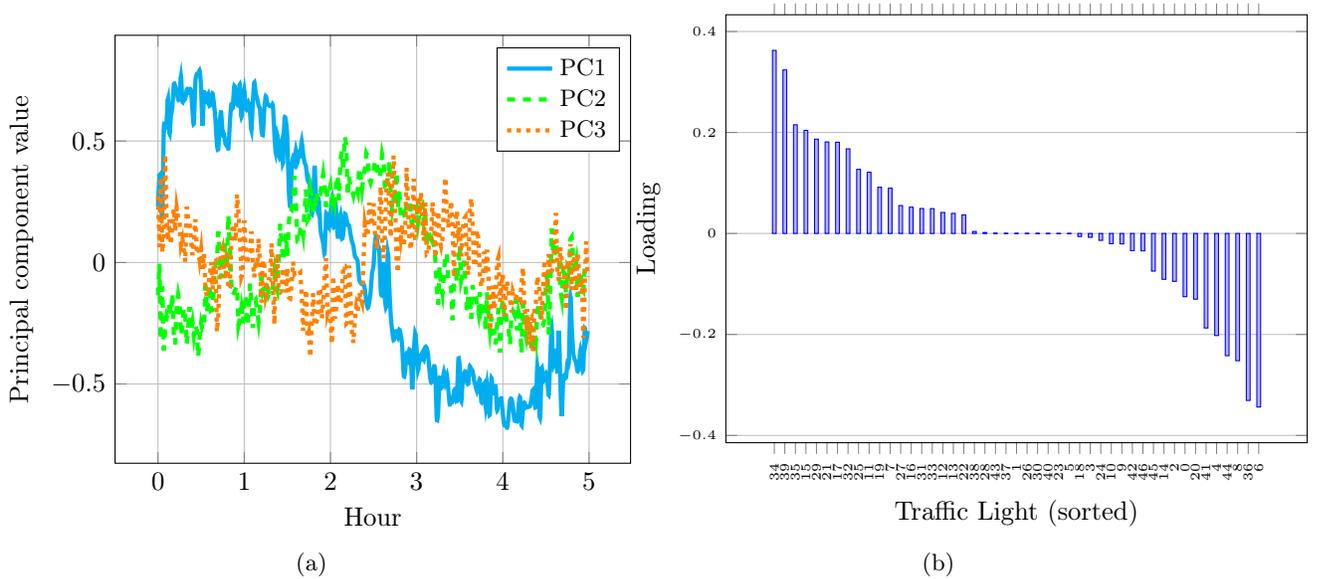

\paragraph{Input analysis}
To illustrate the control decision made by the \gls{DeePC} algorithm, 
we further analyze the time series of the optimal inputs $\bs \lambda (t)$ 
found by \gls{DeePC}. 
To this end we collect the time series of the inputs in a matrix $\Lambda \in \mathbb{R}^{l\times T}$ 
and perform a \gls{PCA} over this matrix using $k = 10$ components. 
\gls{PCA} is used to find patterns high dimensional data, it works by finding 
the eigenvectors, named principal components, of the covariance matrix $\Sigma = \frac{1}{T-1}\Lambda^T\Lambda \in \mathbb{R}^{T\times T}$ constructed from 
data. In this setting each eigenvalue $\textup{eig}_i(\Sigma)$ of the covariance matrix represents
the amount of variance in the data explained by it corresponding principal component. 
The explained variance ratio associated to the $i$-th eigenvalue is defined as 
\begin{equation}
    \text{EVR}_i \coloneqq \frac{\textup{eig}_i(\Sigma)}{\sum^k_j \textup{eig}_j(\Sigma)}
\end{equation}
These ratios represent the percentage of the dataset variance each principal 
components accounts for. 
\begin{table}[ht!]
    \centering
    \begin{tabular}{|c|c|c|c|c|c|c|c|c|c|c|}
        \hline
        Component & PC1 & PC2 & PC3 & PC4 & PC5 & PC6 & PC7 & PC8 & PC9 & PC10  \\
        \hline 
        EVR & $0.477$ & $0.103$ & $0.051$ & $0.037$ & $0.033$ & $0.026$ & $0.022$ & $0.021$ & $0.020$ & $0.018$\\
        \hline
    \end{tabular}
    \caption{Explained variance ratios of the components, each value represents the percentage
    of variation in the control input that can be explained by the corresponding principal component.}
    \label{tab:evr}
\end{table}
Table \ref{tab:evr} shows that the first principal component accounts 
for almost $50\%$ of the total input variation in all the 47 actuators, with 
the second and third components accounting for $10.3\%$ and $5.1\%$ respectively.
In other words, the first component is describing a strong global pattern that the 
controller used to adjust the input to the actuators.
Figure \ref{fig:pca} shows the evolution of the value of the first three principal 
components during the simulation. We observe that, while the second and the 
third components values remain relatively stable, the first component starts with 
positive values and then around two and a half hours experiences a 
rapid change to negative values. This suggests a 
change in the environment in which \gls{DeePC} is operating at that time, 
that led to a change the global pattern. 
Indeed, by observing Figure \ref{fig:ZRHdensity} we can see that exactly around that 
time the traffic density starts to increase rapidly, approaching the line defined by 
the reference trajectory $\rho_{\tup{cr}}$, once the line is crossed the network enters 
a state of congestion, which triggers the observed inversion in the global input pattern.  

\begin{figure}[ht!]
\centering
     \begin{subfigure}[t]{.49\textwidth}
     \centering
        \includegraphics[width=.95\linewidth, trim={30cm 8cm 34cm 18cm}, clip]{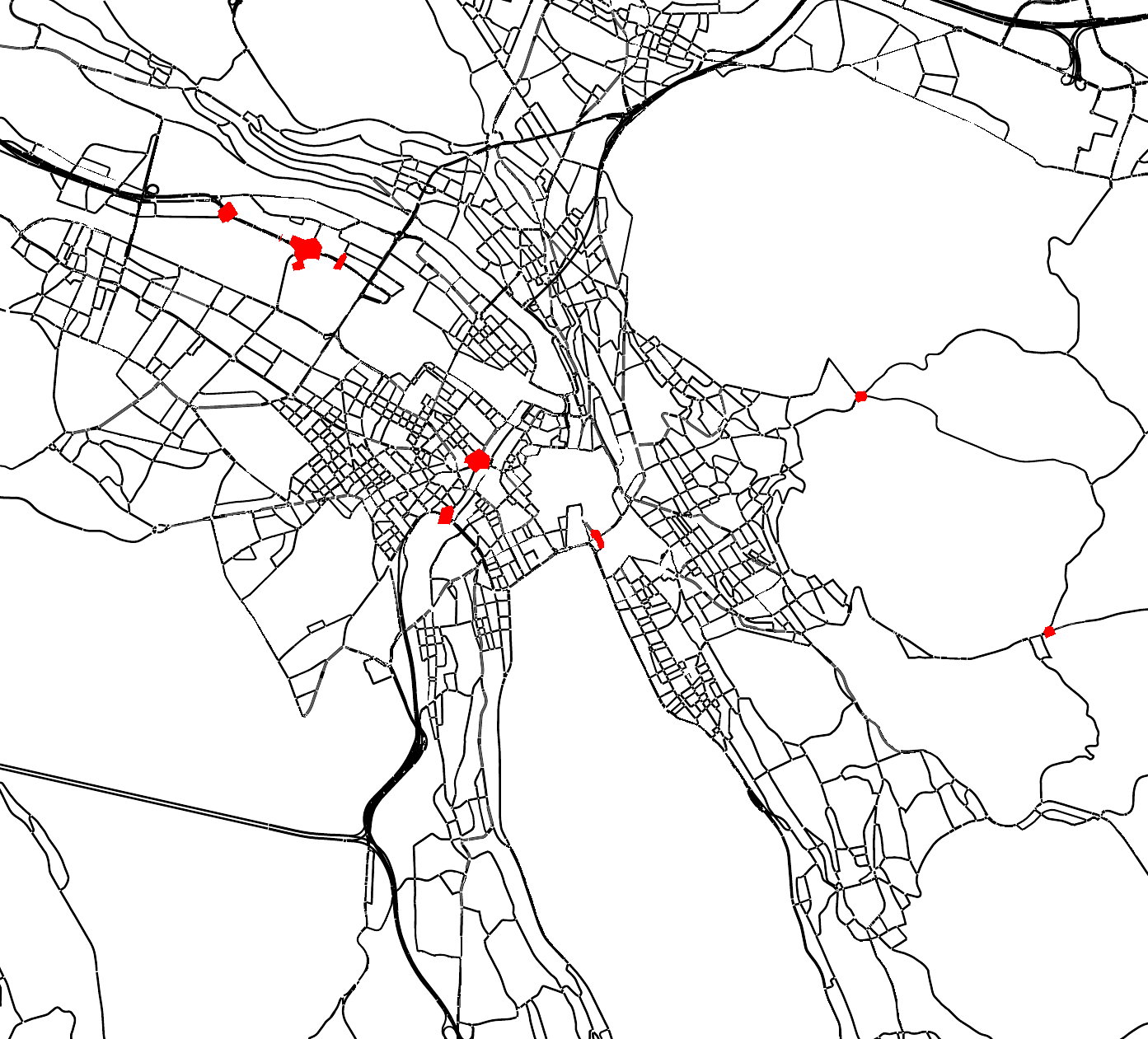};
        \caption{}
        \label{fig:poscorrelated}
    \end{subfigure}
    \hfill
    \begin{subfigure}[t]{.49\textwidth}
     \centering
    \includegraphics[width=.95\linewidth,trim={30cm 8cm 34cm 18cm}, clip]{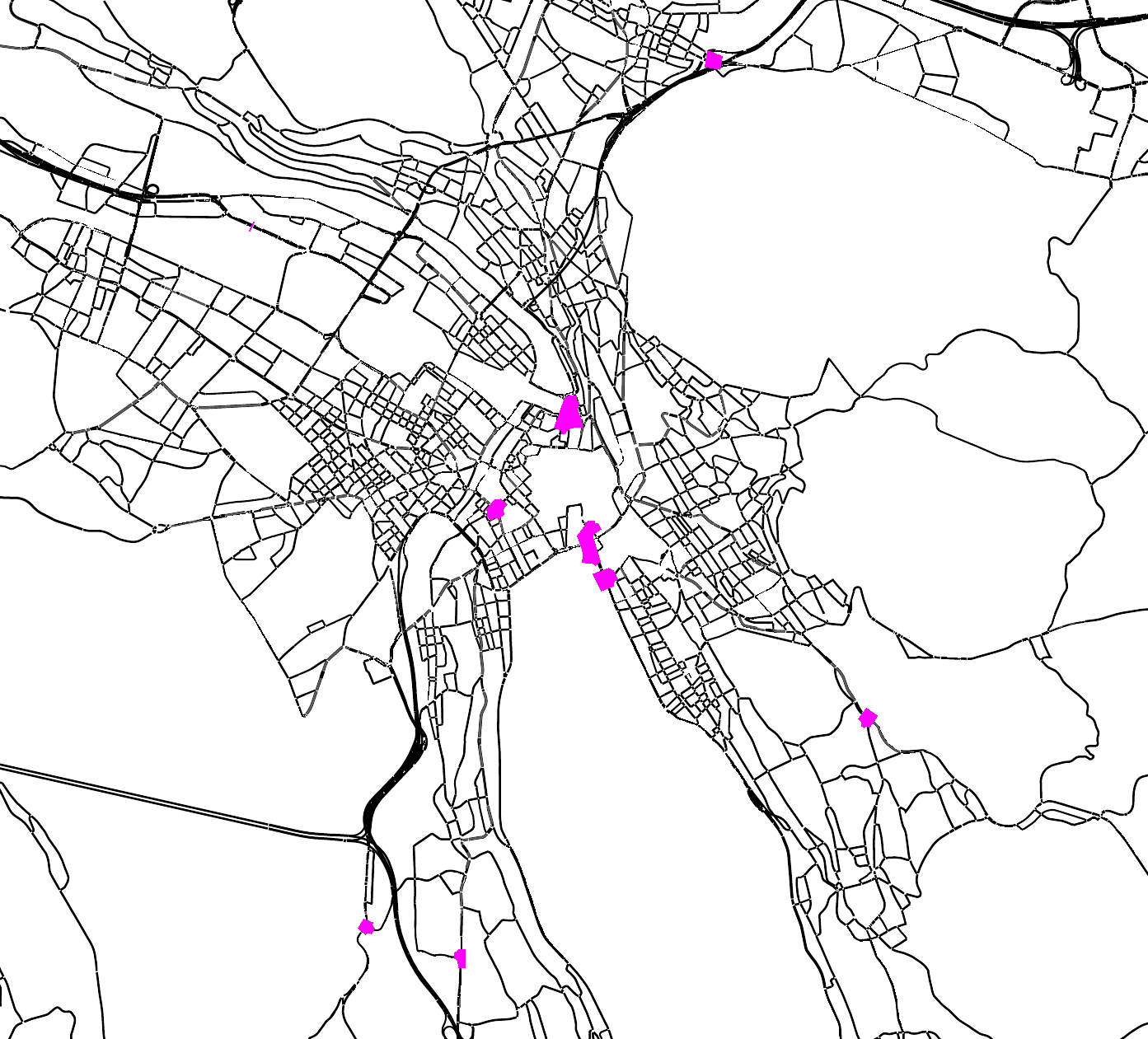}
    \caption{}
    \label{fig:invcorrelated}
    \end{subfigure}
  \caption{Positions of the highly influential actuators. Left panel shows the positions of the ten highest
  loading value (positively correlated) actuators highlighted in red. Right panel shows the positions of the
  ten lowest loading values (inversely correlated) actuators highlighted in magenta.} 
  \label{fig:correlated}
\end{figure}

Figure \ref{fig:loadings}, shows a bar plot of the loading values, or actuator weights, associated to the 
first principal component sorted in descending order. The magnitude of a loading gives information on 
the amount of variation in the input that the corresponding actuator is responsible for, with positive
values indicating positive correlation with the pattern and the negative ones negative correlation.  
This allows us to identify which actuators had the most influence over the traffic conditions. 
We can see that few actuators have large absolute loading value, allowing us to pinpoint 
the geographical locations of these influential actuators; their positions can be seen in 
Figures \ref{fig:correlated}. We observe the positively correlated 
actuators to be positioned on key access points to the city, such as highway exits, while the 
negatively correlated actuators cluster around the city center.

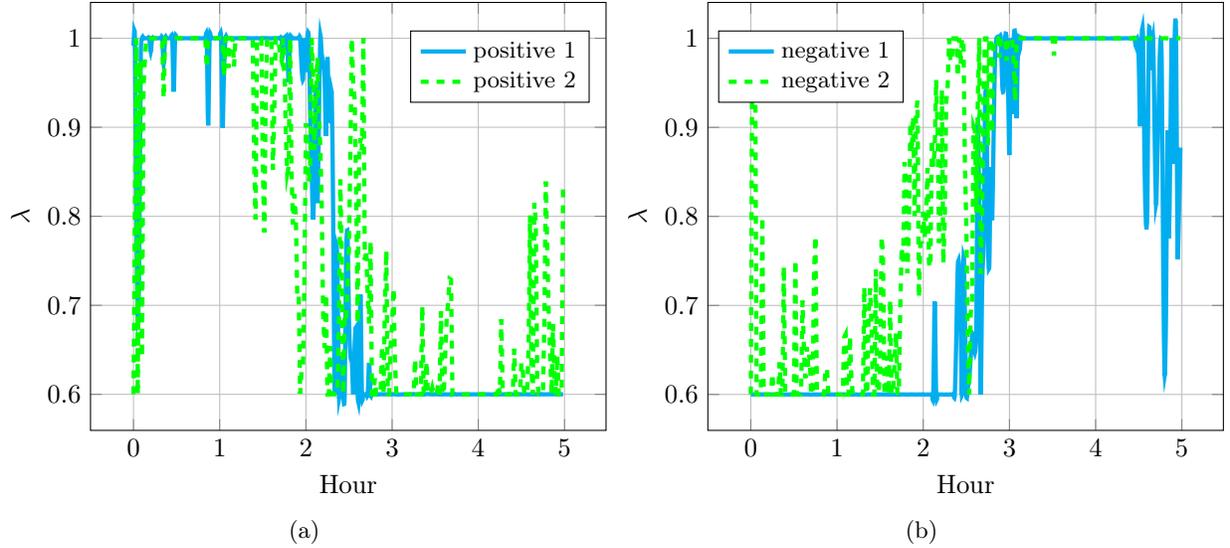
\begin{figure}[ht!]
    \begin{subfigure}[t]{.49\textwidth}
        \centering
            \begin{tikzpicture}
                \begin{axis}[
                black,
                grid= both,
                x coord trafo/.code={
                        \pgfmathparse{\pgfmathresult/60}
                    },
                    x coord inv trafo/.code={
                        \pgfmathparse{\pgfmathresult*60}
                    },
                xticklabels = {0,0,1,2,3,4,5,6},
                xlabel= Hour,
                ylabel= $\lambda$,
                legend style={cells={align=right},at={(0.62,0.85)}, anchor=west},
                legend columns =1,
                legend entries={\small {positive 1},\small {positive 2},\small {actuator 3}}
                ]
                \addplot [cyan, smooth, ultra thick] table [x=0,y=act_1,col sep = comma] {./csv/ZRH_NEW/pos.csv};
                \addplot [green, dashed, ultra thick] table [x=0,y=act_2,col sep = comma] {./csv/ZRH_NEW/pos.csv}; 
                \end{axis}
            \end{tikzpicture}
        \caption{}
        \label{fig:posinput}
    \end{subfigure}
        \begin{subfigure}[t]{.49\textwidth}
        \centering
            \begin{tikzpicture}
                \begin{axis}[
                black,
                grid= both,
                x coord trafo/.code={
                        \pgfmathparse{\pgfmathresult/60}
                    },
                    x coord inv trafo/.code={
                        \pgfmathparse{\pgfmathresult*60}
                    },
                xticklabels = {0,0,1,2,3,4,5,6},
                xlabel= Hour,
                ylabel= $\lambda$,
                legend style={cells={align=right},at={(0.02,0.85)}, anchor=west},
                legend columns =1,
                legend entries={\small {negative  1},\small {negative 2},\small {actuator 3}}
                ]
                \addplot [cyan, smooth, ultra thick] table [x=0,y=act_1,col sep = comma] {./csv/ZRH_NEW/inv.csv};
                \addplot [green, dashed, ultra thick] table [x=0,y=act_2,col sep = comma] {./csv/ZRH_NEW/inv.csv}; 
                \end{axis}
            \end{tikzpicture}
        \caption{}
        \label{fig:invinput}
    \end{subfigure}
    \caption{Input time series of the two most influential actuators respectively positively (left panel) and negatively correlated actuators.}
    \label{fig:input}
\end{figure}

\begin{figure}[h!]
    \centering
    \begin{tikzpicture}
        \begin{axis}[black,
        width = \textwidth,
        height= 8 cm,
        grid = both,
        mark color= gray!50,
        legend style={cells={align=right}},
        legend columns = 1,
        legend entries={Baseline Data, Baseline \gls{MFD}, \gls{DeePC} Data, \gls{DeePC} \gls{MFD}}, 
        ylabel={$\phi_i$},
        xlabel={$\rho_i$},
        xmin= 0,
        ymin=0,
        xmax=70,
        ymax=200
        ]
        \coordinate (a) at (16.24,225);
        \coordinate (b) at (16.24,15);
        \coordinate (c) at (126.4,150);
        \coordinate (d) at (126.4,0);
        \coordinate (e) at (147.6,150);
        \coordinate (f) at (147.6,0);
         
        \draw[dashed,blue] (c) -- (d);
        \draw[dashed,black] (e) -- (f);

        \addplot [only marks, black!50] table[x=density, y=flow,col sep = comma]{./csv/ZRH_NEW/mfd_baseline.csv};
        \addplot [dashed, black, ultra thick] table[x=baseline x,y=baseline y, col sep = comma]{./csv/ZRH_NEW/approximated_mfds.csv}; 
        \addplot [only marks, blue, mark=square*] table[x=density, y=flow, col sep = comma]{./csv/ZRH_NEW/mfd_control.csv};
        \addplot [smooth, blue, ultra thick] table[x=control x,y=control y, col sep = comma]{./csv/ZRH_NEW/approximated_mfds.csv};
         
        \addplot [ only marks,
                        color=blue,
                        mark= star, 
                        mark size=5pt,
                        point meta= explicit symbolic,
                        nodes near coords
                        ] coordinates {(12.64,0)};
        \node [color=blue] at (105,10) {$\rho_{\tup{cr},i}$};
        \addplot [ only marks,
                        color=black,
                        mark= star, 
                        mark size=5pt,
                        point meta= explicit symbolic,
                        nodes near coords
                        ] coordinates {(14.76,0)}; 
        \node [color=black] at (175,10) {$\rho_{\tup{cr},i}$};
 
        \end{axis}
    \end{tikzpicture}
    \caption{Comparison between the density-flow data distribution of Wiedikon obtained under the static control policy 
   (gray dots) and under the \gls{DeePC} control policy (blue squares). The \gls{MFD}s and critical densities $\rho_{\tup{cr}}$
   are shown in the corresponding colors for both distributions.
   } 
    \label{fig:dual-mfd}  
\end{figure}
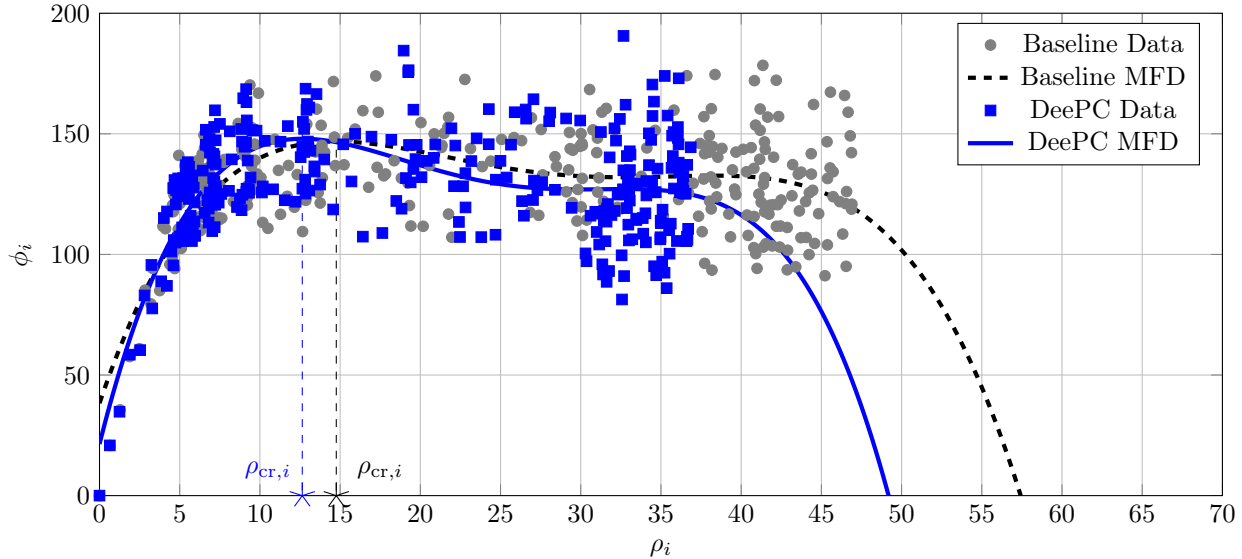

Figures \ref{fig:input} show the control input time series of the two most 
important positively and negatively correlated actuators. Note that the time series are indeed correlated with the first component evolution in Figure \ref{fig:pca}. 
Most importantly, the positively correlated actuators start with higher green times, and then change to lower green times, while the 
opposite is true for the negatively correlated actuators.
This is reasonable, higher green times are allowed until the congestion threshold $\rho_{\tup{cr}}$ is reached,
after that the control regime changes to lower green times. Given that the positively correlated actuators are placed
at key access points to the city, lower green times lead to less people accessing the inner regions city at a time,
giving time for the current demand to clear aided by the higher green times provided by the negatively correlated actuators 
in the city center.
\paragraph{Macroscopic fundamental diagram comparison}
Another way to interpret the effect of the control policy of \gls{DeePC} is to compare the \gls{MFD} obtained under its influence with the one
obtained under the static baseline policy.
Figure \ref{fig:dual-mfd} shows two data distributions, each point representing a density-flow pair for the Wiedikon region.
The first distribution (gray dots) is data collected under the static cycle specified by the \gls{SUMO} configuration with 
no input passed to the simulation, that is the baseline simulation 
used in Figures \ref{fig:ZRHcontrol}, \ref{fig:ZRHcontrol-period}.
The second distribution (blue squares) is data collected under the \gls{DeePC} control policy with period 1 $\Delta_{\textrm{dc}}$, that is
computing the control action for every duty cycle. 
Note that the data distribution collected under \gls{DeePC} remains at lower densities compared to the baseline data distribution while maintaining 
a comparable flow, this is reflected also in Figure \ref{fig:ZRHcontrol}.
Since the average flow of a region can be expressed as $\phi_i=\rho_i v_i$ where $v_i$ is the average speed of the vehicles in the region,
Figure \ref{fig:dual-mfd} suggests that to maintain the same level of flow as the baseline data distribution at a lower density, 
the vehicles must be moving at a higher average speed, leading to a decrease in travel time with respect to the baseline.
The \gls{MFD}s corresponding to the two data distributions are retrieved by fitting two  
polynomials of 4-th degree (black and blue lines), the critical density $\rho_{\tup{cr}}$ is computed for both distributions and shown in Figure \ref{fig:dual-mfd}.
We note that the two values do indeed differ from one another by a factor of 2.5 vehicles per kilometer, with the \gls{DeePC} 
distribution being more weighted towards the origin, this bias did not noticeably influence performance.

In conclusion, using \gls{DeePC} to control large traffic systems presents several advantages.
The formulation requires only the solution of a quadratic problem that can be efficiently solved
in real time using modern solvers. Modeling efforts are substituted by a data collection step
and estimation of the critical densities. 
Thanks to this, it is able to exploit the seemingly linear relation between traffic 
lights control and densities, bypassing the non-linear relation between traffic
light control and flows.

\section{Conclusions and future research}\label{sec:discussion}
In this work, we have shown that the \gls{DeePC} algorithm is able to effectively 
solve the \gls{ITSCP} leveraging data. We modeled traffic dynamics using behavioral 
systems theory and obtained a flexible and scalable formulation useful to optimize 
key metrics (e.g., CO$_2$ emissions and travel times) for large non-linear traffic systems. 
Our methodology has several advantages, first, the modeling efforts that are usually 
required upfront to solve the \gls{ITSCP} are substituted by a much simpler data collection 
procedure. Second, the formulation is flexible, allowing for seamless incorporation of 
different kinds of data and actuators. Third, our methodology does not require that the 
actuators used to be placed at, or close to, the perimeter of the regions. 
In other words, our methodology formally generalizes standard perimeter control, through 
the Hankel matrix an implicit mapping is created between the actuators and the 
aggregate dynamics of the network. Therefore, we resolve to normal perimeter 
control if the actuators are placed at the region boundaries. \\
The method comes with some drawbacks too. First, when dealing with non-linear 
systems, one cannot be sure that the Hankel matrix spans the whole behavior of the 
system, even if it satisfies the rank conditions needed for persistency of 
excitation. This leads to the fact that, depending on the distribution of the data 
collected from the system, we may not be able to obtain a satisfactory control 
policy from every Hankel matrix. This means that a Hankel matrix will only be able 
to represent the system appropriately when the conditions of the system, i.e., 
the exogenous demand, are close enough to the conditions in which the data were collected. 
This leads to a number of new research directions that could be explored in order to 
make this method more robust to different conditions. For example, one could 
design a data collection pipeline aimed at mapping multiple behaviors of the 
non-linear system and then either aggregate behaviors into a single Hankel matrix 
or build a library of Hankel matrixes and choose through some criterion the matrix 
that best matches the current conditions. Another challenging problem regards scalability, 
as the dimension of the optimization problem scales with the number of actuators. 
A possible solution is to distribute the \gls{DeePC} algorithm via alternating projections 
\cite{Split-as-a-pro}. As final words, \gls{DeePC} is a promising 
new tool that has been proven to work in a variety of applications, in this 
paper we explored how it can be applied to perform urban traffic control by 
solving the \gls{ITSCP} problem, while the results are promising many 
research directions have been opened for future work. 

\begin{appendix}
\section{Preliminaries}\label{app:preliminaries}

The following preliminary discussion closely follows the one in 
\cite{rimoldi2023urban} and it is 
reported here to render this work self-contained.

The restriction of a sequence over an interval gives rise to the cut operator. 
Formally, given ${w\in {(\R^q)}^\T}$ and ${L\in\T}$, 
the \textit{cut operator} is defined as
\begin{equation} \nn
w|_L = (w(1),\ldots,w(L)) \in {(\R^q)}^L .
\end{equation}
For infinite trajectories, the definition holds verbatim with 
${w\in{(\R^q)}^\N}$ and ${L\in\N}$. Applied to a set ${\mathcal{W}}$ of 
(finite or infinite) trajectories, the cut operator acts on all elements 
in the set giving rise to the \emph{restricted} set 
${\mathcal{W}|_{L}  =\{w|_L \,:\, w\in \mathcal{W} \}}$. 

Shifting elements of a sequence give rise to the shift operator. 
Formally, given 
${w\in {(\R^q)}^\T}$ and $\tau\in\T$, 
the \textit{shift operator} is defined as 
\begin{equation} \nn
\sigma^{\tau-1}w = (w(\tau),\ldots,w(T)) \in {\R^{q(T-\tau+1)}} .
\end{equation}
For infinite trajectories, the shift operator is defined as 
$w \mapsto \sigma^{\tau-1}w$, with ${\sigma^{\tau-1}w(t) = w(t+\tau-1)}$, 
for any ${\tau \in \N}$. Applied to a set ${\mathcal{W}}$ of (finite or infinite) 
trajectories,the shift operator acts on all elements in the set giving rise to 
the \emph{shifted} set $\sigma^\tau\mathcal{W}  =\{\sigma^\tau w\,:\, w\in\mathcal{W} \}$.
For finite-length time series, the \textit{unit shift operator}, defined as 
$\sigma(\cdot) = \sigma^{1}(\cdot),$  can be applied at most $T-1$ times.

Given ${m\in\N}$ and a permutation matrix ${\Pi\in\R^{q\times q}}$, 
the map $w\mapsto \Pi w = (u,y)$ defines a \textit{partition} of 
${w\in(\R^q)^{\N}}$ into the variables ${u\in(\R^m)^{\N}}$ and 
${y\in(\R^{q-m})^{\N}}$. 
The map is an \textit{input-output partition} of a system 
${\Sigma}$ if $u$ is \textit{free}, \textit{not anticipated}, and 
\textit{causal}~\cite{willems1989models}, in which case $u$ is the 
\textit{input} and $y$ is the output of $\Sigma$. 
The reader is referred to~\cite[Section 6]{willems1986timeI} 
and~\cite{willems1989models} for further detail.

The structure of an LTI system is characterized by a set of integer invariants 
known as \textit{structure indices}~\cite{willems1986timeI}. The most important 
ones are the \textit{number of inputs} $m$, \textit{number of outputs} $p$, 
the \textit{lag} $\ell$, and the \textit{order} $n$, see,
~\cite[Section 7]{willems1986timeI} for definitions. 
Every finite-dimensional LTI system admits a minimal representation and can be 
described by the equations
\beq \label{eq:state-space}
\sigma x = Ax + Bu, \quad y=Cx+Du,
\eeq
where
$\scalebox{0.75}{$\bma\!
\begin{array}{cc}
A & B \\
C & D
\end{array}\!\ema $}
 \in \R^{(n+p)\times (n+m)}$ and  $m$, $n$, and $p$ are the number of inputs, 
the order, and the number of outputs, respectively. 
 
The \textit{order} of an LTI system ${\B\in\Li^q}$ is the smallest ${n\in\N}$ among 
all (minimal) state-space representations~\eqref{eq:state-space} and the 
\textit{lag} is the smallest ${\ell\in\N}$  such  that in a (minimal) state-space 
representation the \textit{observability matrix} 
\beq  \nn
\mathsf{O}_\ell
=
\bma
\begin{array}{cc}
C \\
CA \\
\vdots  \\
CA^{\ell -1} 
\end{array} 
\ema 
\eeq 
is full rank. 

Given a trajectory  ${w_{\tup{d}} \in \R^{qT}}$ of a system 
${\B \in \mathcal{L}^{q}}$, it is possible to derive a non-parametric 
representation of its finite-horizon behavior using raw data. 
We summarize a version of this principle known in the control theory literature 
as the \textit{fundamental lemma}~\cite{willems2005note}.

\begin{lemma}~\cite[Corollary 19]{markovsky2022identifiability} \label{lemma:fundamental_lemma_generalized}
Consider a system ${\B \in \Li^{q}}$ with lag ${\ell\in\N}$ and a trajectory of 
the system ${w_{\tup{d}} \in \B|_{ [1,T] }}$. Assume ${L > \ell}$. Then   
$\B|_{ [1,L] } = \Image H_L(w_{\tup{d}})$
if and only if
\beq  \label{eq:generalized_persistency_of_excitation} 
\rank H_L(w_{\tup{d}})   =  mL+ n,
\eeq
where $n$ and $m$ are the order and the number of inputs of the system, 
respectively.
\end{lemma}

Lemma~\ref{lemma:fundamental_lemma_generalized} is a key result in data-driven 
control~\cite{markovsky2021behavioral}. It characterizes all trajectories of a 
given length of an LTI system in terms of the image of a Hankel matrix, which, in 
turn, can be constructed directly from raw data. 
This foundational principle can be adapted in various ways to suit different 
assumptions, see the recent survey~\cite{markovsky2021behavioral}. 
Remarkably, non-parametric representations have found practical use in 
data-driven control even when dealing with \textit{non-linear} 
systems~\cite{berberich2022linear,padoan2023data}.

The rank condition~\eqref{eq:generalized_persistency_of_excitation} is known as the 
\emph{generalized persistency of excitation} condition
~\cite{markovsky2022identifiability}. Note that upper bounds on the structure 
indices of $\B$ are necessary to check this condition from data. 
Alternatively, the rank condition~\eqref{eq:generalized_persistency_of_excitation} 
can be guaranteed to hold for controllable systems if a certain rank condition 
on the inputs holds~\cite{willems2005note}.

\section{Region partitioning}\label{app:partitioning}
When considering large-scale traffic networks of a whole city, researchers often 
rely on aggregate models describing the flow of vehicles among different regions 
\cite{mfd,linearMPC}. 
We adopt a similar approach, dividing a city into traffic-wise homogeneous regions 
to decrease the number of variables taken into consideration. Aggregate descriptions 
such as the \gls{MFD} also reduce the variability in the data, which in terms 
smoothens the dynamics of the system and renders it more suited to linear 
approximations such as the one performed by \gls{DeePC}. 
In this section we provide the details on how to achieve a partitioning such as the
seen in Figure \ref{fig:partitioning}.

We run a single simulation without any actuation acting on the traffic lights and collect roadwise 
density data. 
We then average the data over the simulation duration and assign to each road its 
average traffic density. We use this data to run the Snake clustering algorithm \cite{gerolinimis2016CsnakeClustering}. 
It is possible to parallelize the first two steps of the algorithms presented in 
\cite{gerolinimis2016CsnakeClustering}, that is \textbf{A.} Running the snakes and 
\textbf{B.} computing similarities. This is especially necessary for step \textbf{B.} as it 
scales with a complexity of $\mathcal{O}(n^{2})$.
Assume to have $n\in \N$ roads in a city collected in a set $\mathcal{X}$ each equipped with its traffic density averaged over time $\hat{\rho}$. Given a depth $m\in \N, m<n$.
We present below the pseudocode for the parallel implementation of the algorithm

\begin{algorithm}
    \caption{Parallel snake clustering}
    \label{alg:efficient_snake_clustering}
    \textbf{Input:}
    A set of roads $\mathcal{X}$ with cardinality $n$, $\mathcal{N}(A)$ the set of neighbouring roads of a set $A$, 
    a maximal snake length $m\leq n$
    \begin{algorithmic}
        \STATE \textbf{A.} Running the snakes
        \STATE \textbf{parallel do}
            \bindent
                \STATE $S_x \leftarrow x$ 
                \WHILE{$\text{Size}(S_x)< m$}
                    \STATE  $S' =\mathcal{N}(S_x)$
                    \STATE  $s^* =\{s| \min_{s\in S'} \text{var}(S_x \cup s) \}$ 
                    \STATE  $S_x \leftarrow [S_x, s^*]$ 
                \ENDWHILE
            \eindent
        \STATE \textbf{end parallel}
        \STATE
        \STATE \textbf{B.} Computing similarities
        \STATE Initialize $\phi$
        \STATE Initialize partitioning $\mathcal{P}(\mathcal{X})$
        \STATE $W \rightarrow \bs 0_{n\times n}$
        \STATE \textbf{parallel do}
        \bindent
            \FORALL{$i,j \in A \subset \mathcal{P}(\mathcal{X})$}
                \STATE $k \leftarrow 1$
                \WHILE{$k \leq m$}
                    \STATE $W[i,j] = W[i,j] + \phi^{n-k}\times intersect(S_{i,k},S_{j,k})$ 
                    \STATE $k\leftarrow k+1$ 
                \ENDWHILE 
            \ENDFOR
        \eindent
        \STATE \textbf{end parallel}
        \STATE
        \STATE \textbf{C.} Symmetric Non-negative Matrix Factorization
        \STATE $D=\text{diag}(d_i)$ \textbf{where} $d_i=\sum_{j=1}^{m}W[i,j]$
        \STATE $\hat{W}= D^{-\frac{1}{2}}WD^{\frac{1}{2}}$
        \STATE $H^*=\min\{H\in \R_+^{n\times n_S}||W-HH^\transpose||^{2} \}$
        \FORALL{$j \in \mathcal{X}$}
            \STATE $j^*=\{j|\max_{j\in\{1,\dots,n_S\}} H^*[i,j] \}$
            \STATE $i \in A_{j^*}$
        \ENDFOR             
    \end{algorithmic}
    \textbf{Output: }
    Set of clusters $(A_1,\dots,A_{n_S})$
\end{algorithm}

\section{Linear model predictive control formulation}\label{app:mpc}

We report the non-linear and linear model of traffic dynamics as well as
the \gls{MPC} formulation used by \cite{linearMPC} to render the paper self-contained.

\subsection*{Non-linear aggregated dynamics for a partitioned city}

Consider an urban network partitioned into $N$ homogeneous regions.
For each region $i \in \mathcal{N} = \{1,\dots,N\}$ define:

\begin{itemize}
    \item $n_i(k)$: total accumulation of region $i$,
    \item $n_{ij}(k)$: vehicles in region $i$ with destination region $j$,
    \item $q_{ij}(k)$: exogenous demand generated in region $i$ for destination $j$,
    \item $\mathcal{N}_i$: set of adjacent regions to $i$,
    \item $M^h_{ij}(k)$: transfer flow from region $i$ to region $h$ for vehicles destined to $j$,
    \item $M_{ii}(k)$: internal trip completion rate of region $i$.
\end{itemize}
The sampling period is denoted by $T_p$.
The state dynamics are modeled as
\begin{align}
n_{ii}(k+1) &= n_{ii}(k)
+ T_p\Big( q_{ii}(k) - M_{ii}(k)
- \sum_{h\in\mathcal{N}_i} M^{h}_{ii}(k)
+ \sum_{h\in\mathcal{N}_i} M^i_{hi}(k) \Big), \label{eq:nonlinear_nii}
\\[2mm]
n_{ij}(k+1) &= n_{ij}(k)
+ T_p\Big( q_{ij}(k)
- \sum_{h\in\mathcal{N}_i} M^{h}_{ij}(k)
+ \sum_{h\in\mathcal{N}_i} M^i_{hj}(k) \Big), \qquad i\neq j. 
\label{eq:nonlinear_nij}
\end{align}

The total accumulation in region $i$ is
\begin{equation}
n_i(k) = \sum_{j \in \mathcal{N}} n_{ij}(k).
\end{equation}

Each region has a production MFD $P_i(n_i(k)) \ [\text{veh}\cdot\text{m}/\text{sec}]$ between accumulation and total production.  
Let $L_i$ be the average trip length and $\theta_{ij}^h(k)$ the routing proportions. The internal trip completion rate is defined as
\begin{equation}
M_{ii}(k)
= \theta_{ii}(k)\,\frac{n_{ii}(k)}{n_i(k)}\,\frac{P_i(n_i(k))}{L_i}.
\end{equation}
while the transfer flows are 
\begin{equation}
M^{h}_{ij}(k)
= \min\!\left\{
C_{ih}(n_h(k)),\;
u_{ih}(k_c)\,
\theta^h_{ij}(k)
\frac{n_{ij}(k)}{n_i(k)}\,
\frac{P_i(n_i(k))}{L_i}
\right\},
\end{equation}
where $k_c = \lfloor k/N_c \rfloor$ is the control time index and
$C_{ih}(n_h(k))$ is the receiving capacity.
The control contraints are
\begin{align}
0 \le u_{ih}(k_c) \le 1, \qquad \forall i \in \mathcal{N},\; h\in \mathcal{N}_i,
\\
|u_{ih}(k_c) - u_{ih}(k_c - 1)| \le u^R_{ih}.
\end{align}
A gridlock avoidance constraint ensures that
\begin{equation}\label{eq:gridlock_avoidance}
0 \le n_i(k) \le n_{i,\max}.
\end{equation}
We can now formulate the non-linear \gls{MPC} problem for prediction horizon $N_p$ as
\begin{align}
\max_{\{n_{ij}(k),\,u_{ih}(k_c)\}}
\quad &
\sum_{k=k_p}^{k_p+N_p - 1}
\sum_{i\in\mathcal{N}}
L_i\!\left[
M_{ii}(k) +  u_{ih}(k_c)\,M^h_{ij}(k)
\right]
\\[1mm]
\text{s.t.} \quad &
\eqref{eq:nonlinear_nii} - \eqref{eq:gridlock_avoidance},\\
&k = k_p, k_p+1,\dots,k_p+N_p-1, \forall i,j \in \mathcal{N}, h\in \mathcal{N}_i \\
&k_c = \left\lfloor \frac{k}{N_c} \right\rfloor, 
\end{align}

\subsection*{Linearised model and linear MPC}
Following the linearisation procedure in \cite{linearMPC}, three approximations are used:
\begin{enumerate}
    \item Accumulation proportions
    \(
    \alpha_{ij}(k) = n_{ij}(k)/n_i(k)
    \)
    are treated as exogenous constants over the horizon.
    \item The MFD is approximated by piecewise-affine (PWA) functions
    \(
    G_i(n_i) = G_i^\ell(n_i),\;\ell = 1,\dots,N_i.
    \)
    \item Linear decision variables are introduced:
    \begin{align}
        f_{ii}(k) &= u_{ii}(k)\,G_i(n_i(k))\,\theta_{ii}(k)\alpha_{ii}(k),
        \\
        f_{ih}(k) &= u_{ih}(k)\,G_i(n_i(k))\sum_{j\in\mathcal{N}}\theta^h_{ij}(k)\,\alpha_{ij}(k).
    \end{align}
\end{enumerate}
The linear approximation of the state dynamics is written as
\begin{align}
n_i(k+1)
= n_i(k)
+ T_p\Big(
q_i(k) - f_{ii}(k)
- \sum_{h\in\mathcal{N}_i} f_{ih}(k)
+ \sum_{h\in\mathcal{N}_i} f_{hi}(k)
\Big).
\end{align}
constrained to
\begin{align}
0 \le f_{ii}(k) \le \theta_{ii}(k)\alpha_{ii}(k)\,G_i^\ell(n_i(k)),
\\
0 \le f_{ih}(k) \le 
G_i^\ell(n_i(k))\sum_{j\in\mathcal{N}}\theta^h_{ij}(k)\alpha_{ij}(k),
\\
0 \le n_i(k) \le n_{i,\max}.
\end{align}
Optional linearised rate constraints:
\begin{align}
f_{ih}(k_p) &\le 
\left(u^{P,R}_{ih} + u^R_{ih}\right)\,
G_i^P(n_i(k_p))\sum_j\theta^h_{ij}(k_p)\alpha_{ij}(k_p),
\\
f_{ih}(k_p) &\ge 
\left(u^{P,R}_{ih} - u^R_{ih}\right)\,
G_i^P(n_i(k_p))\sum_j\theta^h_{ij}(k_p)\alpha_{ij}(k_p).
\end{align}
The linear \gls{MPC} objective is then 
\begin{align}
\max_{\{n_i(k),\,f_{ii}(k),\,f_{ih}(k)\}}
\quad
\sum_{k=k_p}^{k_p + N_p - 1}
\sum_{i\in\mathcal{N}}
L_i\left[ f_{ii}(k) +f_{ih}(k) \right]
\end{align}

subject to the linear dynamics and constraints above.
This is the controller used as comparison to \gls{DeePC} in Section \ref{sec:simulations}.
\section{Simulation and control parameters}\label{app:parameters}
In this section we detail the parameters used in Section \ref{sec:simulations} by the controllers
and the simulation package
\subsection*{Lattice network}
The seed used by \gls{SUMO} is 42.\\
\textbf{\gls{MPC}}: The prediction horizon has value $N_p =4$; the number of piecewise components used in the piecewise 
linear approximation of the \gls{MFD} is $N_i =10$.

\textbf{\gls{DeePC}}:
The prediction horizon has value $\Tf = 4$ and the past trajectory length is $\Tini = 5$.
The parameters used in the cost function have value  

\begin{equation*}
\lambda_1 = 1, \lambda_2 =1, \lambda_y = 0,
    Q = \begin{bmatrix}
        1 & 0 \\
        0 & 1 
    \end{bmatrix}, \quad \begin{bmatrix}
    2 & 0 \\
    0 & 2 
    \end{bmatrix}
\end{equation*}

\subsection*{Z\"urich network}
The seed used by \gls{SUMO} is 2389.\\
\textbf{\gls{MPC}}: The prediction horizon has value $N_p =4$; the number of piecewise components used in the piecewise 
linear approximation of the \gls{MFD} is $N_i =10$.\\
\textbf{\gls{DeePC}}: The prediction horizon has value $\Tf = 4$ and the past trajectory length is $\Tini = 5$.
The parameters used in the cost function have value  

\begin{equation*}
\lambda_1 = 15, \lambda_2 =20, \lambda_y = 0,
    Q = \begin{bmatrix*}
        1&0&0&0&0&0&0&0\\
        0&1&0&0&0&0&0&0\\
        0&0&1&0&0&0&0&0\\
        0&0&0&1&0&0&0&0\\
        0&0&0&0&1&0&0&0\\
        0&0&0&0&0&1&0&0\\
        0&0&0&0&0&0&1&0\\
        0&0&0&0&0&0&0&1\\
    \end{bmatrix*}, \quad R = \begin{bmatrix*}
        2&0&0&0&0&0&0&0\\
        0&2&0&0&0&0&0&0\\
        0&0&2&0&0&0&0&0\\
        0&0&0&2&0&0&0&0\\
        0&0&0&0&2&0&0&0\\
        0&0&0&0&0&2&0&0\\
        0&0&0&0&0&0&2&0\\
        0&0&0&0&0&0&0&2\\
    \end{bmatrix*}  
\end{equation*}
\end{appendix}

\selectlanguage{english}
\FloatBarrier

\bibliographystyle{IEEEtran}
 \bibliography{bibliography/converted_to_latex}

\begin{thebibliography}{10}
\providecommand{\url}[1]{#1}
\csname url@samestyle\endcsname
\providecommand{\newblock}{\relax}
\providecommand{\bibinfo}[2]{#2}
\providecommand{\BIBentrySTDinterwordspacing}{\spaceskip=0pt\relax}
\providecommand{\BIBentryALTinterwordstretchfactor}{5}
\providecommand{\BIBentryALTinterwordspacing}{\spaceskip=\fontdimen2\font plus
\BIBentryALTinterwordstretchfactor\fontdimen3\font minus \fontdimen4\font\relax}
\providecommand{\BIBforeignlanguage}[2]{{%
\expandafter\ifx\csname l@#1\endcsname\relax
\typeout{** WARNING: IEEEtran.bst: No hyphenation pattern has been}%
\typeout{** loaded for the language `#1'. Using the pattern for}%
\typeout{** the default language instead.}%
\else
\language=\csname l@#1\endcsname
\fi
#2}}
\providecommand{\BIBdecl}{\relax}
\BIBdecl

\bibitem{un2018}
\BIBentryALTinterwordspacing
{\relax United Nations}. (2018) 68\% of the world population projected to live in urban areas by 2050, says un. [Online]. Available: \url{https://www.un.org/development/desa/en/news/population/2018-revision-of-world-urbanization-prospects.html}
\BIBentrySTDinterwordspacing

\bibitem{eutransportfigures}
{\relax European Commission and Directorate-General for Mobility and Transport}, \emph{EU transport in figures – Statistical pocketbook 2022}.\hskip 1em plus 0.5em minus 0.4em\relax Publications Office of the European Union, 2022.

\bibitem{transport-in-eu}
R.~Diemer and F.~Dittrich, \emph{Transport in the European Union: current trends and issues}.\hskip 1em plus 0.5em minus 0.4em\relax Directorate-General for Mobility and Transport of the European Union, 2019.

\bibitem{aripollution-health}
\BIBentryALTinterwordspacing
P.~J. Landrigan, ``Air pollution and health,'' \emph{The Lancet Public Health}, vol.~2, no.~1, pp. e4--e5, Jan 2017. [Online]. Available: \url{https://doi.org/10.1016/S2468-2667(16)30023-8}
\BIBentrySTDinterwordspacing

\bibitem{eufututure-of-transport}
M.~Raposo, I.~Vandecasteele, and {\relax et al.}, ``The future of road transport,'' no. KJ-1A-29748-EN-N (online), KJ-1A-29748-EN-C (print), 2019.

\bibitem{braess-paradox}
\BIBentryALTinterwordspacing
D.~Braess, A.~Nagurney, and T.~Wakolbinger, ``On a paradox of traffic planning,'' \emph{Transportation Science}, vol.~39, no.~4, pp. 446--450, 2005. [Online]. Available: \url{https://homepage.rub.de/Dietrich.Braess/Paradox-BNW.pdf}
\BIBentrySTDinterwordspacing

\bibitem{traffic-control-review-papageorgiu}
M.~Papageorgiou, C.~Diakaki, V.~Dinopoulou, A.~Kotsialos, and Y.~Wang, ``Review of road traffic control strategies,'' \emph{Proceedings of the IEEE}, vol.~91, no.~12, pp. 2043--2067, 2003.

\bibitem{gerolimins:2013:optimal_periter_two_regions}
N.~Geroliminis, J.~Haddad, and M.~Ramezani, ``Optimal perimeter control for two urban regions with macroscopic fundamental diagrams: A model predictive approach,'' \emph{IEEE Transactions on Intelligent Transportation Systems}, vol.~14, no.~1, pp. 348--359, 2013.

\bibitem{kouvelas-mpc}
\BIBentryALTinterwordspacing
A.~Kouvelas, M.~Saeedmanesh, and N.~Geroliminis, ``A linear-parameter-varying formulation for model predictive perimeter control in multi-region mfd urban networks,'' \emph{Transportation Science}, vol.~57, no.~6, pp. 1496--1515, 2023. [Online]. Available: \url{https://doi.org/10.1287/trsc.2022.0103}
\BIBentrySTDinterwordspacing

\bibitem{rimoldi2023urban}
A.~Rimoldi, C.~Cenedese, A.~Padoan, F.~Dörfler, and J.~Lygeros, ``\BIBforeignlanguage{en}{Urban traffic congestion control: A deepc change},'' in \emph{\BIBforeignlanguage{en}{2024 European Control Conference (ECC)}}.\hskip 1em plus 0.5em minus 0.4em\relax Piscataway, NJ: IEEE, 2024, Conference Paper, pp. 2909 -- 2914, 22nd European Control Conference (ECC 2024); Conference Location: Stockholm, Sweden; Conference Date: June 25-28, 2024.

\bibitem{arel-rl}
I.~Arel, C.~Liu, T.~Urbanik, and A.~Kohls, ``Reinforcement learning-based multi-agent system for network traffic signal control,'' \emph{Intelligent Transport Systems, IET}, vol.~4, pp. 128 -- 135, 07 2010.

\bibitem{balaji-rl}
B.~Parasumanna~Gokulan, X.~German, and D.~Srinivasan, ``Urban traffic signal control using reinforcement learning agents,'' \emph{Intelligent Transport Systems, IET}, vol.~4, pp. 177 -- 188, 10 2010.

\bibitem{mas-bazzan}
\BIBentryALTinterwordspacing
A.~L.~C. Bazzan, ``A distributed approach for coordination of traffic signal agents,'' \emph{Autonomous Agents and Multi-Agent Systems}, vol.~10, no.~1, pp. 131--164, Jan 2005. [Online]. Available: \url{https://doi.org/10.1007/s10458-004-6975-9}
\BIBentrySTDinterwordspacing

\bibitem{webster}
\BIBentryALTinterwordspacing
F.~V. Webster. (1958) Traffic signal settings. [Online]. Available: \url{https://trid.trb.org/view/113579}
\BIBentrySTDinterwordspacing

\bibitem{transyt}
D.~Robertson, ``Transyt method for area traffic control,'' \emph{Transportation Engineering \& Control}, vol.~11, pp. 276--281, 1969.

\bibitem{traffic-light-review-eom}
\BIBentryALTinterwordspacing
M.~Eom and B.~Kim, ``The traffic signal control problem for intersections: a review,'' \emph{European Transport Research Review}, vol.~12, no.~1, p.~50, Sep 2020. [Online]. Available: \url{https://doi.org/10.1186/s12544-020-00440-8}
\BIBentrySTDinterwordspacing

\bibitem{SimOpt-review}
S.~S. S.~M. Qadri, M.~Gökçe, and E.~Oner, ``State-of-art review of traffic signal control methods: challenges and opportunities,'' \emph{European Transport Research Review}, vol.~12, pp. 1--23, 10 2020.

\bibitem{potts-algorithm}
\BIBentryALTinterwordspacing
M.~C. Dunne and R.~B. Potts, ``Algorithm for traffic control,'' \emph{Operations Research}, vol.~12, no.~6, pp. 870--881, 1964. [Online]. Available: \url{http://www.jstor.org/stable/168173}
\BIBentrySTDinterwordspacing

\bibitem{Panos-oversaturatedI}
\BIBentryALTinterwordspacing
P.~G. Michalopoulos and G.~Stephanopoulos, ``Oversaturated signal systems with queue length constraints—i: Single intersection,'' \emph{Transportation Research}, vol.~11, no.~6, pp. 413--421, 1977. [Online]. Available: \url{https://www.sciencedirect.com/science/article/pii/0041164777900065}
\BIBentrySTDinterwordspacing

\bibitem{Panos-oversaturatedII}
\BIBentryALTinterwordspacing
------, ``Oversaturated signal systems with queue length constraints—ii: Systems of intersections,'' \emph{Transportation Research}, vol.~11, no.~6, pp. 423--428, 1977. [Online]. Available: \url{https://www.sciencedirect.com/science/article/pii/0041164777900077}
\BIBentrySTDinterwordspacing

\bibitem{fuzzy-sets}
\BIBentryALTinterwordspacing
L.~Zadeh, ``Fuzzy sets,'' \emph{Information and Control}, vol.~8, no.~3, pp. 338--353, 1965. [Online]. Available: \url{https://www.sciencedirect.com/science/article/pii/S001999586590241X}
\BIBentrySTDinterwordspacing

\bibitem{fuzzy-application}
\BIBentryALTinterwordspacing
S.~Mohanaselvi and B.~Shanpriya, ``{Application of fuzzy logic to control traffic signals},'' \emph{AIP Conference Proceedings}, vol. 2112, no.~1, p. 020045, 06 2019. [Online]. Available: \url{https://doi.org/10.1063/1.5112230}
\BIBentrySTDinterwordspacing

\bibitem{fuzzy-adaptive}
S.~Chiu and S.~Chand, ``Adaptive traffic signal control using fuzzy logic,'' in \emph{Proceedings. The First IEEE Regional Conference on Aerospace Control Systems,}, 1993, pp. 122--126.

\bibitem{fuzzy-controller}
C.~P. Pappis and E.~H. Mamdani, ``A fuzzy logic controller for a trafc junction,'' \emph{IEEE Transactions on Systems, Man, and Cybernetics}, vol.~7, no.~10, pp. 707--717, 1977.

\bibitem{fuzzy-multi-phased}
\BIBentryALTinterwordspacing
Y.~S. Murat and E.~Gedizlioglu, ``A fuzzy logic multi-phased signal control model for isolated junctions,'' \emph{Transportation Research Part C: Emerging Technologies}, vol.~13, no.~1, pp. 19--36, 2005. [Online]. Available: \url{https://www.sciencedirect.com/science/article/pii/S0968090X04000683}
\BIBentrySTDinterwordspacing

\bibitem{dynamic-transit}
W.~Ekeila, T.~Sayed, and M.~Elesawey, ``Development of dynamic transit signal priority strategy,'' \emph{Transportation Research Record}, vol. 2111, pp. 1--9, 12 2009.

\bibitem{fuzzy-niitty}
\BIBentryALTinterwordspacing
J.~Niittymäki and M.~Pursula, ``Signal control using fuzzy logic,'' \emph{Fuzzy Sets and Systems}, vol. 116, no.~1, pp. 11--22, 2000. [Online]. Available: \url{https://www.sciencedirect.com/science/article/pii/S0165011499000342}
\BIBentrySTDinterwordspacing

\bibitem{dp-sen-head}
S.~Sen and L.~Head, ``Controlled optimization of phases at an intersection,'' \emph{Transportation Science}, vol.~31, pp. 5--17, 02 1997.

\bibitem{dp-mirchandani}
\BIBentryALTinterwordspacing
P.~Mirchandani and L.~Head, ``A real-time traffic signal control system: architecture, algorithms, and analysis,'' \emph{Transportation Research Part C: Emerging Technologies}, vol.~9, no.~6, pp. 415--432, 2001. [Online]. Available: \url{https://www.sciencedirect.com/science/article/pii/S0968090X00000474}
\BIBentrySTDinterwordspacing

\bibitem{dp-cai}
\BIBentryALTinterwordspacing
C.~Cai, C.~K. Wong, and B.~G. Heydecker, ``Adaptive traffic signal control using approximate dynamic programming,'' \emph{Transportation Research Part C: Emerging Technologies}, vol.~17, no.~5, pp. 456--474, 2009, artificial Intelligence in Transportation Analysis: Approaches, Methods, and Applications. [Online]. Available: \url{https://www.sciencedirect.com/science/article/pii/S0968090X09000321}
\BIBentrySTDinterwordspacing

\bibitem{dp-zheng}
\BIBentryALTinterwordspacing
X.~Zheng and W.~Recker, ``An adaptive control algorithm for traffic-actuated signals,'' \emph{Transportation Research Part C: Emerging Technologies}, vol.~30, pp. 93--115, 2013. [Online]. Available: \url{https://www.sciencedirect.com/science/article/pii/S0968090X13000375}
\BIBentrySTDinterwordspacing

\bibitem{white-distributed-control}
\BIBentryALTinterwordspacing
D.~McKenney and T.~White, ``Distributed and adaptive traffic signal control within a realistic traffic simulation,'' \emph{Eng. Appl. Artif. Intell.}, vol.~26, pp. 574--583, 2013. [Online]. Available: \url{https://api.semanticscholar.org/CorpusID:36526311}
\BIBentrySTDinterwordspacing

\bibitem{villabos-game}
\BIBentryALTinterwordspacing
I.~A. Villalobos, A.~S. Poznyak, and A.~M. Tamayo, ``Urban traffic control problem: a game theory approach,'' \emph{IFAC Proceedings Volumes}, vol.~41, no.~2, pp. 7154--7159, 2008, 17th IFAC World Congress. [Online]. Available: \url{https://www.sciencedirect.com/science/article/pii/S1474667016400960}
\BIBentrySTDinterwordspacing

\bibitem{choi-mas}
M.~C. Choy, D.~Srinivasan, and R.~Cheu, ``Cooperative, hybrid agent architecture for real-time traffic signal control,'' \emph{IEEE Transactions on Systems, Man, and Cybernetics - Part A: Systems and Humans}, vol.~33, no.~5, pp. 597--607, 2003.

\bibitem{varaiya_max_pressure}
\BIBentryALTinterwordspacing
P.~Varaiya, ``Max pressure control of a network of signalized intersections,'' \emph{Transportation Research Part C: Emerging Technologies}, vol.~36, pp. 177--195, Nov. 2013. [Online]. Available: \url{https://www.sciencedirect.com/science/article/pii/S0968090X13001782}
\BIBentrySTDinterwordspacing

\bibitem{wongpiromsarn_max_pressure}
\BIBentryALTinterwordspacing
T.~Wongpiromsarn, T.~Uthaicharoenpong, Y.~Wang, E.~Frazzoli, and D.~Wang, ``Distributed traffic signal control for maximum network throughput,'' in \emph{2012 15th {International} {IEEE} {Conference} on {Intelligent} {Transportation} {Systems}}, Sep. 2012, pp. 588--595, iSSN: 2153-0017. [Online]. Available: \url{https://ieeexplore.ieee.org/document/6338817}
\BIBentrySTDinterwordspacing

\bibitem{original_max_pressure}
\BIBentryALTinterwordspacing
L.~Tassiulas and A.~Ephremides, ``Stability properties of constrained queueing systems and scheduling policies for maximum throughput in multihop radio networks,'' \emph{IEEE Transactions on Automatic Control}, vol.~37, no.~12, pp. 1936--1948, Dec. 1992. [Online]. Available: \url{https://ieeexplore.ieee.org/document/182479}
\BIBentrySTDinterwordspacing

\bibitem{mpc-guarantees}
P.~Scokaert, D.~Mayne, and J.~Rawlings, ``Suboptimal model predictive control (feasibility implies stability),'' \emph{IEEE Transactions on Automatic Control}, vol.~44, no.~3, pp. 648--654, 1999.

\bibitem{mfd}
N.~Geroliminis and D.~C. F., ``Macroscopic modelling of traffic in cities,'' \emph{86th Annual Meeting Transportation Research Board}, vol.~42, 2007.

\bibitem{existencemfd}
\BIBentryALTinterwordspacing
N.~Geroliminis and C.~F. Daganzo, ``Existence of urban-scale macroscopic fundamental diagrams: Some experimental findings,'' \emph{Transportation Research Part B: Methodological}, vol.~42, no.~9, pp. 759--770, 2008. [Online]. Available: \url{https://www.sciencedirect.com/science/article/pii/S0191261508000180}
\BIBentrySTDinterwordspacing

\bibitem{geroliminis:2011:properties_MFD}
\BIBentryALTinterwordspacing
N.~Geroliminis and J.~Sun, ``Properties of a well-defined macroscopic fundamental diagram for urban traffic,'' \emph{Transportation Research Part B: Methodological}, vol.~45, no.~3, pp. 605--617, 2011. [Online]. Available: \url{https://www.sciencedirect.com/science/article/pii/S0191261510001372}
\BIBentrySTDinterwordspacing

\bibitem{geroliminig:2018:MPC_perimeter_control}
I.~I. Sirmatel and N.~Geroliminis, ``Economic model predictive control of large-scale urban road networks via perimeter control and regional route guidance,'' \emph{IEEE Transactions on Intelligent Transportation Systems}, vol.~19, no.~4, pp. 1112--1121, 2018.

\bibitem{lin2012efficient}
\BIBentryALTinterwordspacing
S.~Lin, B.~{De Schutter}, Y.~Xi, and H.~Hellendoorn, ``Efficient network-wide model-based predictive control for urban traffic networks,'' \emph{Transportation Research Part C: Emerging Technologies}, vol.~24, pp. 122--140, 2012. [Online]. Available: \url{https://www.sciencedirect.com/science/article/pii/S0968090X12000150}
\BIBentrySTDinterwordspacing

\bibitem{markovsky2021behavioral}
\BIBentryALTinterwordspacing
I.~Markovsky and F.~Dörfler, ``Behavioral systems theory in data-driven analysis, signal processing, and control,'' \emph{Annual Reviews in Control}, vol.~52, pp. 42--64, 2021. [Online]. Available: \url{https://www.sciencedirect.com/science/article/pii/S1367578821000754}
\BIBentrySTDinterwordspacing

\bibitem{willems1986timeI}
\BIBentryALTinterwordspacing
J.~C. Willems, ``From time series to linear system—part i. finite dimensional linear time invariant systems,'' \emph{Automatica}, vol.~22, no.~5, pp. 561--580, 1986. [Online]. Available: \url{https://www.sciencedirect.com/science/article/pii/000510988690066X}
\BIBentrySTDinterwordspacing

\bibitem{coulson2019data}
J.~Coulson, J.~Lygeros, and F.~Dörfler, ``Data-enabled predictive control: In the shallows of the deepc,'' in \emph{2019 18th European Control Conference (ECC)}, 2019, pp. 307--312.

\bibitem{carlet2020data}
P.~G. Carlet, A.~Favato, S.~Bolognani, and F.~Dörfler, ``Data-driven predictive current control for synchronous motor drives,'' in \emph{2020 IEEE Energy Conversion Congress and Exposition (ECCE)}, 2020, pp. 5148--5154.

\bibitem{huang2019data}
L.~Huang, J.~Coulson, J.~Lygeros, and F.~Dörfler, ``\BIBforeignlanguage{en}{Data-enabled predictive control for grid-connected power converters}.''\hskip 1em plus 0.5em minus 0.4em\relax Piscataway, NJ: IEEE, 2019-12, Conference Paper, pp. 8130 -- 8135, 58th IEEE Conference on Decision and Control (CDC 2019); Conference Location: Nice, France; Conference Date: December 11-13, 2019; Conference lecture held on December 13, 2019.

\bibitem{wang2023deep}
J.~Wang, Y.~Zheng, K.~Li, and Q.~Xu, ``Deep-lcc: Data-enabled predictive leading cruise control in mixed traffic flow,'' \emph{IEEE Transactions on Control Systems Technology}, vol.~31, no.~6, pp. 2760--2776, 2023.

\bibitem{zhu2023data}
P.~Zhu, G.~Ferrari-Trecate, and N.~Geroliminis, ``Data-enabled predictive control for empty vehicle rebalancing,'' in \emph{2023 European Control Conference (ECC)}, 2023, pp. 1--6.

\bibitem{deng-phd}
G.~Deng, ``Simulation-based optimization doctoral dissertation,'' Ph.D. dissertation, University of Wisconsin-Madison, 2007.

\bibitem{AimsunManual}
\BIBentryALTinterwordspacing
Aimsun, \emph{Aimsun Next 24 User’s Manual}, aimsun next 24.0.0~ed., Barcelona, Spain, Accessed on: Month, Day, Year 2024. [Online]. [Online]. Available: \url{https://docs.aimsun.com/next/24.0.0}
\BIBentrySTDinterwordspacing

\bibitem{corsim}
L.~Owen, Y.~Zhang, L.~Rao, and G.~McHale, ``Street and traffic simulation: traffic flow simulation using corsim.'' 01 2000, pp. 1143--1147.

\bibitem{matsim}
A.~Horni, K.~Nagel, and K.~Axhausen, \emph{MATSim}.\hskip 1em plus 0.5em minus 0.4em\relax London: Ubiquity Press, 2016.

\bibitem{paramics}
\BIBentryALTinterwordspacing
P.~Kachroo and K.~Ozbay, \emph{Paramics}.\hskip 1em plus 0.5em minus 0.4em\relax Boston, MA: Springer US, 2003, pp. 293--327. [Online]. Available: \url{https://doi.org/10.1007/978-1-4419-8961-1_12}
\BIBentrySTDinterwordspacing

\bibitem{sumo}
\BIBentryALTinterwordspacing
P.~A. Lopez, M.~Behrisch, L.~Bieker-Walz, J.~Erdmann, Y.-P. Fl{\"o}tter{\"o}d, R.~Hilbrich, L.~L{\"u}cken, J.~Rummel, P.~Wagner, and E.~Wie{\ss}ner, ``Microscopic traffic simulation using sumo,'' in \emph{The 21st IEEE International Conference on Intelligent Transportation Systems}.\hskip 1em plus 0.5em minus 0.4em\relax IEEE, 2018. [Online]. Available: \url{https://elib.dlr.de/124092/}
\BIBentrySTDinterwordspacing

\bibitem{vissim}
M.~Fellendorf and P.~Vortisch, \emph{Microscopic traffic flow simulator VISSIM}, 06 2011, pp. 63--93.

\bibitem{simopt-singleinter-1}
S.~Chen and D.~J. Sun, ``An improved adaptive signal control method for isolated signalized intersection based on dynamic programming,'' \emph{IEEE Intelligent Transportation Systems Magazine}, vol.~8, no.~4, pp. 4--14, 2016.

\bibitem{simopt-singleinter-2}
M.~Miletić, B.~Kapusta, and E.~Ivanjko, ``Comparison of two approaches for preemptive traffic light control,'' in \emph{2018 International Symposium ELMAR}, 2018, pp. 57--62.

\bibitem{simopt-singleinter-3}
D.~Garg, M.~Chli, and G.~Vogiatzis, ``Deep reinforcement learning for autonomous traffic light control,'' in \emph{2018 3rd IEEE International Conference on Intelligent Transportation Engineering (ICITE)}, 2018, pp. 214--218.

\bibitem{simopt-singleinter-4}
K.~Jintamuttha, B.~Watanapa, and N.~Charoenkitkarn, ``Dynamic traffic light timing optimization model using bat algorithm,'' in \emph{2016 2nd International Conference on Control Science and Systems Engineering (ICCSSE)}, 2016, pp. 181--185.

\bibitem{simopt-singleinter-5}
E.~K.~E. Ahmed, A.~M.~A. Khalifa, and A.~Kheiri, ``Evolutionary computation for static traffic light cycle optimisation,'' in \emph{2018 International Conference on Computer, Control, Electrical, and Electronics Engineering (ICCCEEE)}, 2018, pp. 1--6.

\bibitem{simopt-lattice-1}
\BIBentryALTinterwordspacing
S.~Araghi, A.~Khosravi, D.~Creighton, and S.~Nahavandi, ``Influence of meta-heuristic optimization on the performance of adaptive interval type2-fuzzy traffic signal controllers,'' \emph{Expert Systems with Applications}, vol.~71, pp. 493--503, 2017. [Online]. Available: \url{https://www.sciencedirect.com/science/article/pii/S0957417416305966}
\BIBentrySTDinterwordspacing

\bibitem{simopt-lattice-2}
\BIBentryALTinterwordspacing
F.~Ahmed and Y.~Hawas, ``An integrated real-time traffic signal system for transit signal priority, incident detection and congestion management,'' \emph{Transportation Research Part C: Emerging Technologies}, vol.~60, pp. 52--76, 2015. [Online]. Available: \url{https://www.sciencedirect.com/science/article/pii/S0968090X15002727}
\BIBentrySTDinterwordspacing

\bibitem{simopt-lattice-3}
A.~Hajbabaie and R.~F. Benekohal, ``A program for simultaneous network signal timing optimization and traffic assignment,'' \emph{IEEE Transactions on Intelligent Transportation Systems}, vol.~16, no.~5, pp. 2573--2586, 2015.

\bibitem{simopt-lattice-4}
\BIBentryALTinterwordspacing
S.~Araghi, A.~Khosravi, and D.~Creighton, ``Intelligent cuckoo search optimized traffic signal controllers for multi-intersection network,'' \emph{Expert Systems with Applications}, vol.~42, no.~9, pp. 4422--4431, 2015. [Online]. Available: \url{https://www.sciencedirect.com/science/article/pii/S095741741500086X}
\BIBentrySTDinterwordspacing

\bibitem{simopt-lattice-5}
Z.~Li, M.~Shahidehpour, S.~Bahramirad, and A.~Khodaei, ``Optimizing traffic signal settings in smart cities,'' \emph{IEEE Transactions on Smart Grid}, vol.~8, no.~5, pp. 2382--2393, 2017.

\bibitem{simopt-roundabout-1}
S.~S. Sultan Mohiuddin~Qadri, M.~Ali~Gökçe, E.~Öner, and E.~G. Gökçe, ``Analysis of various scenarios to mitigate congestion at a signalized roundabout using microsimulation,'' in \emph{2019 Innovations in Intelligent Systems and Applications Conference (ASYU)}, 2019, pp. 1--6.

\bibitem{simopt-roundabout-2}
\BIBentryALTinterwordspacing
M.~A. Gökçe, E.~Öner, and G.~Işık, ``Traffic signal optimization with particle swarm optimization for signalized roundabouts,'' \emph{SIMULATION}, vol.~91, no.~5, pp. 456--466, 2015. [Online]. Available: \url{https://doi.org/10.1177/0037549715581473}
\BIBentrySTDinterwordspacing

\bibitem{simopt-arterial-1}
S.~Dabiri and M.~Abbas, ``Arterial traffic signal optimization using particle swarm optimization in an integrated vissim-matlab simulation environment,'' in \emph{2016 IEEE 19th International Conference on Intelligent Transportation Systems (ITSC)}, 2016, pp. 766--771.

\bibitem{simopt-arterial-2}
I.~Dakic, J.~Stevanovic, and A.~Stevanovic, ``Backpressure traffic control algorithms in field-like signal operations,'' in \emph{2015 IEEE 18th International Conference on Intelligent Transportation Systems}, 2015, pp. 137--142.

\bibitem{simopt-changsha-china}
\BIBentryALTinterwordspacing
L.~Zheng, X.~Xue, C.~Xu, and B.~Ran, ``A stochastic simulation-based optimization method for equitable and efficient network-wide signal timing under uncertainties,'' \emph{Transportation Research Part B: Methodological}, vol. 122, pp. 287--308, 2019. [Online]. Available: \url{https://www.sciencedirect.com/science/article/pii/S019126151830403X}
\BIBentrySTDinterwordspacing

\bibitem{simopt-skopje}
D.~Pavleski, D.~Koltovska-Nechoska, and E.~Ivanjko, ``Evaluation of adaptive traffic control system utopia using microscopic simulation,'' in \emph{2017 International Symposium ELMAR}, 2017, pp. 17--20.

\bibitem{simopt-bologna}
P.~T.~M. Nguyen, B.~N. Passow, and Y.~Yang, ``Improving anytime behavior for traffic signal control optimization based on nsga-ii and local search,'' in \emph{2016 International Joint Conference on Neural Networks (IJCNN)}, 2016, pp. 4611--4618.

\bibitem{simopt-lausanne}
\BIBentryALTinterwordspacing
X.~Chen, C.~Osorio, and B.~F. Santos, ``Simulation-based travel time reliable signal control,'' \emph{Transportation Science}, vol.~53, no.~2, pp. 523--544, 2019. [Online]. Available: \url{https://doi.org/10.1287/trsc.2017.0812}
\BIBentrySTDinterwordspacing

\bibitem{linearMPC}
\BIBentryALTinterwordspacing
A.~Kouvelas, M.~Saeedmanesh, and N.~Geroliminis, ``A linear formulation for model predictive perimeter traffic control in cities,'' \emph{IFAC-PapersOnLine}, vol.~50, no.~1, pp. 8543--8548, 2017, 20th IFAC World Congress. [Online]. Available: \url{https://www.sciencedirect.com/science/article/pii/S2405896317319572}
\BIBentrySTDinterwordspacing

\bibitem{willemsBehavioural}
J.~C. Willems, ``The behavioral approach to open and interconnected systems,'' \emph{IEEE Control Systems Magazine}, vol.~27, no.~6, pp. 46--99, 2007.

\bibitem{review-clustering}
\BIBentryALTinterwordspacing
A.~E. Ezugwu, A.~M. Ikotun, O.~O. Oyelade, L.~Abualigah, J.~O. Agushaka, C.~I. Eke, and A.~A. Akinyelu, ``A comprehensive survey of clustering algorithms: State-of-the-art machine learning applications, taxonomy, challenges, and future research prospects,'' \emph{Engineering Applications of Artificial Intelligence}, vol. 110, p. 104743, 2022. [Online]. Available: \url{https://www.sciencedirect.com/science/article/pii/S095219762200046X}
\BIBentrySTDinterwordspacing

\bibitem{gerolinimis2016CsnakeClustering}
\BIBentryALTinterwordspacing
M.~Saeedmanesh and N.~Geroliminis, ``Clustering of heterogeneous networks with directional flows based on “snake” similarities,'' \emph{Transportation Research Part B: Methodological}, vol.~91, pp. 250--269, 2016. [Online]. Available: \url{https://www.sciencedirect.com/science/article/pii/S0191261515302605}
\BIBentrySTDinterwordspacing

\bibitem{zhanfeng:2001:single_loop_detector}
Z.~Jia, C.~Chen, B.~Coifman, and P.~Varaiya, ``The pems algorithms for accurate, real-time estimates of g-factors and speeds from single-loop detectors,'' in \emph{ITSC 2001. 2001 IEEE Intelligent Transportation Systems. Proceedings (Cat. No.01TH8585)}, 2001, pp. 536--541.

\bibitem{lagrangian-estimation}
\BIBentryALTinterwordspacing
J.~C. Herrera and A.~M. Bayen, ``Incorporation of lagrangian measurements in freeway traffic state estimation,'' \emph{Transportation Research Part B: Methodological}, vol.~44, no.~4, pp. 460--481, 2010. [Online]. Available: \url{https://www.sciencedirect.com/science/article/pii/S0191261509001222}
\BIBentrySTDinterwordspacing

\bibitem{claudel:2008:lagrangian-eulerian_sensing}
C.~G. Claudel and A.~M. Bayen, ``Guaranteed bounds for traffic flow parameters estimation using mixed lagrangian-eulerian sensing,'' in \emph{2008 46th Annual Allerton Conference on Communication, Control, and Computing}, 2008, pp. 636--645.

\bibitem{kumar_short-term_2015}
\BIBentryALTinterwordspacing
S.~V. Kumar and L.~Vanajakshi, ``\BIBforeignlanguage{en}{Short-term traffic flow prediction using seasonal {ARIMA} model with limited input data},'' \emph{\BIBforeignlanguage{en}{European Transport Research Review}}, vol.~7, no.~3, p.~21, Jun. 2015. [Online]. Available: \url{https://doi.org/10.1007/s12544-015-0170-8}
\BIBentrySTDinterwordspacing

\bibitem{foulds:2013:origin_destination_matrix}
L.~R. Foulds, H.~A. {do Nascimento}, I.~C. Calixto, B.~R. Hall, and H.~Longo, ``A fuzzy set-based approach to origin–destination matrix estimation in urban traffic networks with imprecise data,'' \emph{European Journal of Operational Research}, vol. 231, no.~1, pp. 190--201, 2013.

\bibitem{olmos:2018:urban_traffic}
L.~E. Olmos, S.~Çolak, S.~Shafiei, M.~Saberi, and M.~C. González, ``Macroscopic dynamics and the collapse of urban traffic,'' \emph{Proceedings of the National Academy of Sciences}, vol. 115, no.~50, pp. 12\,654--12\,661, 2018.

\bibitem{mixed-integer-phase-specs}
\BIBentryALTinterwordspacing
N.~H. Gartner, J.~D.~C. Little, and H.~Gabbay, ``Optimization of traffic signal settings by mixed-integer linear programming: Part i: The network coordination problem,'' \emph{Transportation Science}, vol.~9, no.~4, pp. 321--343, 1975. [Online]. Available: \url{http://www.jstor.org/stable/25767802}
\BIBentrySTDinterwordspacing

\bibitem{willems1986timeII}
\BIBentryALTinterwordspacing
J.~C. Willems, ``From time series to linear system - part ii. exact modelling,'' \emph{Automatica}, vol.~22, pp. 675--694, 1986. [Online]. Available: \url{https://api.semanticscholar.org/CorpusID:7620875}
\BIBentrySTDinterwordspacing

\bibitem{willems1986timeIII}
\BIBentryALTinterwordspacing
------, ``From time series to linear system—part iii: Approximate modelling,'' \emph{Automatica}, vol.~23, no.~1, pp. 87--115, 1987. [Online]. Available: \url{https://www.sciencedirect.com/science/article/pii/0005109887901208}
\BIBentrySTDinterwordspacing

\bibitem{first-order-traffic-model}
\BIBentryALTinterwordspacing
M.~Kontorinaki, A.~Spiliopoulou, C.~Roncoli, and M.~Papageorgiou, ``First-order traffic flow models incorporating capacity drop: Overview and real-data validation,'' \emph{Transportation Research Part B: Methodological}, vol. 106, pp. 52--75, 2017. [Online]. Available: \url{https://www.sciencedirect.com/science/article/pii/S0191261516302351}
\BIBentrySTDinterwordspacing

\bibitem{second-order-traffic-model}
\BIBentryALTinterwordspacing
C.~Wagner, C.~Hoffmann, R.~Sollacher, J.~Wagenhuber, and B.~Sch\"urmann, ``Second-order continuum traffic flow model,'' \emph{Phys. Rev. E}, vol.~54, pp. 5073--5085, Nov 1996. [Online]. Available: \url{https://link.aps.org/doi/10.1103/PhysRevE.54.5073}
\BIBentrySTDinterwordspacing

\bibitem{DeePC-regularized}
F.~Dörfler, J.~Coulson, and I.~Markovsky, ``Bridging direct and indirect data-driven control formulations via regularizations and relaxations,'' \emph{IEEE Transactions on Automatic Control}, vol.~68, no.~2, pp. 883--897, 2023.

\bibitem{transcality}
\BIBentryALTinterwordspacing
E.~Spin-Off. (2024) Transcality. [Online]. Available: \url{https://transcality.com/}
\BIBentrySTDinterwordspacing

\bibitem{lukas-two-layer}
L.~Amb{\"u}hl and K.~Axhausen, ``\BIBforeignlanguage{English (US)}{A case study of zurich{\textquoteright}s two-layered perimeter control},'' 2018, transportation Research Arena (TRA) ; Conference date: 01-04-2018.

\bibitem{Split-as-a-pro}
Y.~Tang, C.~Cenedese, A.~Rimoldi, F.~Dörfler, J.~Lygeros, and A.~Padoan, ``Split-as-a-pro: behavioral control via operator splitting and alternating projections,'' in \emph{2025 European Control Conference (ECC)}, 2025, pp. 1495--1501.

\bibitem{willems1989models}
\BIBentryALTinterwordspacing
J.~C. Willems, \emph{Models for Dynamics}.\hskip 1em plus 0.5em minus 0.4em\relax Wiesbaden: Vieweg+Teubner Verlag, 1989, pp. 171--269. [Online]. Available: \url{https://doi.org/10.1007/978-3-322-96657-5_5}
\BIBentrySTDinterwordspacing

\bibitem{willems2005note}
\BIBentryALTinterwordspacing
J.~C. Willems, P.~Rapisarda, I.~Markovsky, and B.~L. {De Moor}, ``A note on persistency of excitation,'' \emph{Systems \& Control Letters}, vol.~54, no.~4, pp. 325--329, 2005. [Online]. Available: \url{https://www.sciencedirect.com/science/article/pii/S0167691104001434}
\BIBentrySTDinterwordspacing

\bibitem{markovsky2022identifiability}
I.~Markovsky and F.~Dörfler, ``Identifiability in the behavioral setting,'' \emph{IEEE Transactions on Automatic Control}, vol.~68, no.~3, pp. 1667--1677, 2023.

\bibitem{berberich2022linear}
J.~Berberich, J.~Köhler, M.~Muller, and F.~Allgöwer, ``Linear tracking mpc for nonlinear systems part ii: The data-driven case,'' \emph{IEEE Transactions on Automatic Control}, vol.~67, pp. 1--1, 09 2022.

\bibitem{padoan2023data}
A.~Padoan, F.~Dörfler, and J.~Lygeros, ``\BIBforeignlanguage{en}{Data-driven representations of conical, convex, and affine behaviors},'' in \emph{\BIBforeignlanguage{en}{2023 62nd IEEE Conference on Decision and Control (CDC)}}.\hskip 1em plus 0.5em minus 0.4em\relax Singapore: IEEE, 2023, Conference Paper, pp. 596 -- 601, 62nd IEEE Conference on Decision and Control (CDC 2023); Conference Location: Singapore; Conference Date: December 13-15, 2023.

\end{thebibliography}
\end{document}